\documentclass[pdflatex,sn-mathphys-num]{sn-jnl}

\usepackage{graphicx}%
\usepackage{multirow}%
\usepackage{amsmath,amssymb,amsfonts}%
\usepackage{amsthm}%
\usepackage{mathrsfs}%
\usepackage[title]{appendix}%
\usepackage{xcolor}%
\usepackage{textcomp}%
\usepackage{manyfoot}%
\usepackage{booktabs}%
\usepackage{algorithm}%
\usepackage{algorithmicx}%
\usepackage{algpseudocode}%
\usepackage{listings}%
\usepackage{blindtext}%
\usepackage{hyperref}%
\usepackage{subcaption}%
\usepackage{caption}
\usepackage{adjustbox}
\usepackage{subcaption}
\usepackage{natbib}
\usepackage{bm}
\usepackage{booktabs}
\usepackage{tabularx}
\usepackage{float}
\usepackage{rotating}
\usepackage{listings}
\usepackage{xcolor}
\usepackage{pdflscape}
\usepackage{longtable}
\theoremstyle{thmstyleone}%
\theoremstyle{thmstyletwo}%

\theoremstyle{thmstylethree}%

\begin{document}

\title[Multilayer Perceptron Neural Network Models in Asset Pricing: An Empirical Study on Large-Cap US Stocks
]{Multilayer Perceptron Neural Network Models in Asset Pricing: An Empirical Study on Large-Cap US Stocks}

\author*[1]{\fnm{Shanyan}\sur{Lai}}\email{shanyan.lai@york.ac.uk, annieyanyan125@gamil.com}



\affil*[1]{\orgdiv{Department of Economics and Related Studies}, \orgname{Univiersity of York}, \orgaddress{\street{Heslington}, \city{York}, \postcode{YO10 5DD}, \country{UK}}}




\abstract{In this study, MLP models with dynamic structure are applied to factor models for asset pricing tasks. Concretely, the MLP pyramid model structure was employed on firm characteristic-sorted portfolio factors for modelling the large-cap US stocks. It was further developed as a practical factor investing strategy based on the predictions. The main findings were evaluated from 2 angles: model predictive power and backtesting performance, which were compared for the periods with and without COVID-19. The empirical results indicated that, given the constraints of the data size, the MLP models no longer perform ‘deeper, better’ in terms of predictive power, whereas the proposed MLP models with 2 and 3 hidden layers have greater flexibility in modelling the factors in this case. This study also verified the idea from previous work that MLP models for factor investing are more meaningful for downside risk control than for pursuing absolute annual returns.\\

Note: An earlier version is available at Zenodo (DOI: 10.5281/zenodo.15333718)}

\keywords{Asset Pricing, MLP, Neural Network, Factor Investment}



\maketitle

\vspace{0.5em}
\noindent
\textbf{Current Version:} June 2, 2026\\
Following insightful comments from Dr Mark Hallam and Dr Chrysovalantis Vaslakis, this version introduces several improvements. First, I correct an error in the out-of-sample $R^2$ calculation caused by the `r2\_score()' function in scikit-learn, which incorrectly aligns predicted returns when computing the denominator. This led to a systematic underestimation of the average OOS $R^2$. I address this by manually computing the mean squared error (MSE) and OOS $R^2$. The denominator now correctly uses the historical mean based on both training and validation samples, consistent with standard asset pricing literature (e.g., Campbell and Thompson, 2008). The updated code is provided in the Appendix. Second, I incorporate a heteroskedasticity and autocorrelation consistent (HAC) estimator into the Diebold-Mariano (DM) test to account for serial correlation in the difference between the two models’ forecast errors. Third, I conduct factor importance analysis and a dynamic transaction-cost robustness examination on the portfolio-wise backtesting. I am grateful to Dr Mark Hallam and Dr Chrysovalantis Vaslakis for their valuable feedback. All remaining errors are my own.

\vspace{1em}

\section{Introduction}\label{sec:introduction}
Since the beginning of the Capital Asset Pricing Model (CAPM) \citep{Treynor1961MarketRisk,sharpe1964capital,lintner1965security,mossin1966equilibrium}, the empirical asset pricing research has been dominated by the linear factor framework. This includes the most notable Arbitrage Pricing Theory (APT) \citep{Ross1978TheCAPM} models, such as the Fama-French three-factor model \citep{Fama1993CommonBonds}. Their linear factor framework modelled the multi-dimensional risk factors such as size and value. Although these models established the foundation for understanding risk premiums, their linear assumptions become a challenge later when an increasing number of factors are explored by enthusiastic researchers. The factor exploration was separated into two groups: researchers in academia were attempting to explore more factors that can give the asset excess returns a better interpretation, which shows as lower or insignificant alpha, while the practitioners were developing the factors for searching for abnormal returns ($\alpha$, also known as anomaly) beyond factors, which shows as higher alpha. Due to the motivation from these two groups, the number of factors explored increased dramatically. Since then, four hundred more factors have been explored for stock pricing or factor investing \citep{Harvey2019AZoo}. In this work, a ‘factor zoo’ was developed. However, a challenge was that the increasing complexity of factors led to a ‘dimensionality disaster’ for the naïve linear model.  Although in the early stage, researchers noticed the non-linearity of the factor models, for instance, the consumption CAPM model \citep{breeden1979intertemporal} and its advanced forms \citep{Bansal2004RisksPuzzles,Lettau2001ResurrectingTime-varying} or ‘disaster as a factor’ \citep{Gabaix2012VariableMacro-finance}, none of them attempt to solve the issue of dimensional reduction for factors. Nonetheless, the development of machine learning (ML) algorithms over the past two decades has sparked a revolution in factor models for asset pricing, shedding light on the dimensionality-reduction task in factor models. Since 2017, ML algorithms have attracted significant interest among researchers in finance and economics due to their remarkable predictive power and flexibility. For example, \citet{Baek2020MachineMarkets,Chang2021PairsLearning,Sarmento2020EnhancingLearning} all utilized the ML technique for constructing financial trading strategies. In the asset pricing domain, \citet{Gu2020EmpiricalLearning} tested the most classic ML algorithms such as Elastic Net, random forest (RF), gradient-boosted regression trees and Multilayer Perceptron (MLP), benchmarked with OLS, principal components regression (PCR), and partial least square (PLS) on asset pricing tasks of U.S. stocks. It is considered the seminal work of ML application for asset pricing (AP) assignment, which made it the most cited paper in that domain. Their models are also known as ‘GKX2020’ in later papers \citep{Gu2021AutoencoderModels}. They found that the MLP model with 3 hidden layers with the fixed pyramid structure (32,16,8 neurons in each hidden layer respectively) has the highest out-of-sample average fit ($R^2$). The study also managed the balance of computational costs of ML algorithms and the predictive fit of the ML models, which gave a taste of ML application on asset pricing. It is a great work that inspires practitioners to use machine learning techniques with an applicable computational cost to improve their factor investing efficiency and profitability. Building on this seminal work, subsequent studies have further extended ML applications in asset pricing by incorporating alternative architectures and penalty functions  \citep{Bagnara2022AssetReview,Giglio2022FactorPricing,Nagel2021MachinePricing,Wang2021CryptocurrenciesLearning}.\\

Their research sparked my curiosity to pursue research on ML-based factor investing strategies and to explore the application of deep MLP models in the asset pricing domain. Specifically, how to efficiently use the ML technique to solve the `dimensional disaster’ of the `factor zoo’ and increase the predictability of the factor models via MLP structures.  The main research questions in this chapter are:
\begin{itemize}
  \item ‘Is that ML model for factor investing under suspicion of pure data mining without any economic or financial meanings, or easy to be trapped by data snooping?’\\
  \item ‘Is there any benefit for using the firm characteristics sorted portfolio factors instead of individual firm characteristics for pricing the individual stocks?’\\
  \item ‘How can we decide the suitable MLP structure instead of the fixed pre-assumed ones?’\\
  \item ‘According to the general MLP theory of “Deeper, Better”, does that still hold in asset pricing models?’\\
  \item ‘Could the MLP models' forecasting performance maintain the stability under the extreme market conditions (e.g. extreme market fluctuation caused by COVID-19)?’\\
\end{itemize}
To answer these questions, the classic MLP models with dynamic structure are examined on the firm characteristics- sorted portfolio factors. As one of the developers of the GKX2020’s work, this study attempts to upgrade four aspects of the GKX2020’s MLP models. 1) increasing the economic meanings of the MLP structure for factor asset pricing models. It employs 182 factors organized by firm characteristic-sorted returns from \citet{AndrewY.Zimmermann2020OpenPricing}. They serve as predictors, or features in ML terminology, to replace GKX2020’s approximately 900 predictors, which extended from the 94 firm characteristics and 8 macroeconomic indicators via Kronecker product, and industry dummy variables. 2) I select 420 large capital stocks with the assumption of  ‘going concern’ from NYSE and NASDAQ, which covers 21.38\% of aggregate market capitalization. Concretely, the ‘going concern’ applies, as data are available for a full testing period to circumvent the systematic risk of abrupt delisting, which could be noticeably harmful for practitioners. 3) employing the dynamic MLP structure introduced by \citet{Coqueret2020MachineVersion}, where the MLP structure varies according to its depth. 4) Comparing MLP models' performance under consideration of the COVID-19 effect to further evaluate the model suitability and investment performance for factor models. \\

The main contributions of this chapter appear in proposing the innovative MLP structure for asset pricing factor models, moderating the overfitting issue that existed in GKX2020 models via the firm characteristic-sorted factors, introducing a novel factor investing strategy based on classic MLP models, exploring the truth that how the COVID-19 affects the model fitness and factor investing performance on the large-cap U.S. stocks and verifying some results of the previous factor investing literature.  \\

This chapter contains six main sections. Section~\ref{sec:related works} presents the literature review.  Section ~\ref{sec:Data Description} presents the data that have been used, while Section ~\ref{sec:models_ch2} discusses the models employed in this chapter. Section ~\ref{sec:Empirical_Results_ch2} shows the empirical experiment and results, and Section ~\ref{sec:conclusion_ch2} gives the conclusion and further discussion.\\

\section[Related works of Chapter 2]{Related works}\label{sec:related works}
Before \citet{Treynor1961MarketRisk,sharpe1964capital,lintner1965security,mossin1966equilibrium} created the CAPM model, people had fuzzy notions about how the risks could affect the expected returns of assets. He proposed a single-factor linear model for interpreting expected stock returns via a factor generated from the difference between expected market return and the risk-free rate. Specifically, it can be written as:
\begin{equation}
E(R_i) = R_f + \beta_i \left(E(R_m) - R_f\right)
\end{equation}
Where $E(R_i)$ is the expected return of asset $i$, $R_f$ is the risk-free rate, $\beta_i$ is the asset $i$'s factor loading, representing its sensitivity to the market factor, and $E(R_m)$ is the expected return of the market. Although it is the cornerstone of asset pricing theory, people cast doubt on the market factor's explainability for individual assets or portfolios. Thus, \citet{Ross1978TheCAPM} insists asset returns were driven by the combined effect of factors, hence proposed a generalized factor model’s framework:
\begin{equation}
R_i = \alpha_i + \beta_{i1} F_1 + \beta_{i2} F_2 + \cdots + \beta_{ik} F_k + \epsilon_i
\end{equation}
Where $F_1$ to $F_k$ are factors, $\beta_{i1}$ to $\beta_{ik}$ are factor loadings which measure the sensitivity of the factors. $\alpha_i$ is the intercept of the linear model, which represents the pricing error (namely, abnormal return for investors or anomaly in the empirical asset pricing context), and the residual $\epsilon_i$ represents the firm-specific risks. This theory suggests that if pricing errors exist, investors can take advantage of the existing market arbitrage opportunities. It is known as the Arbitrage Pricing Theory (APT). The profit from the arbitrage is equal to $\alpha_i$, which means that the returns cannot be explained fully by the factors. The model frame relaxed the single-factor assumption of CAPM, thereby improving the model's flexibility. However, with such a frame, scholars still struggle with questions such as what factors should be present and how many factors should appear in a single frame. The most well-known multi-factor model is the Fama-French 3-factor model (FF3) \citep{Fama1993CommonBonds}. It deployed the market factor together with size and value (book-to-market ratio) factors to the APT frame, which  improved the explanation of the sorted portfolios. The FF3 model led to an increased enthusiasm for research on factor exploration afterwards. Researchers developed new factors and models based on the APT framework, and had the massive competitive models on their innovative factors and factor combinations, which pursue the insignificant $\alpha$ for proving a better interpretability of the factors. For example, \citet{Carhart1997OnPerformance} extended FF3 to 4 factors by adding the momentum factor, while \citet{Novy-Marx2013ThePremium} proposed adding a gross profit factor to the FF3 frame. Further development led to the Fama-French 5-factor (FF5) model \citep{Fama2015AModel}, which added profitability and investment factors to the FF3.  \citet{Hou2015DigestingApproach} optimized the FF5’s investment factor by Tobin's Q theory \citep{Tobin1969ATheory}, which utilized the ratio of the market value of the company and replaced the cost of capital to measure the quality of investment. They offered great ideas for interpreting asset or portfolio returns. However, with continually emerging factors and countless combinations among them, it remains a key research challenge to assess the reliability of different factors. There are 400+ factors that play a role in stock pricing \citep{Hou2020ReplicatingAnomalies}. Handpicking factors may cause the ‘omitted factor’ issue, where important variables are intentionally or unintentionally neglected. However, inserting all factors in the unpenalized linear APT frame causes the ‘dimensionality disaster’ since it leads to problems, such as overfitting and instability of estimation, which cause the false factor detection and enlarged pricing errors. \\ 

Therefore, some researchers devoted themselves to developing non-linear model forms or adding constraints to the linear model form to satisfy the increasing factor size. \citet{Stock2002ForecastingPredictors} introduced the use of principal components analysis (PCA) to extract factors from a large number of original input factors. PCA projects the predictors into the new coordinate system via a linear transformation and reorganizes the predictors into components. It extracts components that explain the most variance as the new factors for the APT frame. The entire process is named principal components regression (PCR), and the predefined parameter, which indicates how many new factors should be retained, is known as the hyperparameter of PCR. PCA is the most widely applied technique for dimension reduction tasks in different domains, but it has limitations, such as the linear combination assumption of factors, and new factors subtracted from the original factors may not be easy to interpret and require normalization. Later, \citet{Kelly2019CharacteristicsReturn} utilised the instrumented PCA (IPCA), which was upgraded from the PCA for factor extraction. IPCA employs a dynamic process on PCA that analyses only a small batch of the dataset each time and updates the factors gradually, ultimately converging on the optimal factors for the full dataset. While IPCA optimizes the extraction process for large-scale factors through incremental updates, it does not resolve PCA's core theoretical weaknesses. Concretely, the extracted latent factors remain linear combinations of the original predictors, and the model still struggles with low interpretability and the loss of information that does not align with maximum variance. In contrast, the Neural Network approach, MLP models, proposed in this chapter, addresses these limitations by employing non-linear activation functions (e.g. ReLU) for a better mapping of original factors. This allows for a more flexible dimensionality reduction process that captures complex factor structures, therefore minimizing the loss of financially or economically significant information that traditional linear methods usually overlook.\\

Furthermore, the Least Absolute Shrinkage and Selection Operator (LASSO) and ridge regression are two commonly applied constraints for linear models, known as L1 and L2 regularization, respectively. They are employed for controlling the oversized factors in literature. \citet{Kelly2013MarketValues} proposed a new pricing method via market expectation, which relies on LASSO for noise reduction.  \citet{Feng2020TamingFactors} introduced the concept of ‘factor zoo’ and applied an LASSO-based examination method. \citet{Chinco2019SparseReturns} has also verified LASSO’s ability to denoise datasets. They proved that LASSO regression can significantly reduce the noise created by ineffective factors and improve model fitness and robustness. Whereas the classic LASSO has limitations, such as extracting only one factor from a group of highly correlated factors, its limitation is that it cannot jointly measure multiple correlated or equally important factors. This leads to information loss. Therefore, some papers turn to the constraint of classic ridge regression. For example, \citet{Giglio2021AssetFactors} applied ridge regression to the study of how investor beliefs have an impact on portfolio decisions from a survey of wealthy retail investors. It matched the survey data to investors' portfolios, trading activities, and online behaviors data. It proved that the $l2$ penalty can identify which factors should be the prioritized factors, identifying potential multicollinearity and moderating model overfitting in the dataset. Nonetheless, classic ridge regression has limitations, such as it is impossible to eliminate the ineffective factors, since it never fails to shrink the coefficients to zero, and it penalizes all correlated factors when it addresses the multicollinearity, which makes it difficult to interpret the individual effect of factors with high correlation. Moreover, for sparse factors, where only a few variables have significant effects, ridge regression shows less priority than LASSO due to the effect of non-zero coefficient shrinkage. Both classic LASSO and ridge regression face the same limitations of being computationally expensive and time-consuming for searching hyperparameters, and are therefore not suitable for non-linear data and sensitive to noise. \citet{Yuan2006ModelVariables,Freyberger2020DissectingNonparametrically,Bertelsen2022TheZoo} later extended LASSO to group LASSO, adaptive group LASSO and prior adaptive LASSO, respectively, for moderating the aforementioned issues of classic LASSO, but it still has not been fundamentally resolved. \\

Fortunately, the development of machine learning (ML) methods has led to the progression of the asset pricing factor model structures. The earliest neural network (NN) models for financial applications can be traced back to the 1990s, where \citet{Hutchinson1994ANetworks} used a new non-parametric NN model for simulating Black-Scholes option prices. They proved that the model recovered the Black-Scholes formula from a two-year training set of daily option prices and that it can be used on both out-of-sample price and delta-hedge options, which gave hope to subsequent research that NN models may overcome some limitations of traditional econometric methods. Later, \citet{Heaton2016DeepFinance} tested classic deep NN structures on financial time series prediction tasks, from MLP to Autoencoder, and found that deep neural network structures significantly improved the performance of financial series prediction. Whereas, unlike financial data, which includes higher frequency options, such as daily data and hourly data with a larger number of observations, stock pricing typically employs monthly data, which aligns with the release frequency of financial reports. Thus, the limited data size is a challenge for the research of stock pricing via ML.\\

The seminal work of asset pricing in ML models is \citet{Gu2020EmpiricalLearning}. They examined classic ML methods, including MLP, decision trees, and random forests, on 30,000 U.S. stocks over 60 years, from March 1957 to December 2016. The work deployed 94 firm characteristics \citep{Green2017TheReturns}, 8 macroeconomic indicators \citep{Welch2008APrediction} and 74 industry dummy variables as factors pricing on individual stocks' excess return series. They also applied the extending window method to their work, which means they re-estimated the parameters every 12 months after the first in-sample window. And it enlarged the factor dimension via the Kronecker product method to ensure the data size for the ML frameworks. The final factor dimension for each stock is approximately 850-900. Their fixed-neuron pyramidical MLP structures borrowed the idea of \citet{Masters1993PracticalC++}. Their models in later works are also named ‘GKX2020’ \citet{Gu2020EmpiricalLearning}. They found MLP with 3-4 hidden layers outperformed all alternative benchmark models (e.g. random forest, MLP with 1-2 and 5 hidden layers and Elastic net) by comparing the average out-of-sample (OOS) fitness indicator $R^2$. However, as an early work of asset pricing, it has inherent limitations: they adopted the cross-sectional rank-winsorization method for denoising the noise that exists in the firm characteristics, but the method artificially truncates fat tails, stripping the model of critical signals needed to predict extreme market events or crash risks. Also, they did not consider the transaction costs, such as variable borrowing rates for short positions and liquidity risks. From a modelling perspective, they left a question of why the simple fixed pyramidical MLP structures can be the optimal selections for stock pricing, as well as it does not present the difference between the MLP models in large-cap stocks and small-cap stocks. Inspired by their work, I apply the firm characteristics-sorted stock return portfolios as factors for moderating their noisy factor data, and examining these factors on large-cap stocks to prevent liquidity risks in practice. \\

By following GKX2020's footpath, \citet{Wang2021CryptocurrenciesLearning} applied GKX2020 ML models on cryptocurrencies, and their empirical results agree with the findings that GKX2020 MLP models, even with different asset categories and different configurations of factors. This extended investigation proved that MLP models generally improve the model fitness and forecasting performance in the context of asset pricing. However, they emphasise the excellence in projecting the factors' nonlinear influence on asset prices, but provide no backtesting to further support investment practice.  \citet{Gu2021AutoencoderModels} extended the MLP models of GKX2020 by adopting an MLP autoencoder architecture for dimensionality reduction of factors, which was inspired by \citet{Kelly2019CharacteristicsReturn}'s instrumented principal component analysis (IPCA) model. Concretely, they employed a dual-MLP structure for pricing the U.S. stock returns, which was used in \citet{Gu2020EmpiricalLearning}. In their model structure, the left side is an MLP autoencoder (conditional autoencoder) structure, which maps high-dimensional lagged firm characteristics to lower-dimensional factor loadings, while the right side is another MLP autoencoder extracting the information from firm characteristic-sorted portfolios as input factors and reducing the dimension of these factors. They linearly mapped the outputs of these two MLP autoencoders to price the individual stock returns. Although their model achieved an annualized value-weighted Sharpe ratio of 1.53 in long-short value-weighted portfolio back testing, the predictive power of the model was nearly identical to pure machine learning prediction models that do not assume a factor structure, even if their model improved the model's economic interpretability in the ML context. Moreover, their model doubled the parameter size, which increases the risk of model overfitting and unnecessary computational costs. Thus, the proposed model adopts a simplified dynamic structure to improve the computational efficiency and moderate the overfitting risk.\\

Furthermore, \citet{Avramov2021MachinePredictability} provided the first large-scale examination of whether ML signals in stock-return prediction retain economic significance when realistic investment frictions are imposed. By comparing four prominent ML approaches: \citet{Gu2020EmpiricalLearning}'s three-layer neural network (GKX2020), \citet{Chen2024DeepPricing}'s adversarial deep-learning framework for stochastic discount factor estimation (CPZ), \citet{Kelly2019CharacteristicsReturn}'s IPCA model, and \citet{Gu2021AutoencoderModels}'s conditional autoencoder (CA), they constructed value-weighted long-short portfolios from 1987 to 2017. Although the unrestricted full-sample strategies deliver strong raw and risk-adjusted returns (e.g., FF6 alphas of 0.92\% to 1.87\% per month), imposing common economic restrictions, excluding microcaps, non-rated firms, or stocks around credit-rating downgrades, dramatically attenuates profitability, especially for deep-learning methods; many $\alpha$s lose statistical significance. They further document extremely high portfolio turnover (87\% to 163\% monthly), extreme positions in the tangency portfolio, and the concentration of gains in high limits-to-arbitrage states, yet they note that ML signals remain profitable in long-only positions, recent years, and crisis periods, and are economically interpretable as they largely aggregate well-known anomaly characteristics. This motivates backtesting long-only portfolios in this study. The paper concluded by proposing a practical protocol for future ML research that emphasized value-weighting, transaction costs, long positions, and recent-sample performance, thereby bridging the gap between statistical predictability and implementable investment value. Similarly, \citet{Giglio2022FactorPricing} conducted a comprehensive literature review of ML-based asset pricing research. They organized the literature around core empirical objectives, such as estimating expected returns, latent factors and risk exposures, risk premia, the stochastic discount factor (SDF), model comparison, and alpha testing. They contrasted classical low-dimensional approaches (e.g., time series and cross-sectional regressions, Fama-MacBeth procedures, and the Gibbons-Ross-Shanken test) with ML models designed for high-dimensional individual stocks or firm characteristics. They highlighted innovations such as instrumented principal components analysis (IPCA), risk-premia PCA, conditional autoencoders, penalized SDF estimation, double ML methods for factor selection, and Bayesian model comparison, while emphasizing regularization, overfitting controls, and asymptotic theory under long time series (large T), large cross-sectional (large N), or extended panel (large T and N) regimes. The survey emphasizes how ML methods address the challenges of high-dimensional predictors, time-varying loadings, weak factors and omitted variables, and provides a rigorous guide for researchers seeking to harness these instruments for asset pricing discoveries in the background of ML. It also identifies promising directions for future work at the intersection of economic theory, statistical efficiency, and practical implementation.\\
 
In addition, \citet{Kelly2023FinancialLearning} also contributed a general literature review on machine learning in financial markets. It summarized the core challenges of asset pricing, for instance, large information sets, ambiguous functional forms, and the low signal-to-noise environment that distinguishes finance from forecasting tasks in other domains. They organize the discussion around four pillars: the case for financial ML (emphasizing prices as predictions, the benefits of economic structure amid small data and weak signals, and the complementary roles of structural hypothesis testing versus atheoretical prediction models); the virtues of complex, overparameterized models (introducing tools like ridge regression with generated features and random matrix theory to show that, under realistic mis-specification, higher model complexity improves out-of-sample $R^2$, expected returns, volatility, and Sharpe ratios via the `complexity wedge'); return prediction (covering data construction, recursive cross-validation designs, benchmarks such as Fama-MacBeth linear models, penalized linear regressions, dimension reduction, decision trees, vanilla and sophisticated neural networks, and models for alternative data); and extensions to risk-return trade-offs (unconditional/conditional/complex factor models, high-frequency settings, and alphas) and optimal portfolios (plug-in versus integrated estimation, maximum Sharpe ratio regression, SDF estimation, and reinforcement learning with trading costs). Throughout, the authors highlight empirical gains from ML methods while stressing best practices for regularization, out-of-sample evaluation, and the integration of economic theory, offering a practical guide for researchers seeking robust, implementable discoveries at the intersection of financial asset pricing and ML.\\

\section[Data Description of Chapter 2]{Data Description}\label{sec:Data Description} 
The responses, or labels in ML terminology, used in this research contain 420 large-capital stock excess returns from January 1957 to December 2021. These stocks come from the National Association of Securities Dealers Automated Quotations (NASDAQ) and the New York Stock Exchange (NYSE), which are derived from CRSP of Wharton Research Data Services (WRDS). These 420 stocks originally are filtered from the top 15\% of the largest market capitalization (MC), which covers 85\% of total MC for the two markets. They satisfy the conditions of having no missing data in the testing period and less than 50\% of values missing during the training period, hence they can meet the requirement of the ‘going concern’ \citep{Fama1970EfficientMarkets} and data representability. Selecting the large-cap stocks is for avoiding the liquidity risks of market trading, which happens as investors cannot close the positions to stop loss or confirm the trading profit since no counterparty takes their positions. And as a benefit of the assumption of ‘too-big-to-fail’ \citep{Mishkin2006HowBailouts} and ‘going concern’, this setting is more realistic for practitioners since it protects the investors from systemic risks such as companies that are delisted unexpectedly due to insolvency (e.g. Lehman Brothers) or other unexpected issues. The overall coverage of market capitalization from these stocks is 21.38\% in the U.S. stock markets of the NASDAQ and the NYSE. Additionally, excessive missing values can significantly compromise model reliability. They reduce the effective sample size, enlarge estimation variance, and introduce bias if the missingness is not completely at random \citep{little2019statistical}. In this case, missing observations may disrupt the temporal dependency structure,  affect predictive performance and cause the unstable out-of-sample evaluation. Thus, I remove the factors with excessively missing values. The risk-free rate for constructing the excess returns and evaluating the Sharpe ratio (SR) and Sortino ratio (SO) is derived from the Kenneth R. French Data Library. \\

The predictors, also known as features for the ML terminology, are organized by 182 explored firm characteristic-sorted portfolio factors provided by \citet{AndrewY.Zimmermann2020OpenPricing} in their Open Source Asset Pricing (OSAP) database, with at least 60\% of available values in the training period. The 60\% data availability threshold during the training period ensures the historical continuity of the time series from 1957 to 2021. Although the full OSAP dataset provided 202 factors that authors confirmed as high-quality factors, numerous of  them, especially those labeled as `indirect' or `insignificant' ones, suffer from massive data gaps in the early decades. The selected 182 factors provides a high-density information set that covers nearly all `Clear Predictors' across six core clusters: Momentum, Value, Investment, Profitability, Intangibles, and Trading Frictions.\\ 

\begin{figure}[htbp!]
\begin{center}
\includegraphics[width=0.9\columnwidth]{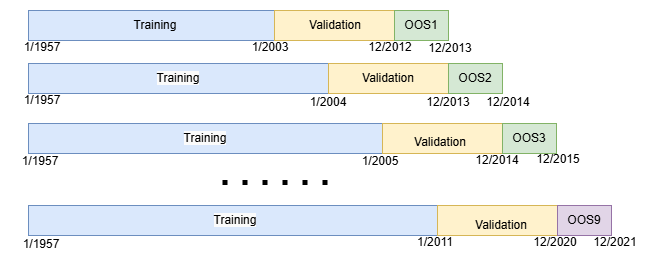}
\end{center}
\caption[The mechanism of the extended window method.]{The mechanism of the extended window method. The blue bars show the sample section of the training period with an extended base. The yellow bars represent the validation period on a fixed rolling basis. The `$OOS (i)$’, $i=1,2,3,\dots$ are the out-of-sample iterations, and each row points to one iteration. It moves forward following the timeline. The black dots between the 3rd line and the 4th line are the ellipses, which indicate that more iterations are involved.}
\label{fig:data_des1}
\end{figure}

There are three reasons for employing the firm characteristic-sorted portfolios as factors instead of firm characteristics directly as input factors, which is adopted by literature such as \citet{Gu2020EmpiricalLearning,Ma2023AttentionApproach}. Firstly, compared with the individual firm characteristics as factors, the sorted portfolio factors consider the effects between assets, stocks in this case, and cross-sectionally evaluate the asset excess returns, which improves the economic interpretability of the factor model. Secondly, compared with the factor method of GKX2020 \citep{Gu2020EmpiricalLearning}, the factor method in this study highly reduce the dimension of the input, which potentially causes model overfitting. Thirdly, if firm characteristics are used directly as factors, the data exhibit large disparities across variables, and even after standardization, such heterogeneity remains problematic for model estimation. The details of these 182 factors are presented in the appendix of Chapter 2. The entire data set is split into training (70\% of the entire data length), validation (15\% of the entire data length) and testing periods (15\% of the entire data length for Period 2112) as shown in Table~\ref{fig:data_des1}. It is consistent with the general machine learning data split rule of thumb that the training data length should normally exceed 70\% of the entire data length \citep{Goodfellow2016DeepLearning}. And the validation window and testing period in this case are adjusted to satisfy the research target of the COVID-19 impact investigation. The training window is on an extension base with a step size of 12 months, while the validation window is fixed with 119 observations. The training window, together with a fixed validation window, moves forward 12 months at a time until it exhausts the entire data length. Thus, the out-of-sample, which contains the full testing period, has 9 years with one update per year. This follows the convention of \citet{Gu2020EmpiricalLearning}, and this approach ensures all models operate on the identical information set and avoids the excessive computational workload of monthly retraining. From an economic perspective, since the input predictors are organized by firm characteristics that a substantial number of them update on a quarterly or annual basis, annual re-estimation aligns the model's learning frequency with the natural turnover of these fundamental signals. This prevents the networks from overfitting to short-term monthly trading noise that often lacks long-term predictive power. Between parameter updates, the models generate monthly out-of-sample predictions by applying the most recent estimated parameters to the updated monthly factors. Figure~\ref{fig:data_des1} shows the mechanism of the extensive window method in this chapter. \\

As for the robustness checking, the study compares the models’ testing periods depending on whether including the COVID-19 period, which is denoted as ‘2112’ for the testing period covering COVID-19, which contains a sharp uptrend with deep drawdowns, and ‘1911’ for the testing period without the COVID-19 period, which contains a mild uptrend. Concretely, the training extension window and validation window are identical for both methods; the only difference is the length of the testing period. It is demonstrated in Table~\ref{tab:data_split}.\\

\begin{table}[ht]
\centering
\begin{tabular}{lccc}
\hline
\textbf{Name} & \textbf{Start date} & \textbf{End date} & \textbf{Observation No.} \\
\hline
Training & 1/1957 & 1/2003 & 553 \\
Validation & 2/2003 & 12/2012 & 119 \\
Testing (OOS) for ‘1911’ & 1/2013 & 11/2019 & 83 \\
Testing (OOS) for ‘2112’ & 1/2013 & 12/2021 & 108 \\
\hline
\end{tabular}
\vspace{0.5em}
\raggedright
\caption[Data split table for initial setting of training, validation, and testing period]{Data split table for initial setting of training, validation, and testing period. `Testing (OOS) for 2112’ refers to the full out-of-sample evaluation period, while `Testing (OOS) for 1911’ excludes the pandemic period from December 2019 to December 2021.}
\label{tab:data_split}
\end{table}

\section[Models for Chapter 2]{Models}\label{sec:models_ch2}
As in most ML-based asset pricing literature, it starts from the general additive prediction error model commonly used for asset pricing assignment:
\begin{equation}
r_{i,t+1} = \mathbb{E}_t \left[ r_{i,t+1} \right] + \varepsilon_{i,t+1}
\label{eq:return2.4}
\end{equation}
Where $r_{i,t+1}$ is the asset excess returns at time t+1. $E_t[r_{i,t+1}]$ for $t\in[1, T-1]$ is the expected return, which is the `fair' compensation investors demand at time $t$ for bearing the systematic risk associated with the asset. It is determined by the asset's sensitivity (or `Beta') to broader market factors, such as the market risk premium, size, or value factors, and is calculated using all information available up to that moment. $t$ is the time index in the range from 1 to the entire time length $T$. $i$ is the asset index in the range from 1 to the entire asset number $N$. $\varepsilon_{i,t+1}$ is the residual (or idiosyncratic shock in asset pricing). While the expected return is the reward for staying exposed to the market's structural risks, the idiosyncratic shock is the result of firm-specific events, such as a sudden lawsuit or a breakthrough invention, that deviate from the original forecast. Concerning the work of GKX2020 \citep{Gu2020EmpiricalLearning}, the conditional expectation of $r_{i,t+1}$ can also be expressed as a function of $K$ number of factors. If the number of factors is small, it could be suitable for traditional linear structures. Whereas if the number of factors is large, it raises the ‘factor zoo’ issue \citep{Harvey2019AZoo}, the additional dimension reduction process, such as principal components analysis (PCA), is required for shrinking the dimension of factors. This process is also known as ‘feature engineering or abstracting’ in ML terminology. Equation~\eqref{eq:return2.5} provides the general idea of a factor model. The early literature on empirical asset pricing generally applies the ordinary least squares (OLS) linear function with fewer than 10 factors, but later, alternative non-linear function forms are applied due to the increasing number of factors. Equation~\eqref{eq:return2.5} is a general function form of the asset pricing factor model in the context of machine learning, where $g(\cdot)$ represents the variable model structures in this chapter. $f_{j,t}$ is the $j$th factor at time $t$ and the $\theta$ is the parameters of the factors.    
\begin{equation}
r_{i,t+1} = g\left(f_{j,t}; \theta\right) + \varepsilon_{i,t+1}
\label{eq:return2.5}
\end{equation}
\subsection[Multilayer Perceptron Algorithms (MLPs)]{Multilayer Perceptron Algorithms (MLPs)}
\label{subsec:MLPs_ch2}
From Equation~\eqref{eq:return2.5}, it is clear that the function form is variable for adapting the factors. In this chapter, we employ the MLP models with a pyramid structure as a substitution for the traditional linear form of the factor model. The mechanism of MLP will be presented with an example, while the application of the factor model for asset pricing will be explained in line with the work of GKX2020. \\

\begin{figure}[htbp!]
\begin{center}
\includegraphics[width=0.9\columnwidth]{"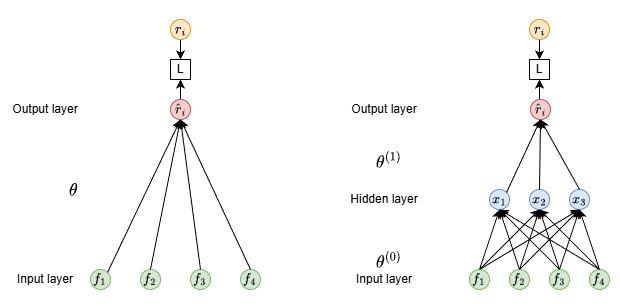"}
\end{center}
\caption[An example of simple MLP structures with (right) and without (left) hidden layers.]{A example of simple MLP structures with (right panel) and without (left panel) hidden layers. $\theta$ are the learnable parameters for the factors $f_j(j= 1,2,3,4)$ and the $x_1,x_2,x_3$ are values calculated from an activation function $\sigma(\cdot)$. L represents the loss function, and $\hat{r}_i$, $r_i$ represent the estimated and actual excess return of stock $i$, respectively. The layer with green neurons is the input layer, assumed with 4 factors, and the layer with 3 blue neurons is the hidden layer, which is calculated from the input factors with the activation function. The layer with one single red neuron is the output layer, which derives the estimated stock excess return $\hat{r}_i$ that is adjusted according to the actual stock excess return $r_i$ through the loss function.}
\label{fig:MLPs1_ch2}
\end{figure}

Influenced by biological brain mechanisms, vanilla neural network algorithms (alternatively multilayer perceptron) were developed in the 1960s, and a multitude of literature contributed to developing the modern concept of MLPs \citep{Rosenbaltt1957TheAutomation,Rumelhart1986LearningErrors,Werbos1974BeyondSciences}. A complete simple MLP structure constitutes one input layer, a single hidden layer or multiple hidden layers and one output layer. Each layer contains a single neuron or multiple neurons. Neurons are connected to the neurons from the layers next to them. Figure~\ref{fig:MLPs1_ch2} demonstrates an example of fully connected MLPs with and without hidden layers. 
The left panel of Figure~\ref{fig:MLPs1_ch2} exhibits the MLP without a hidden layer. It can be as simple as a linear model, which can be presented as:
\begin{equation}
g(f_{t}; \theta) = \theta_0 + \sum_{j=1}^{K} \theta_{j} f_{j,t}
\end{equation}
Where $g(\cdot)$ is a function form for mapping the factors to the excess returns $r_{i,t}, f_{j}$ is the $j$th of the original factors. The $\theta_0$ performs as the intercept of the linear model, and it is interpreted as the abnormal return in the context of factor investing or pricing error in the context of asset pricing. $\theta_j$ is the parameter for factor $j$. \\

As the number of factors $K$ increases, the linear model may fail to provide reliable estimates due to its inherent limitations. Firstly, the model encounters the curse of dimensionality, where the number of parameters to be estimated approaches or exceeds the number of observations, leading to a loss of degrees of freedom and singular covariance matrices. Secondly, high-dimensional factor sets often exhibit severe multicollinearity, causing the Ordinary Least Squares (OLS) estimators to become highly unstable with inflated variances. Thirdly, the model begins to `fit the noise' rather than the signal, resulting in significant overfitting, which manifests as the sharp decline in out-of-sample predictive power. Finally, a rigid linear specification is incapable of capturing non-linearities and high-order interactions between characteristics. Thus, MLP with hidden layers are the more flexible solution for the empirical asset pricing model.\\

The right panel of Figure~\ref{fig:MLPs1_ch2} exhibits an example of a simple pyramidical MLP structure with 1 hidden layer containing 3 neurons and an input of 4 factors. The algorithm of a neuron `$m$' in the first hidden layer can be presented as:
\begin{equation}
x^{(1)}_{m,t} = \sigma\left( \theta^{(0)}_{m,0} + \sum_{j=1}^{K} f_{j,t} \theta^{(0)}_{m,j} \right)
\end{equation}
Where $\sigma(\cdot)$ is the activation function, $x^{(1)}$ represents the neuron $m$ in the first hidden layer, $\theta^{(0)}_{m,0}$ and $\theta^{(0)}_{m,j}$, $j\neq0$ are known as `bias' and `weights' in the context of machine learning, respectively. The number in parentheses is the sequence notation of layers, in which `$(0)$' means the input layer and `$(1)$' means the first hidden layer.\\

From the second hidden layer to the final hidden layer, the general form of mathematical presentation is shown as:
\begin{equation}
x^{(l)}_{m,t} = \sigma\left( \theta^{(l-1)}_{m,0} + \sum_{h=1}^{M_{l-1}} x^{(l-1)}_{h,t} \theta^{(l-1)}_{m,h} \right)
\end{equation}
Where $l$ indicates the index of the layer, $l>1$, and $M$ is the number of neurons in a hidden layer. Thus, $M_{l-1}$ means the neuron number in the hidden layer $l-1$, $h$ is the index of neurons from the previous hidden layer. The rectified linear unit (ReLU) function is deployed as the activation function by following the work of GKX2020 \citep{Gu2020EmpiricalLearning}. It shows as:
\begin{equation}
ReLU(x) = \max\{0, x\}
\end{equation}
The economic rationale for this functional form lies in its ability to model asymmetric payoff structures and threshold-based pricing anomalies without imposing the rigid constraints of a global linear model. By nullifying negative activations, ReLU inherently introduces an internal feature (factor) selection process, which is crucial for navigating the low signal-to-noise ratio of financial data. This acceleration of convergence, coupled with the avoidance of saturation regions, enables MLP models to robustly extract non-linear predictive signals from a vast set of $K$ input factors. Thus, it effectively improves the computational efficiency and mitigates the vanishing gradient problem during the backpropagation.\\

A linear output layer is assembled for the asset return value prediction context. It is mathematically presented as:
\begin{equation}
\hat{r}_{i,t+1} = \theta^{(L)}_{0} + \sum_{m=1}^{M_L} x^{(L)}_{m,t} \theta^{(L)}_{m}
\end{equation}
Where $L$ is the notation of the final hidden layer.\\

By following the work of GKX2020, the pyramid structure is applied in this chapter, which follows the sequence:
\begin{equation}
\text{input dimension} > h_1 > h_2 > \cdots > \text{output dimension}
\end{equation}
Where the input dimension is equal to the number of features, $h(i),i=1,2,3,\dots$ indicates the dimension of hidden layers. The output dimension is the number of output dimensions. GKX2020 simply used a fixed pyramid structure with $h_1 =32, h_2=16, h_3=8, h_4=4, h_5=2$ as suggested by \citet{Masters1993PracticalC++}. However, the fixed structure may not be adaptive for the variable input dimensions, and \citet{Coqueret2020MachineVersion} proposed a dynamic pyramid structure design method for MLP, which is:
\begin{equation}
h_k \approx \left\lfloor O\left( \frac{I}{O} \right)^{\frac{L+1-k}{L+1}} \right\rfloor
\end{equation}
Where $I$ is the input dimension, $L$ is the number of hidden layers. It provides a structure with a more reasonable shrinkage of input factors. The MLP models implement the \citet{Coqueret2020MachineVersion}’s method for constructing the pyramid structure, with the output dimension of one, which represents an individual stock’s excess return series. A factor counts as one dimension of the input layer.\\

The choice of this pyramidal architecture is not merely an empirical convention inspired by the work of \citet{Masters1993PracticalC++}, but is grounded in two primary considerations. First, from an information-theoretic perspective, the declining neuron count functions as an informational bottleneck. Given the high dimensionality of $I$ and the inherently low signal-to-noise ratio in financial markets, this structure forces the network to distill redundant factors into a parsimonious set of latent pricing signals, therefore filtering out idiosyncratic noise. Second, the geometric decay in $h_k$ serves as a form of implicit regularization. By systematically reducing the parameter space in deeper layers, the model mitigates the risk of overfitting, which is a crucial failure reason for linear models when the factor number $K$ is large. Unlike the fixed configuration in \citet{Gu2020EmpiricalLearning}, the dynamic approach of \citet{Coqueret2020MachineVersion} ensures that the compression ratio remains proportional to the specific input dimension of the dataset, maintaining a consistent level of abstraction across different model specifications. Thus, it is generalized for the variable dimension feature (factor) input in other economic or financial scenarios.\\

\subsection[Loss Function and Model Estimation]{Loss Function and Model Estimation}
\label{subsec:lossEst_ch2}
The loss function is a function that compares the predicted stock excess return values with actual stock excess return values. By minimizing the loss function, which converges the predicted values to the actual values, the neural network models are able to detect the parameter sets for the best prediction performance. Different loss functions for different tasks. For example, classification tasks employ Cross-Entropy, Hinge, or Focal loss functions. For regression tasks, it could be Mean Absolute Error (MAE), Mean Squared Error (MSE), Huber or Mean Squared Logarithmic Error (MSLE). It is interesting that, in the mainstream of the literature on ML-based financial prediction and asset pricing domain, authors agreed to deploy the MSE as the loss function \citep{Gu2020EmpiricalLearning,Kelly2019CharacteristicsReturn,Freyberger2020DissectingNonparametrically,Bianchi2021BondLearning,Chen2024DeepPricing}. By following their pathway, the MSE with L1 regularization is adopted as the loss function in this chapter and later chapters. This loss function can be presented as:
\begin{equation}
\mathcal{L}_{MSE_{L1}} = \frac{1}{T} \sum_{t=1}^{T} \left( r_{i,t} - \hat{r}_{i,t} \right)^2 + \lambda |\Theta|_1
\label{eq:MSE_ch2}
\end{equation}
Where $\mathcal{L}_{MSE_{L1}}$ represents the loss function with L1 regularization. $r_{i,t}$ and $\hat{r}_{i,t}$ are the stock $i$’s actual and predicted excess returns, respectively. $T$ is the time length of excess returns or estimated excess returns. $\lambda$ is a hyperparameter named Regularization Parameter, and $\Theta$ is the estimated parameters. The term of $\lambda |\Theta|_1$ represents the L1 regularization. This configuration also follows the work of GKX2020 \citep{Gu2020EmpiricalLearning}. Compared to $L_2$ regularization (Ridge), $L_1$ regularization (Lasso) is uniquely suited for asset pricing due to its ability to induce sparsity and perform automated variable selection. While $L_2$ merely shrinks coefficients toward zero, the $L_1$ penalty can force coefficients of irrelevant predictors to be exactly zero. This is essential for navigating the `Factor Zoo' as it effectively filters out noise and identifies the most parsimonious set of factors. Consequently, $L_1$ enhances model interpretability and provides more robust out-of-sample predictions by mitigating the risk of overfitting in high-dimensional financial datasets.\\

Intuitively, since the data used in this chapter is effectively mitigated via data pre-processing (e.g.missing value filtering, normalization), and factors are structured by firm characteristic-sorted portfolios instead of firm characteristics, this naturally reduces the noise of the data, thereby the advantages of robustness for alternative loss functions such as Huber loss or MSLE are insignificant\citep{Gu2020EmpiricalLearning}. Additionally, from the statistical perspective, MSE targets the conditional mean $\mathbb{E}_t[r_{i,t+1}]$, aligning with the expected-value framework of asset pricing theory, whereas MAE targets the median. This means MSE is a smooth, second-order differentiable function with a gradient proportional to the error size, which ensures stable convergence by allowing the model to decelerate near the optimum, and avoiding the numerical oscillations inherent in the constant-gradient MAE.\\

Unlike traditional statistical models such as OLS, which typically rely on in-sample fitting and out-of-sample (OOS) testing, ML frameworks introduce a validation period to facilitate automated model selection and hyperparameter tuning. While the non-linear architecture of the MLP inherently relaxes classical assumptions like linearity and homoscedasticity, the validation set is crucial for preventing overfitting in such highly parameterized models. Specifically, it utilizes Stochastic Gradient Descent (SGD), or its advanced variants like the Adaptive Moment Estimation (Adam) \citep{Kingma2014Adam:Optimization}, to minimize the model's parameters on the training data, simultaneously monitoring performance (the lowest MSE) on the validation data (simulation OOS dataset) to determine the best model for testing data (actual OOS dataset).\\

Equation~\eqref{eq:gra_ch2} and Equation~\eqref{eq:sgd_ch2} shows the GD process, where $\nabla_{\boldsymbol{\theta}} \mathcal{L}(\boldsymbol{\theta})$ gradient, it is the partial derivative of the loss function and the individual model parameter, $\theta$ are the model parameters. $\varphi$ is the pre-set hyperparameter named learning rate, which determines how fast or accurately to find the parameters that let the loss function reach the global minimum. Normally, when $\nabla_{\boldsymbol{\theta}} \mathcal{L}(\boldsymbol{\theta})$ is approaching zero, the loss function is considered as ‘reaching the global minimum’.
\begin{align}
\nabla_{\boldsymbol{\theta}} \mathcal{L}(\boldsymbol{\theta}) = 
\left( 
\frac{\partial \mathcal{L}(\boldsymbol{\theta})}{\partial \theta_1},\ 
\frac{\partial \mathcal{L}(\boldsymbol{\theta})}{\partial \theta_2},\ 
\dots,\ 
\frac{\partial \mathcal{L}(\boldsymbol{\theta})}{\partial \theta_n} 
\right)\label{eq:gra_ch2}\\
\boldsymbol{\theta} := \boldsymbol{\theta} - \varphi \nabla_{\boldsymbol{\theta}} \mathcal{L}(\boldsymbol{\theta})\label{eq:sgd_ch2}
\end{align}
Concerning the computational cost of ML algorithms, the most prevailing parameter searching process is the Stochastic GD (SGD) which updates parameters for each sample in each iteration. However, it introduces unexpected noise which could trigger the unstable convergence of the loss function. That is the reason for adding the optimizer of Adaptive Moment Estimation (Adam) \citep{Kingma2014Adam:Optimization} for stabilizing the SGD process. Adam introduced the first and second-moment estimation of gradients to achieve the acceleration of the loss function convergence and the self-adaptation of the learning rate. The pseudocode of Adam described in the appendix of Chapter 2.\\

The backpropagation algorithm is the algorithm for adjusting the weights in the network by calculating the gradient of the loss function with respect to each parameter from the output direction. It applies the “chain rule” of derivative computation to propagate gradients from the output layer backwards through the network. The process can be separated into five steps: obtaining results from forward propagation, computing the loss, backpropagation, updating weights, and iteration. The mechanism of the MLP backpropagation is exhibited in the appendix of Chapter 2.\\

\subsection[Early Stopping and Batch Normalization]{Early Stopping and Batch Normalization}
\label{subsec:ESBN_ch2}
One parameter searching for all training data points in a batch is defined as one epoch. Epoch is a one-time gradient descent process. Normally, when the epoch number increases, the MSE of training and validation will decrease. However, as epochs are a hyperparameter of the MLP model, it is difficult to define before attempting the model. Oversize epochs will cause the overfitting issue as well because the difference of MSE for both training and validation data increases after epochs reach a critical minimum value. Early-stopping is a process for stopping the epochs when the performance on the validation data cannot be improved further, then logging the parameters for prediction. \\

Batch normalization is a vertical normalization, where each observation in the batch removes the batch’s mean and is divided by its standard deviation; the training data in one batch’s mean is zero, and the variance is equal to 1. Batch normalization can accelerate and stabilize the training process, moderate the issue of gradient vanishing and explosion \citep{Glorot2010UnderstandingNetworks}, which is explained in Section~\ref{subsec:gra_ve_ch2}, and reduce the sensitivity towards the initial weights.\\

\subsection[Gradient Vanishing and Explosion]{Gradient Vanishing and Explosion}
\label{subsec:gra_ve_ch2}
Gradient vanishing usually happens during the backpropagation process. When the hidden layer number of the MLP model increased to a certain level, the gradient was extremely small or equal to zero. Subsequently, when it is passed back to the first hidden layer, the gradient vanishes, causing the parameters to pause updating, leading to poor performance of the model. This was especially prevalent when the activation function was using Sigmoid or Tanh. Whereas ReLU is the activation function which can prevent this issue since its derivative is most likely to be 1, batch normalization is another approach that also helps with this. Opposite to the scenario of gradient vanishing, if the initial learning rate is not properly set or an inappropriate depth to the MLP structure is deployed, the gradient could be extremely large. In that sense, if the learning rate causes excessively large parameter updating, the parameters do not converge on the loss function to the global minimum but diverge from the global minimum.  This phenomenon is known as gradient explosion, and the loss function encountering gradient explosion tends to be infinite. Testing the initial learning rate and batch normalization can be applied as adjustment methods for this issue. \\

\subsection[Benchmark Models]{Benchmark Models}
Five proposed dynamic MLP models are implemented as the function forms of the empirical asset pricing model. They are compared with six benchmark models: Ordinary Least Squares (OLS), Partial Least Squares (PLS), Principal Component Regression (PCR), and GKX2020 with 3, 4, and 5 hidden layers, respectively. GKX2020 MLP models as benchmark models utilizes their original predictor configurations. The GKX2020 MLP models are similar to the proposed MLP models, but with fixed pyramidical neuron structures. In accordance with Equation \eqref{eq:return2.5}, the linear benchmark models are defined by their specific functional forms of $g(f_{i,t}; \theta)$, which include a constant intercept $\alpha$ to capture the mean return not explained by the factors.\\

The OLS benchmark assumes a global linear mapping with an intercept $\alpha$:
\begin{equation}
g(f_{i,t};\alpha, \theta_{OLS}) = \alpha + f_{i,t}^\top \theta_{OLS}
\end{equation}
where the parameters $\alpha, \theta_{OLS}$ are estimated by minimizing the sum of squared residuals over the training sample. The intercept $\alpha$ represents abnormal return or pricing error in the context of factor investing and asset pricing, respectively.\\

PCR \citep{massy1965principal} applies dimensionality reduction through an unsupervised learning framework, where the predictive function $g(\cdot)$ is defined as a linear mapping of the top $K$ principal components:
\begin{equation}
g(f_{i,t}; \alpha, \theta_{PCR}, \Omega_K^{PCR}) = \alpha + (f_{i,t}^\top \Omega_K^{PCR}) \theta_{PCR}
\end{equation}
The projection matrix $\Omega_K^{PCR} = [v_1, \dots, v_K]$ consists of the $K$ eigenvectors corresponding to the largest eigenvalues of the factor covariance matrix $\Sigma_{ff} = E[(f_{i,t} - \bar{f})(f_{i,t} - \bar{f})^\top]$, where $v_1, \dots, v_K$ are the loading vectors, and $\bar{f}$ is the mean of the factors. The intercept $\alpha$ represents the same economic meaning as the one in OLS. The $j$-th principal direction $v_j$ is determined by maximizing the variance of the projected factors:
\begin{equation}
v_j = \arg\max_{v} \text{var}(f_{i,t}^\top v) \quad \text{s.t.} \quad ||v||=1, v^\top v_m = 0 \text{ for } m < j
\end{equation}

Unlike PCR, the PLS model \citep{wold1966estimation} utilizes a supervised dimensionality reduction framework, where the predictive function is defined as:
\begin{equation}
g(f_{i,t}; \alpha, \theta_{PLS}, \Omega_K^{PLS}) = \alpha + (f_{i,t}^\top \Omega_K^{PLS}) \theta_{PLS}
\end{equation}
In contrast to PCR, the PLS model identifies the latent space $\Omega_K^{PLS}$ by accounting for both the variance of factors and their predictive relationship with future stock excess returns $r_{i,t+1}$. Specifically, the $j$-th direction $w_j$ in $\Omega_K^{PLS}$ is determined by maximizing the squared covariance between the linear combination of factors and the target returns:
\begin{equation}
w_j = \arg\max_{w} \text{cov}^2 (f_{i,t}^\top w, r_{i,t+1}) \quad \text{s.t.} \quad ||w||=1, \text{cov}(f^\top w, f^\top w_m) = 0 \text{ for } m < j
\end{equation}
The inclusion of the intercept $\alpha$ ensures that the covariance maximization is performed on the innovations of the factors, and prevents systematic mean levels from biasing the orientation of the latent directions.\\

\subsection[Models for Chapter 2]{Model performance evaluation}
\label{subsec:perform_evalu_ch2}
In this section, the proposed models and benchmark models are compared from two angles: model performance (predictive power) and backtesting performance. Model performance is measured by the mean of 420 stocks' the out-of-sample (OOS) coefficient of determination ($R_{OOS}^2$), mean square error (MSE), residual $\alpha$ and its statistics, which are shown in Table~\ref {tab:model_perform_ch2}. The algorithms of these indicators are presented in Equation~\eqref{eq:r2_ch2} to Equation~\eqref{eq:alpha_ch2}. The research in this chapter still employs the one-step-ahead forecasting method.
\begin{equation}
R^2_{\text{oos}} = 1 - \frac{\sum \left( r_{i,t+1} - \hat{r}_{i,t+1} \right)^2}{\sum \left( r_{i,t+1} - \overline{r}_{\text{in}} \right)^2}
\label{eq:r2_ch2}
\end{equation}
\begin{equation}
MSE = \frac{1}{TN} \sum_{t=1}^{T} \left( r_{i,t} - \hat{r}_{i,t} \right)^2
\label{eq:mse_ch2}
\end{equation}
\begin{equation}
\alpha_{avg} := \frac{1}{N} \left[ E(r_{i}) - E(\hat{r}_{i}) \right]\label{eq:alpha_ch2}
\end{equation}
Where $r_{i,t+1}$ are the real out-of-sample excess returns, $\hat{r}_{i,t+1}$ is the predicted excess returns, which is equal to $\hat{y}_{t+1}$ in Section~\ref{sec:Models_ch2} and $\overline{r}_{\text{in}}$ is the mean of the excess returns in the training and validation (in-sample) period. $i$ indicates the stock $i$, and $TN$ is the time length observation number multiplied by the number of stocks. Here, $\alpha$ denotes the OOS average estimation error in model evaluation, but it has a specific meaning in the asset pricing context, as introduced in the general introduction section.\\

In the traditional asset pricing concept, $\alpha$ is the intercept of a linear factor model. It means the risk-adjusted return of a stock. If the $\alpha=0$, the stock excess return is perfectly explained by the selected or constructed factors; if $\alpha>0$, it indicates the extra capital gain exists after controlling the risks brought by factors, which is also known as the `arbitrage opportunities'; if $\alpha<0$, it implies the stock underperforms the model's prediction. The latter two circumstances indicate the existence of pricing errors, but practitioners chase the positive $\alpha$ as the abnormal returns beyond the factors. In traditional asset pricing literature, those with a linear model frame, factor exploration research pursues $\alpha$ approaching zero to verify that their factors can perfectly explain the asset returns, which satisfies the non-arbitrage hypothesis. That is slightly different in ML-based asset pricing literature, especially in this case. The linear output layer still makes the asset pricing model explainable, but forces the restriction of the non-arbitrage condition by removing the intercept term of $\alpha$. The computation of $\alpha$ for the ML-based asset pricing is also known as residual $\alpha$, which is employed by \citet{Gu2020EmpiricalLearning,Ma2023AttentionApproach} and is shown in Equation~\eqref{eq:alpha_ch2}. That means if the assumption stands for the linear output layer form, the mean of both pricing error $\alpha$ and model bias $\bm{b}$ should be equal to zero; otherwise, it implies that the $\alpha$, the abnormal return or the arbitrage opportunity exists, or there are unexplored factors. Exploring the structure of $\alpha$, how to minimise $\alpha$, and the relation to the bias of the ML frame are interesting directions for future research. In that sense, the residual $\alpha$ includes the traditional $\alpha$ and the risk premium of the linear and non-linear model structures. \\

The Diebold-Mariano (DM) \citep{Diebold1995ComRacy} test with the HAC estimator (preventing loss differential series autocorrelation causes spurious statistical significance) is adopted for evaluating the difference between models in this study. Equation~\eqref{eq:dm1_ch2} to Equation~\eqref{eq:dm4_ch2} presents the mechanism of the DM test. DM test here using the Mean Absolute Error (MAE) as the foundation for its robustness compared with the MSE option. MAE is less sensitive to outliers than MSE, which is particularly suitable for financial time series in AI models that contain more complicated and variable structure in their prediction errors (e.g. non-Gaussian, heavy-tailed, etc.). Therefore, MAE-DM here reflects typical difference forecasting performance and avoids being affected by a few large deviations, which causes false detection of the difference between models. Hence, MAE-DM statistics provide a more realistic and reliable assessment of model superiority in this study.
\begin{align}
d_{t+1} &= \frac{1}{h} \sum_{i=1}^{h} \left( \left| e^{(m)}_{i,t+1} \right| - \left| e^{(n)}_{i,t+1} \right| \right) \label{eq:dm1_ch2} \\
\bar{d} &= \frac{1}{T} \sum_{t=1}^{T} d_{t+1} \label{eq:dm2_ch2} \\
DM_{\text{statistics}} &= \frac{\bar{d}}{d_{\text{standard error}}} \label{eq:dm3_ch2} \\
d_{\text{standard error}} &= \sqrt{\widehat{V}_{HAC} / T} \label{eq:dm4_ch2}
\end{align}
where $|e^{(m)}_{i,t+1}|$, $|e^{(n)}_{i,t+1}|$ are the absolute forecasting errors derived from $(r_{i,t+1} - \hat{r}_{i,t+1})$ of model $m$ and $n$ respectively. $i$ indicates the stock $i$. $h$ is the number of stocks, which is 420 in this case. $d_{t+1}$ is the cross-sectional average difference between model $m$ and model $n$’s absolute forecasting errors, $\bar{d}$ is the mean of $d_{t+1}$. The denominator, $d_{\text{standard error}}$, is the HAC-consistent standard error of the loss differential series $\{d_t\}$. $\widehat{V}_{HAC}$ denotes the heteroskedasticity and autocorrelation consistent (HAC) variance of the loss differential series $d_{t+1}$.\\

For a higher economic and financial interpretation of the model, this study also calculates the factor importance for the MLP autoencoder abstracted factors and the correlation between the abstracted factors and the original factors. Mathematically, the permutation variable importance (VI) can be exhibit as:
\begin{align}
L_{\text{baseline}} &= \mathcal{L}(r_{i,t}, f(X)) \label{eq:perm1_ch2} \\
L_{\text{perm}}(X_i) &= \mathcal{L}(r_{i,t}, f(X_{\text{perm}})) \label{eq:perm2_ch2} \\
\text{Importance}(X_i) &= L_{\text{perm}}(X_i) - L_{\text{baseline}} \label{eq:perm3_ch2}
\end{align}
where $L_{baseline}$ is baseline loss or original out-of-sample loss, $X$ is the original feature matrix, and $f(\cdot)$ is the model forms. $L_{perm(X_i)}$ is the loss that an individual feature $X_i$ is permutated. Then, the importance of $X_i$ is equal to the difference between the permutated loss and the baseline loss. \\

This study uses the trend-following sign signal trading strategy as the market timing approach, and the transaction cost is configured as 50 base points via a traditional static transaction cost calculation method \citep{jegadeesh1993returns,lesmond2004illusory,grundy2001understanding}, and employs a 20 bps \citep{frazzini2018trading,novy2016taxonomy,Hou2020ReplicatingAnomalies} concerning turnover rate (shows in Equation ~\eqref{eq:TC_turnover}) as a robustness examination for exploring how transaction cost and turnover rate affect the backtesting performance of the models in this chapter. For measuring the backtesting performance, it follows the traditional asset pricing indicators, which are annualized returns, Sharpe ratio, Sortino ratio and maximum drawdown (MDD). Equation~\eqref{eq:ar_ch2} to Equation~\eqref{eq:mdd_ch2} presents the calculation of these indicators.
\begin{align}
Annualized\ Return &= \left\{ 1 + \left[ \prod_{i=1}^{n} (1 + r_{i,t}) - 1 \right]^{\frac{12}{n}} \right\} - 1 \label{eq:ar_ch2} \\
Sharpe\ Ratio &= \frac{E(r_p) - r_f}{\sigma_p}\label{eq:sr_ch2} \\
Sortino\ Ratio &= \frac{E(r_p) - r_f}{\sigma_d}\label{eq:so_ch2} \\
Maximum\ Drawdown &= \max_{t \in [0,T]} \left( \frac{c_{\max}(t) - c(t)}{c_{\max}(t)} \right)\label{eq:mdd_ch2}
\end{align}
where $n$ is the number of the observations, $E(r_p)$ is the expected portfolio return and $r_f$ represents the risk-free rate. $\sigma_p$ and $\sigma_d$ are the portfolio and portfolio downside standard deviation, respectively. $C_{\max}(t)$ is the highest value during time $t$, and $C(t)$ is the value at time $t$. For a better comparison of models that perform under the extreme market fluctuations caused by the COVID-19 pandemic, all results are displayed in three periods: Pre–COVID-19 Period (1911), COVID-19–Inclusive Period (2112) and Period Including COVID-19 and One-Year After (2212). \\

The models in this chapter were implemented in Python (v3.8) using the PyTorch deep learning framework \citep{paszke2019pytorch}. Training was conducted on NVIDIA A40 and H100 GPUs provided by the Viking HPC cluster at the University of York. Unlike traditional central processors (CPUs) that handle tasks sequentially, GPUs are engineered for massive parallel computing, allowing the model to process thousands of financial signals and factor interactions simultaneously. This high-performance hardware was essential for research in this chapter since the `factor zoo' presents a high-dimensional nature in individual stock calculations. That would be computationally prohibitive, potentially requiring weeks of processing time, on standard office equipment. Leveraged by these specialized chips, the extensive iterative optimization process becomes feasible. \\

As discussed in Section 4, the proposed models and benchmark models are evaluated via OOS mean squared error (MSE), OOS $R^2$ and DM tests. Table~\ref{tab:model_perform_ch2} shows the MSE and OOS $R^2$ for the main models. Benchmark models for the data period include and exclude the pandemic which is from December 2019 to December 2021. We can see from the period of ‘2112’, which contains the full testing period from January 2013 to December 2021, that all proposed models outperform the benchmark models. The average OOS $R^2$ implies that the proposed MLP model with 2 hidden layers which has 3.66\% as the average OOS $R^2$, outperforms all alternative models and is followed by the one with 3 hidden layers, which achieved 2.86\% of OOS interpretability, but apart from the proposed MLP model with 1,2,3,4 hidden layers, the rest of the models all face the overfitting issue to some extent. Deep overfitting implies model misspecification issues, which are defined as model form mismatch in the features and labels. In this sense, Model OLS, PLS, PCR and GKX2020 with 3 hidden layers potentially have model misspecification suspect under the full testing period condition. \\

\section[Empirical application and results]{Empirical application and results}\label{sec:Empirical_Results_ch1}
This section mainly presents the empirical comparison from the model function and portfolio performance perspectives. Section~\ref{subsec:Emp_application_ch2} exhibits the empirical methods, Section~\ref{subsec:Model_eva_ch2} discusses the model-wise performance, and Section~\ref{subsec:back_testing_eva_ch2} shows the performance of equal-weighted and value-weighted portfolios organised via the selected 420 stocks from the Pre–COVID-19 Period (1911) and COVID-19–Inclusive Period (2112) angles.\\

\subsection[Empirical Application and Results]{Empirical application}\label{subsec:Emp_application_ch2}
The entire experiment follows seven steps: 1) data preparation. 2) modelling; 3) prediction; 4) model evaluation and comparison; 5) market timing (trading signal generation); 6) back testing; and 7) back-testing performance assessment.\\

The model structures of the main models and GKX2020 MLP models are shown in Table~\ref{tab:mlp_architecture}. The input factors are 182 selected firm characteristics-sorted portfolio factors, and the output is the excess return series of individual stocks. Thus, the input dimension is 182, and the output dimension is one. The benchmark models of PLS and PCA with a hyperparameter of 0.95 indicate the components explain 95\% of the variance from the original features. \\
\begin{table}[htbp!]
\centering
\begin{tabular}{lcc}
\hline
\textbf{} & \textbf{Main MLP models} & \textbf{GKX2020} \\
\hline
1\_layers & 15 & 32 \\
2\_layers & 36, 6 & 32, 16 \\
3\_layers & 56, 15, 4 & 32, 16, 8 \\
4\_layers & 73, 25, 9, 3 & 32, 16, 8, 4 \\
5\_layers & 87, 36, 15, 6, 3 & 32, 16, 8, 4, 2 \\
\hline
\end{tabular}

\vspace{0.5em}
\noindent\raggedright
\caption[MLP architecture comparison between our main models and GKX2020]{MLP architecture comparison between our main models and GKX2020. ‘$(i)\_layers$’, $i=1,2,3,\dots,5$, means an MLP model contains $i$ hidden layers. Each number defines the number of neurons in a single hidden layer. The largest number in each table row is the neuron number of the hidden layer closest to the input layer. One neuron indicates one dimension. The smallest number in the row is the one in the hidden layer closest to the output layer.}
\label{tab:mlp_architecture}
\end{table}

A common trading signal detection technique is applied to generate trading signals from the actual stock excess returns and estimated stock excess returns of each stock, which is named the sign signal generation system. It is utilized for market timing. Concretely, in a stock, when the sign of the actual return at time $t$ matches the sign of the estimated return at time $t$ in a back-testing process, it is considered a trading signal. If both the actual return and the estimated return have a positive sign, it is time to open a long position of that stock. Investors may hold this position until the opposite trading signal appears, namely, both of them have a negative sign, which is a signal for investors to close that position. This is also known as long side trading. The short side trading is in contrast to the long side trading, which searches for the negative sign matching for opening the short position, then detects the positive sign matching for closing the position. This trading signal detection technique is commonly used in the trend-following trading strategy, which allows investors to achieve profitability by following the price trend of a stock. \\

All selected stocks are grouped as an equal-weight portfolio or a value-weighted portfolio, since a complete practical trading strategy includes trading individual stocks according to trading signals and deciding how many positions should be held for each stock. Since the asset allocation research direction is a bit out of scope of the empirical asset pricing research, in this study, the simple asset allocation strategies are implemented to complete the backtesting process, which is commonly conducted by the relevant literature \citep{Gu2020EmpiricalLearning}. The equal-weight portfolio is to distribute equal weights to each stock included in the portfolio, and the value-weighted portfolio is to distribute the weights according to the proportion of the market capitalization of each stock. Additionally, the unstable borrowing rates and limited short side trading shares provided by brokers for short side stock trading practice are used to restrict the short side trading of the stocks and bring systematic risks to investors. To prevent the systematic risks brought by the extreme cases, the backtesting of this chapter excludes the short side trading. Following the methodology in existing literature of \citet{jegadeesh1993returns,lesmond2004illusory,grundy2001understanding}, which statically configured as 50bps. For robustness examination of how turnover influences the profitability of these models, a dynamic transaction cost is applied (turnover and 20bps for large-cap stocks), following the work of \citet{Ma2023AttentionApproach,frazzini2018trading,novy2016taxonomy,Hou2020ReplicatingAnomalies}, which shows in Equation ~\eqref{eq:TC_turnover}. The total transaction cost ($TC_t$) is proportional to the absolute change in portfolio weights during rebalancing, which can be expressed as:
\begin{equation}
TC_t = \phi \cdot \sum_{i=1}^{N} |w_{i,t} - w_{i,t-}^{+}|
\label{eq:TC_turnover}
\end{equation}
where $\phi = 0.002$ represents the cost rate (20 basis points), $w_{i,t}$ is the target weight of asset $i$ at time $t$, and $w_{i,t-}^{+}$ is the weight immediately before rebalancing. \\

\subsection[Model performance evaluation]{Model performance evaluation}
\label{subsec:Model_eva_ch2}
As discussed in Section~\ref{subsec:perform_evalu_ch2}, the proposed models and benchmark models are evaluated via OOS mean squared error (MSE), OOS $R^2$ and DM tests. Table~\ref{tab:model_perform_ch2} shows the MSE and OOS $R^2$ for the main models. Benchmark models for the data period include (2112) and exclude (1911) the COVID-19 pandemic (from November 2019 to December 2021). In both periods, we observe that all proposed models (from fw1 to fw5) significantly outperform the benchmark models. The average OOS $R^2$ implies that the proposed MLP model with 2 hidden layers which has 4.45\% as the average OOS $R^2$, outperforms all alternative models and is followed by the one with 3 hidden layers, which achieved 3.65\% of average OOS $R^2$, but apart from the proposed MLP model with 1,2,3,4 hidden layers in Period 2112, and the proposed MLP model with 2,3,4 hidden layers, the alternative models have negative OOS $R^2$, which indicates the model overfitting issue to some extent. Model overfitting refers to a situation in which a predictive model captures not only the underlying patterns in the training data set but also the random noise. As a result, it exhibits excellent in-sample performance but poor out-of-sample performance. Deep overfitting implies the possibility of model mis-specification, which means the model form mismatch. In this sense, Model OLS, PLS, PCR and GKX2020 with 3 hidden layers may not perform as expected in this case. For the testing period, excluding the pandemic, which was from January 2013 to November 2019, the best-proposed models had the highest OOS $R^2$, which are 3.36\% and 2.5\% for 2 and 3 hidden layers, respectively. It is consistent that the proposed MLP model with 2 hidden layers is the best-performing, followed solidly by the one with 3 hidden layers, and it maintains its advantage even during the pandemic-inclusive period (2112).\\

Comparing the fitness of the models in pre-COVID-19 and COVID-19–Inclusive periods, the OOS fitness of MLP models increased when the COVID-19 testing period was included. This means the MLP models present higher compatibility under extreme market conditions. On the other hand, some models are insensitive to pandemic-induced market fluctuations, particularly traditional statistical models (OLS, PCR, PLS) and deep MLP models. Figure~\ref{fig:price_index_ch2} shows the market capital-weighted price index of 420 selected stocks, which provides an intuition of the market fluctuation during the pandemic. The grey shadow-covered period in Figure~\ref{fig:price_index_ch2} is the COVID-19 period. Together with Table~\ref{tab:model_perform_ch2}, they prove the excellence of shallow dynamic MLP models' forecasting capability during the extreme market turbulence. This finding is consistent with the GKX2020's finding that shallow MLP factor models have higher performance on stock return forecasting in the background of asset pricing.\\

\begin{table}[htbp!]
\centering
\small
\begin{tabular}{lrr@{\hskip 0.5in}lrr} 
\toprule
\textbf{Model} & \textbf{Avg\_R2} & \textbf{Avg\_MSE} & \textbf{Model} & \textbf{Avg\_R2} & \textbf{Avg\_MSE} \\
\midrule
\multicolumn{3}{c}{\textbf{Pre--COVID-19 Period (1911)}} & \multicolumn{3}{c}{\textbf{COVID-19--Inclusive Period  (2112)}} \\
\midrule
OLS          & -0.2251 & 0.0066 & OLS          & -0.2464 & 0.0086 \\
PCR          & -0.0380 & 0.0058 & PCR          & -0.0367 & 0.0074 \\
PLS          & -0.2250 & 0.0066 & PLS          & -0.2464 & 0.0086 \\
GKX2020\_fw3 & -0.3419 & 0.0092 & GKX2020\_fw3 & -0.2982 & 0.0118 \\
GKX2020\_fw4 & -0.0331 & 0.0062 & GKX2020\_fw4 & -0.0310 & 0.0083 \\
GKX2020\_fw5 & -0.0057 & 0.0061 & GKX2020\_fw5 & -0.0059 & 0.0081 \\
fw1          & -0.0025 & 0.0058 & fw1          & 0.0183  & 0.0075 \\
fw2          & 0.0336  & 0.0057 & fw2          & 0.0445  & 0.0074 \\
fw3          & 0.0250  & 0.0058 & fw3          & 0.0365  & 0.0076 \\
fw4          & 0.0071  & 0.0060 & fw4          & 0.0106  & 0.0079 \\
fw5          & -0.0047 & 0.0061 & fw5          & -0.0023 & 0.0080 \\
\bottomrule
\end{tabular}
\caption[Model performance during Pre-COVID-19 and COVID-19-Inclusive periods]{Model performance comparison between Pre-COVID-19 (1911) and COVID-19-Inclusive (2112) periods. The ‘$fw(i)$’, $i=1,2,\dots,5$ represents the proposed MLP models with 1 to 5 hidden layers, while ‘$GKX2020\_fw(i)$’ denotes the reference models. Avg\_R2 and Avg\_MSE represent the average R-squared and Mean Squared Error, respectively. The values of $R^2$ are expressed in decimals; for instance, $-0.2251$ in the OLS model represents $-22.51\%$.}
\label{tab:model_perform_ch2}
\end{table}

\begin{figure}[htbp!]
\begin{center}
\includegraphics[width=0.9\columnwidth]{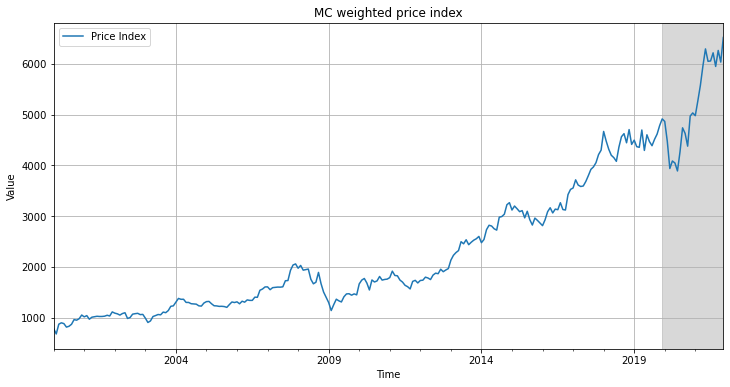}
\end{center}
\caption[Value-weighted price index of selected 420 stocks]{Value-weighted price index of selected 420 stocks. The grey shadow covers the COVID-19 period from December 2019 to December 2021. ‘MC’ means market capital.}
\label{fig:price_index_ch2}
\end{figure}

\begin{figure}[htbp]
    \centering
    
    \begin{subfigure}{0.48\textwidth}
        \centering
        \includegraphics[width=\linewidth]{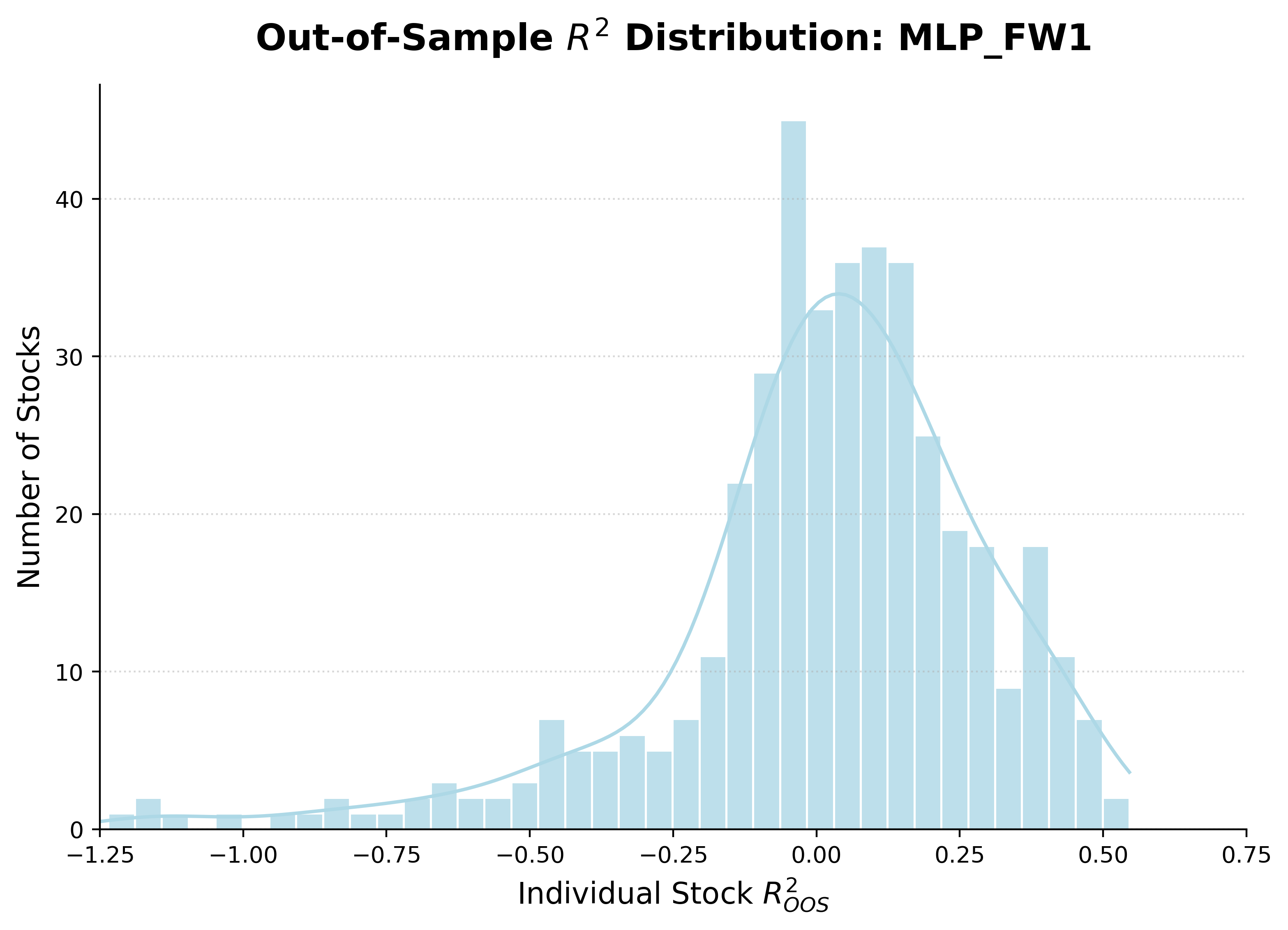}
        \caption{fw1(1911)}
    \end{subfigure}
    \hfill 
    \begin{subfigure}{0.48\textwidth}
        \centering
        \includegraphics[width=\linewidth]{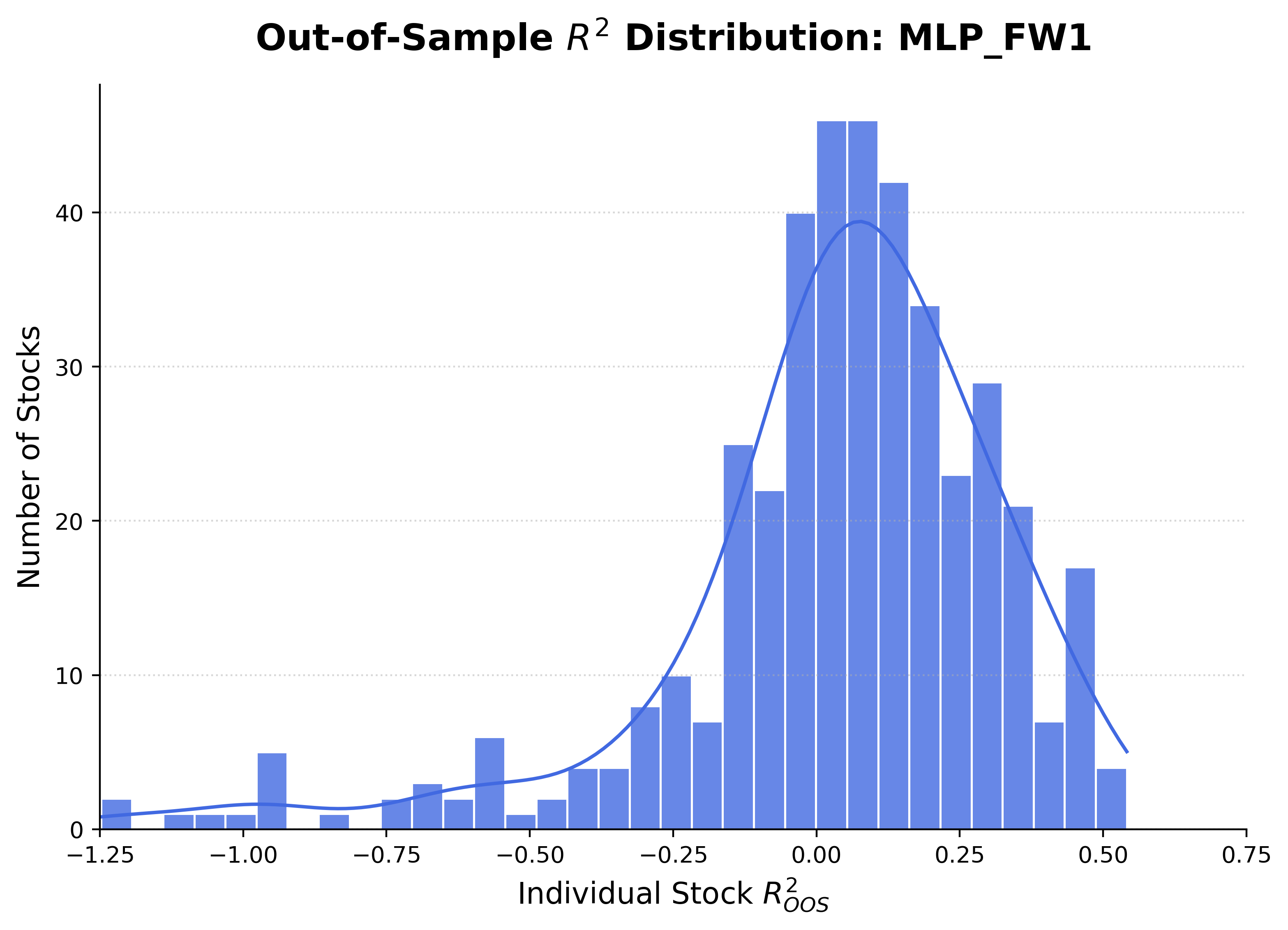}
        \caption{fw1(2112)}
    \end{subfigure}

    \vspace{0.5cm} 

    \begin{subfigure}{0.48\textwidth}
        \centering
        \includegraphics[width=\linewidth]{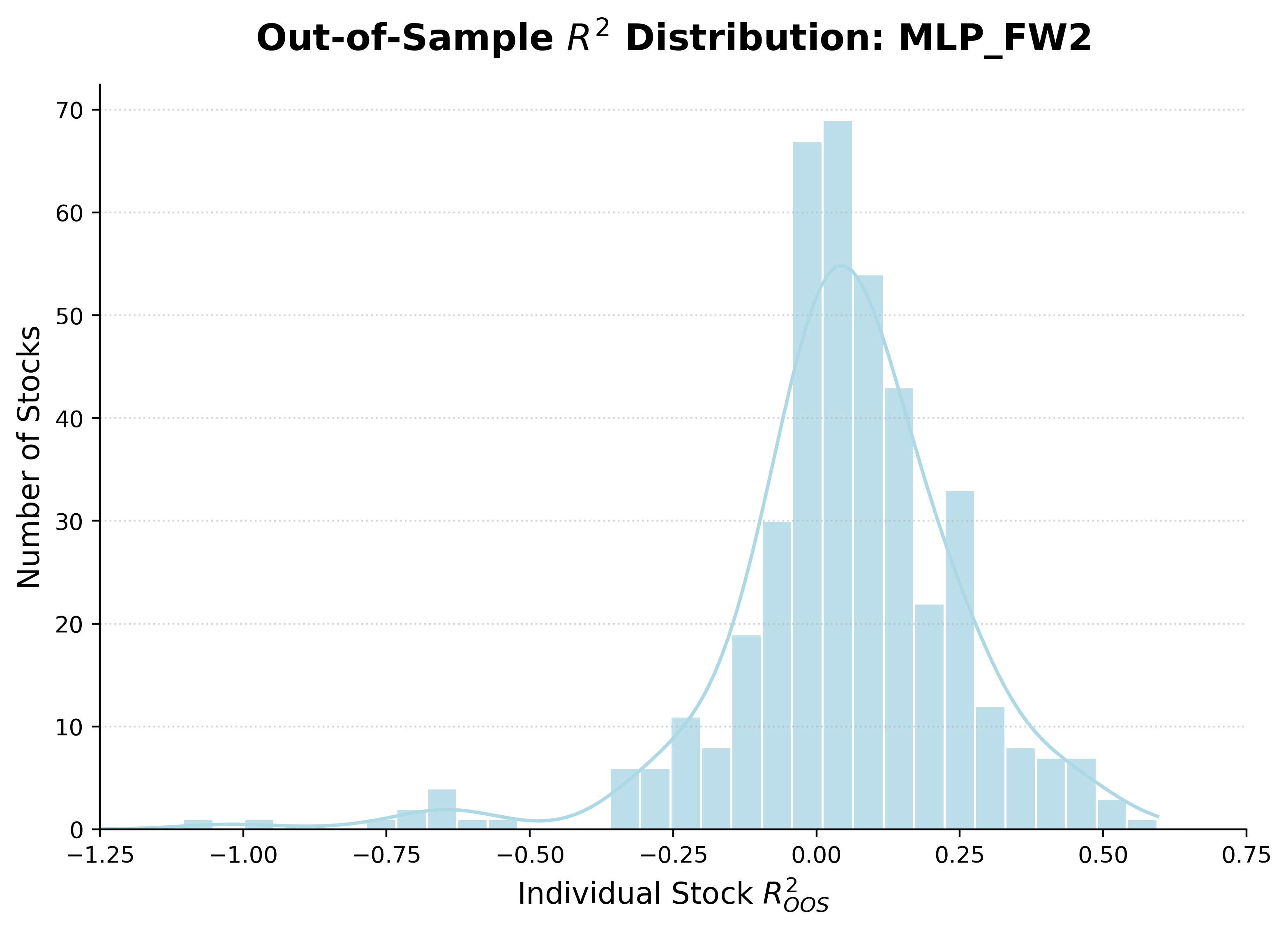}
        \caption{fw2(1911)}
    \end{subfigure}
    \hfill
    \begin{subfigure}{0.48\textwidth}
        \centering
        \includegraphics[width=\linewidth]{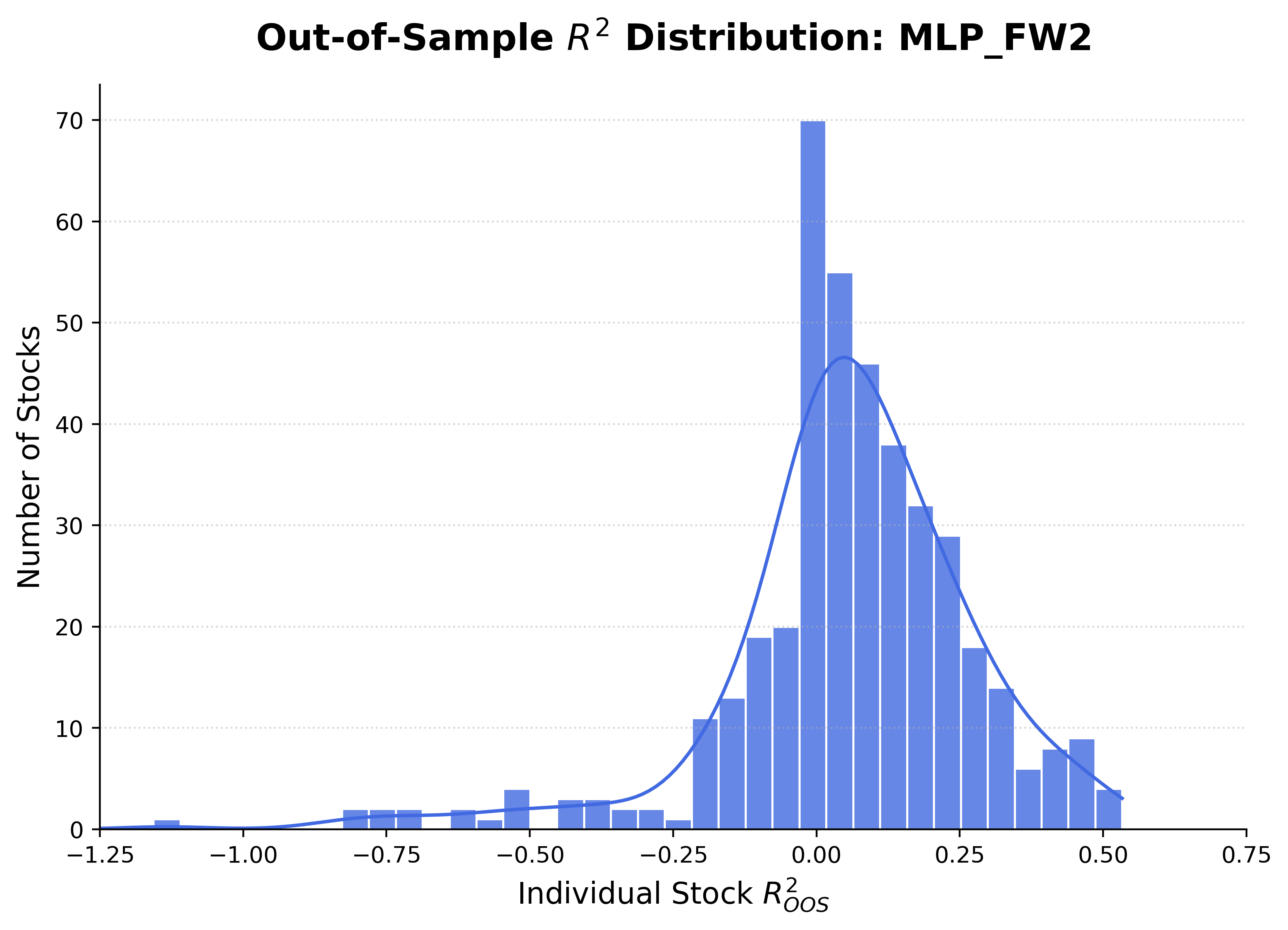}
        \caption{fw2(2112)}
    \end{subfigure}

    \vspace{0.5cm}

    \begin{subfigure}{0.48\textwidth}
        \centering
        \includegraphics[width=\linewidth]{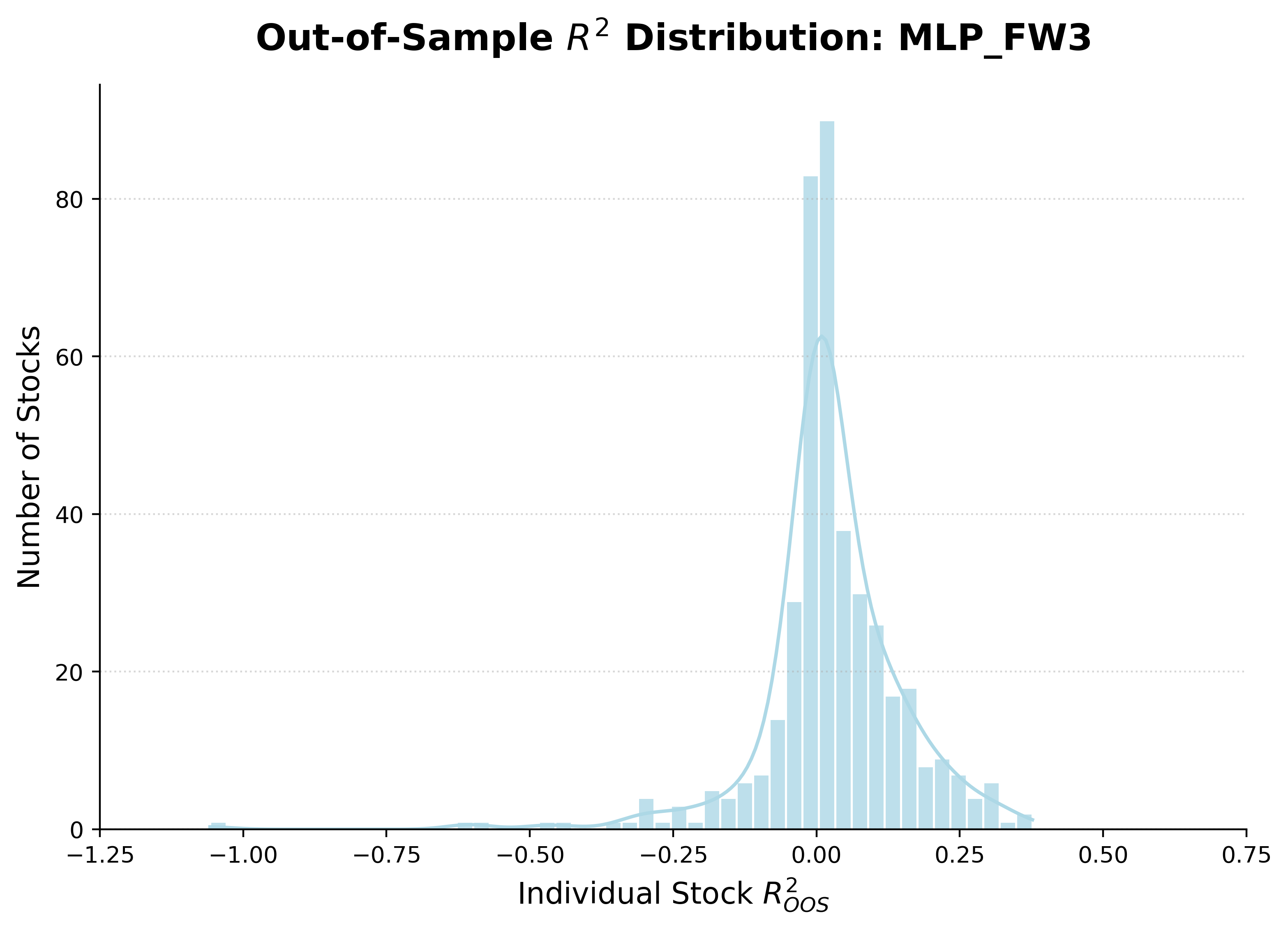}
        \caption{fw3(1911)}
    \end{subfigure}
    \hfill
    \begin{subfigure}{0.48\textwidth}
        \centering
        \includegraphics[width=\linewidth]{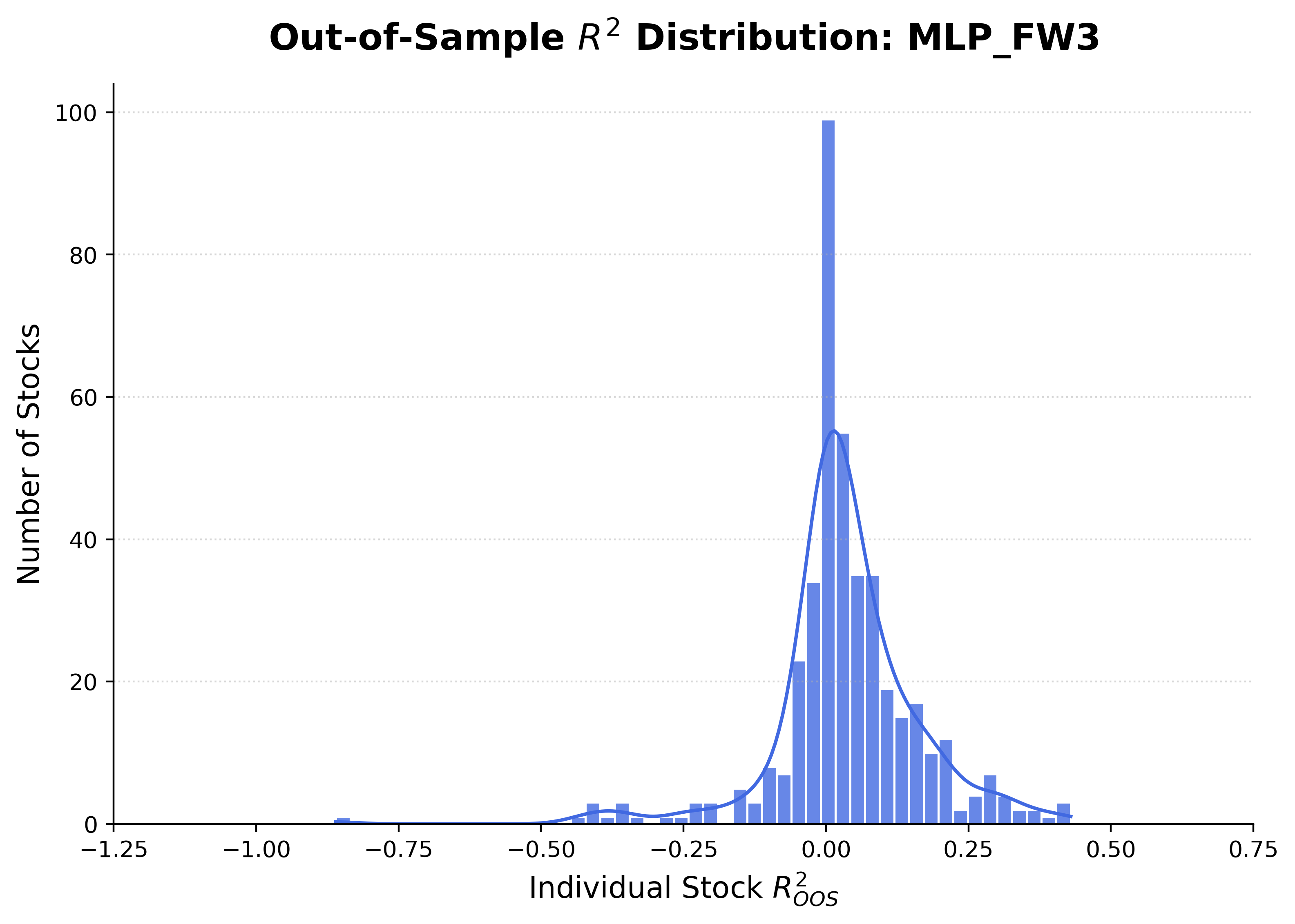}
        \caption{fw3(2112)}
    \end{subfigure}

    \caption[OOS $R^2$ distribution comparison for fw1, fw2 and fw3.]{OOS $R^2$ distribution comparison for fw1, fw2 and fw3.The X-axis exhibits the OOS $R^2$ values, and the Y-axis exhibits the number of stocks. The ‘$fw(i)(2112)$, $i=1,2,\dots,5$’ means the proposed MLP models with the COVID-19 effect, and the '$fw(i)(1911)$', $i=1,2,…,5$ means the proposed MLP models without the COVID-19 effect.}
    \label{fig:r2_dis_ch2_fw1_fw3}
\end{figure}

\begin{figure}[htbp]
    \centering
    
    \begin{subfigure}{0.48\textwidth}
        \centering
        \includegraphics[width=\linewidth]{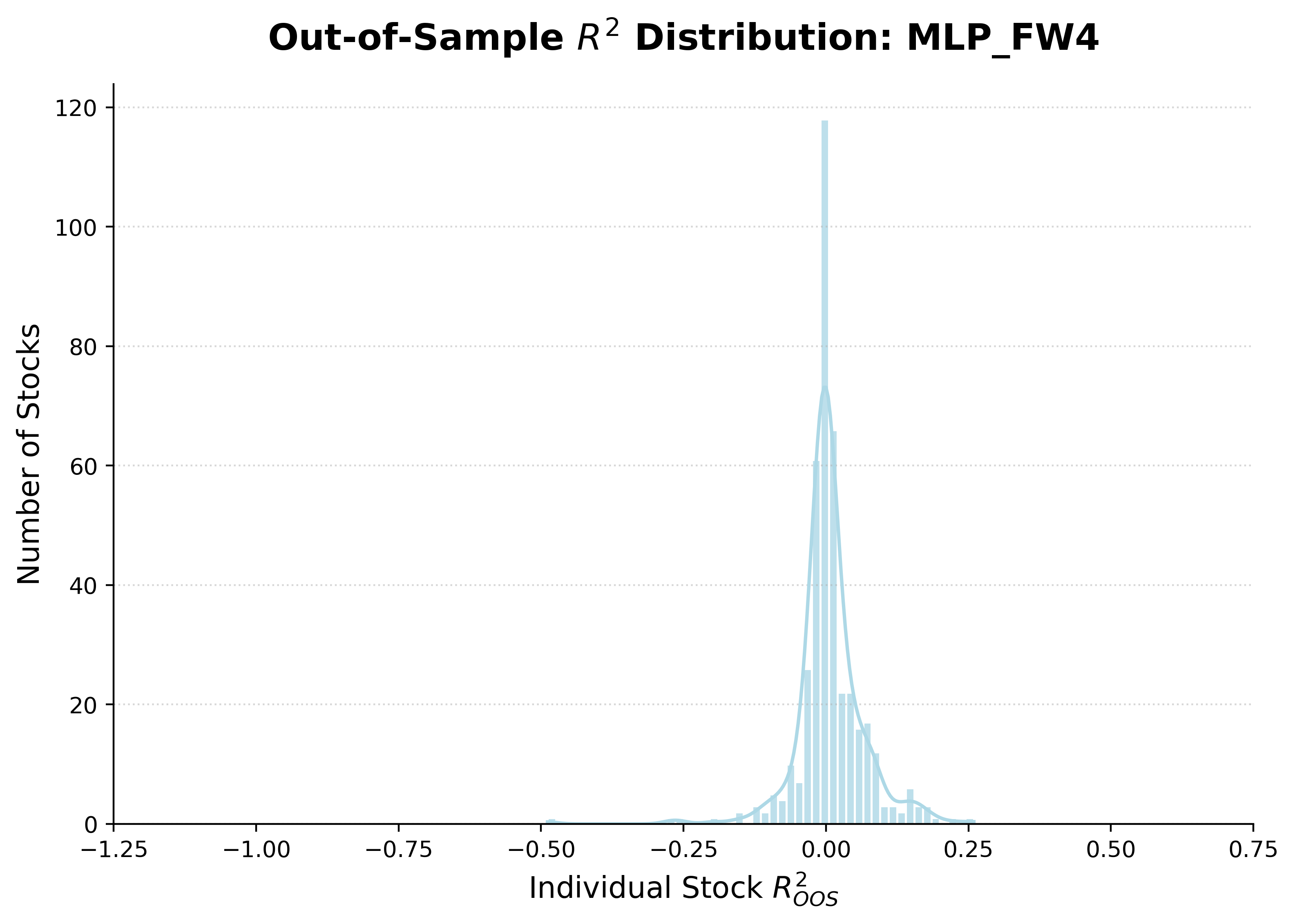}
        \caption{fw4(1911)}
    \end{subfigure}
    \hfill 
    \begin{subfigure}{0.48\textwidth}
        \centering
        \includegraphics[width=\linewidth]{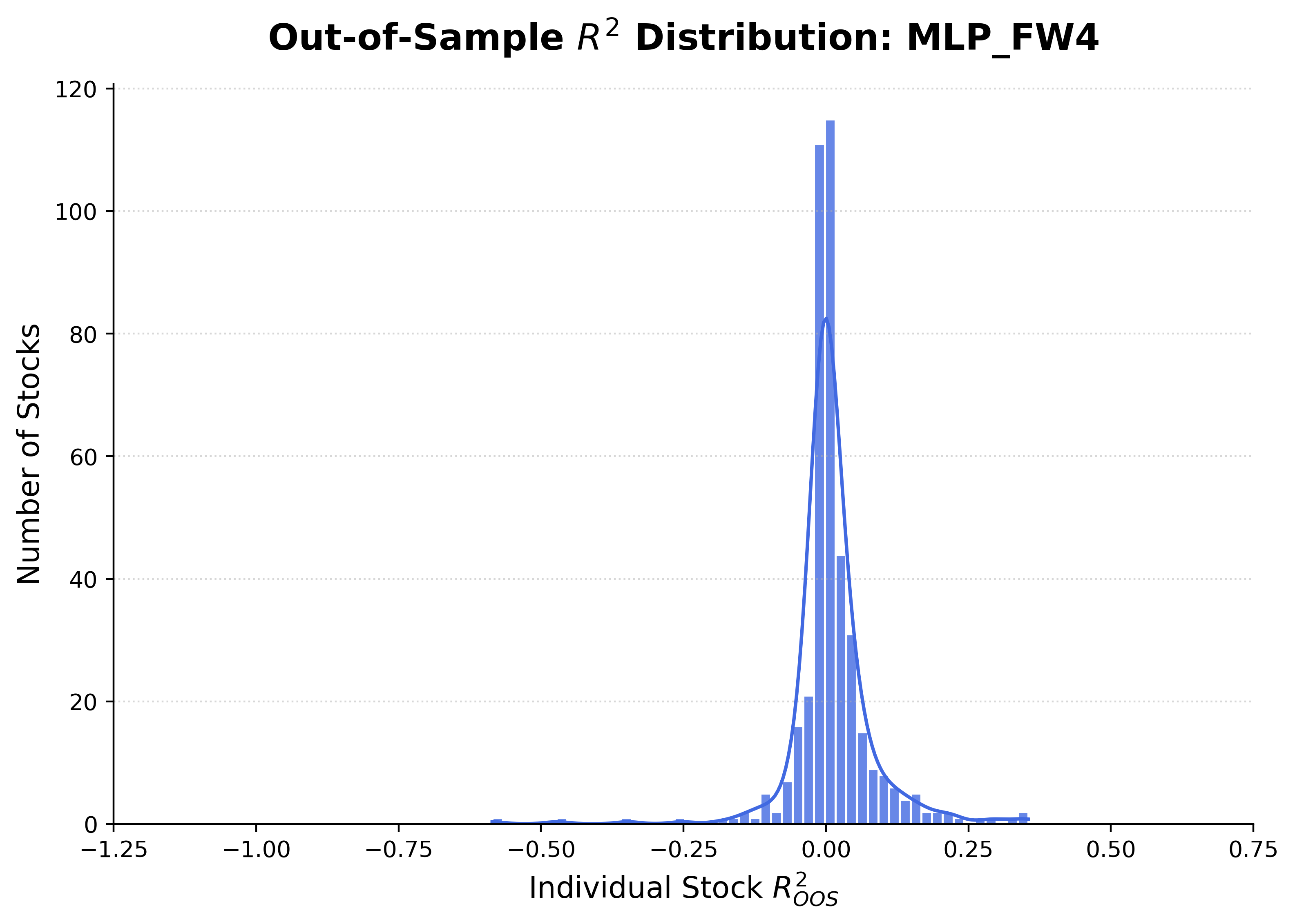}
        \caption{fw4(2112)}
    \end{subfigure}

    \vspace{0.5cm} 

    \begin{subfigure}{0.48\textwidth}
        \centering
        \includegraphics[width=\linewidth]{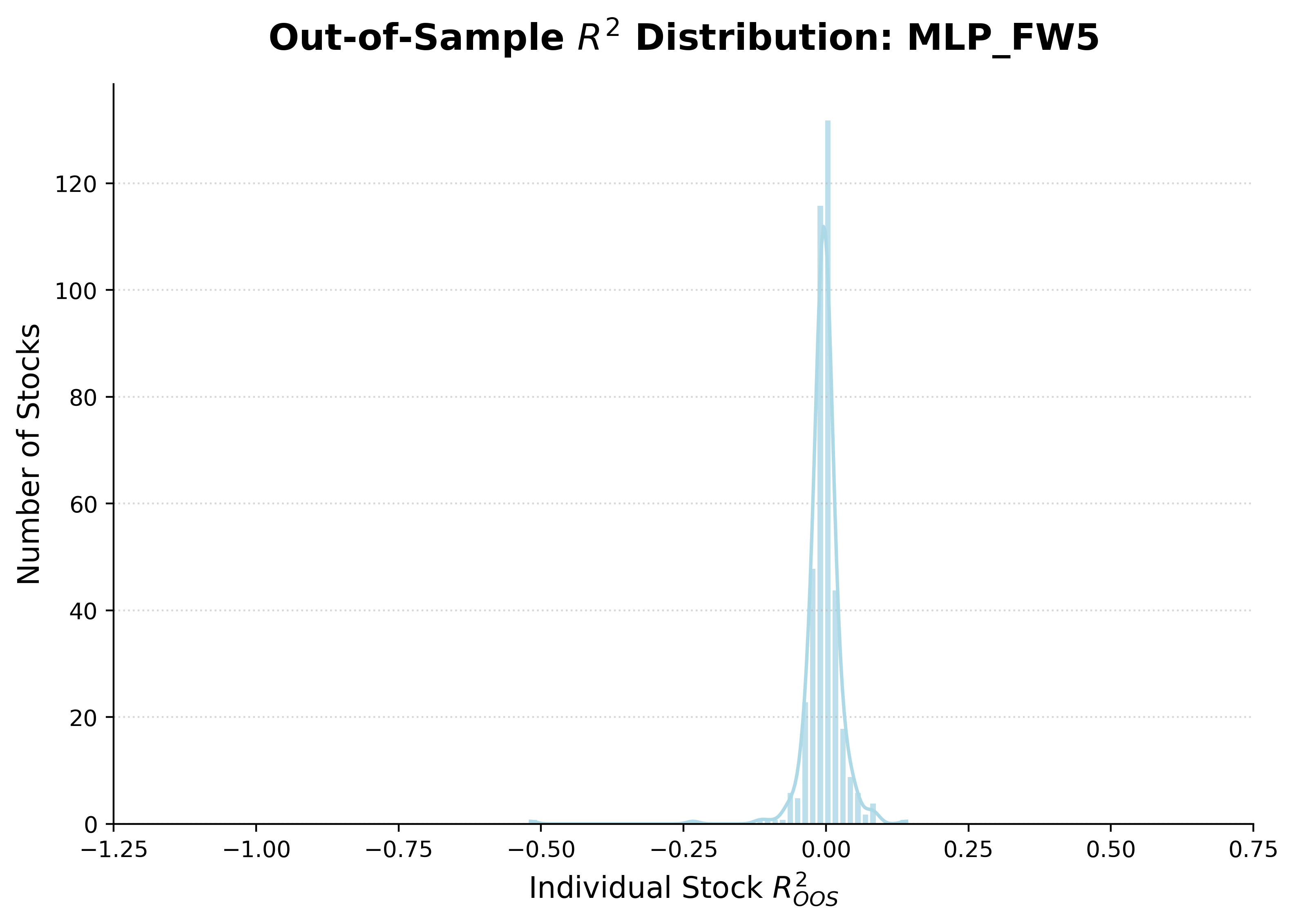}
        \caption{fw5(1911)}
    \end{subfigure}
    \hfill
    \begin{subfigure}{0.48\textwidth}
        \centering
        \includegraphics[width=\linewidth]{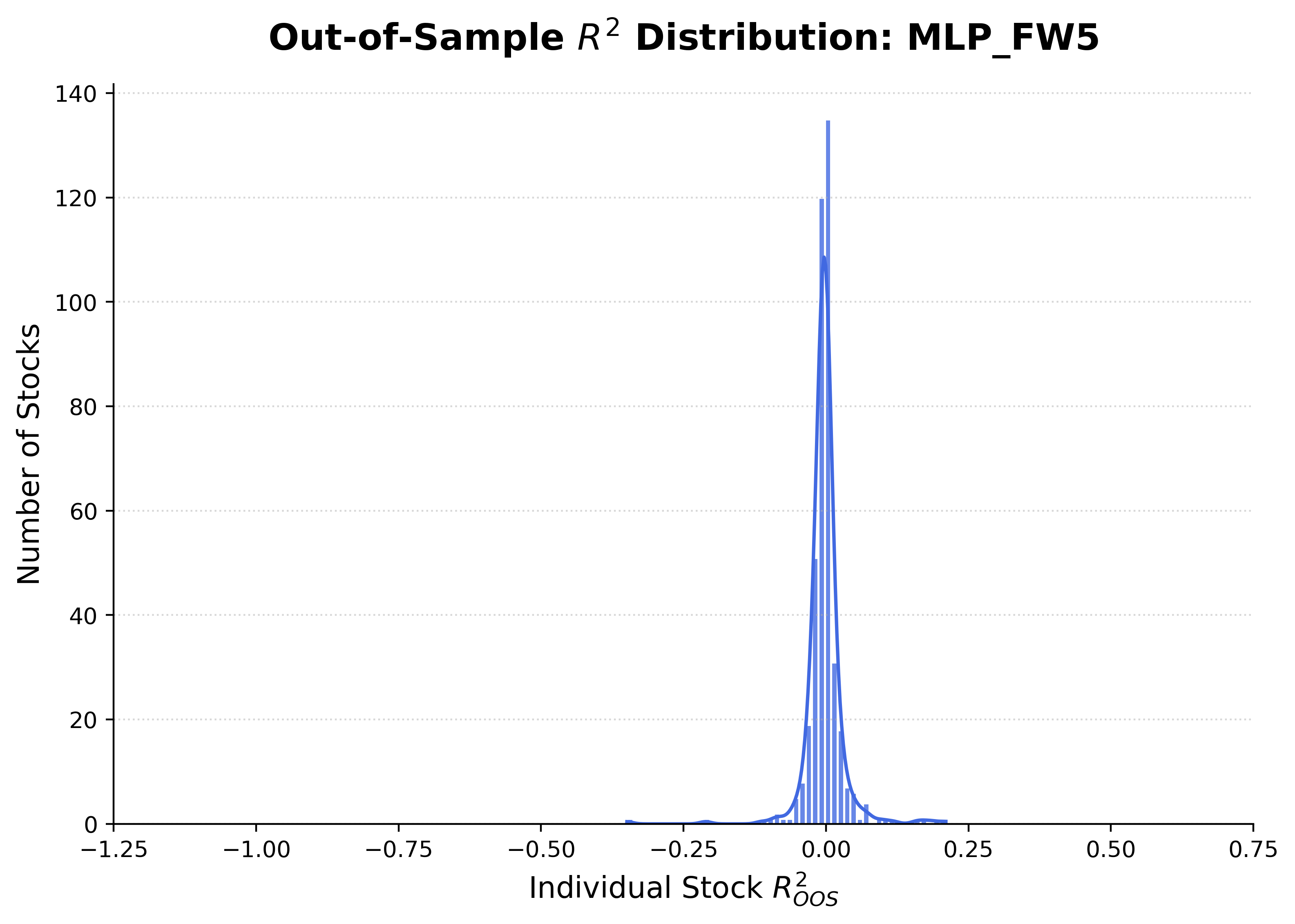}
        \caption{fw5(2112)}
    \end{subfigure}

    \vspace{0.5cm}

    \begin{subfigure}{0.48\textwidth}
        \centering
        \includegraphics[width=\linewidth]{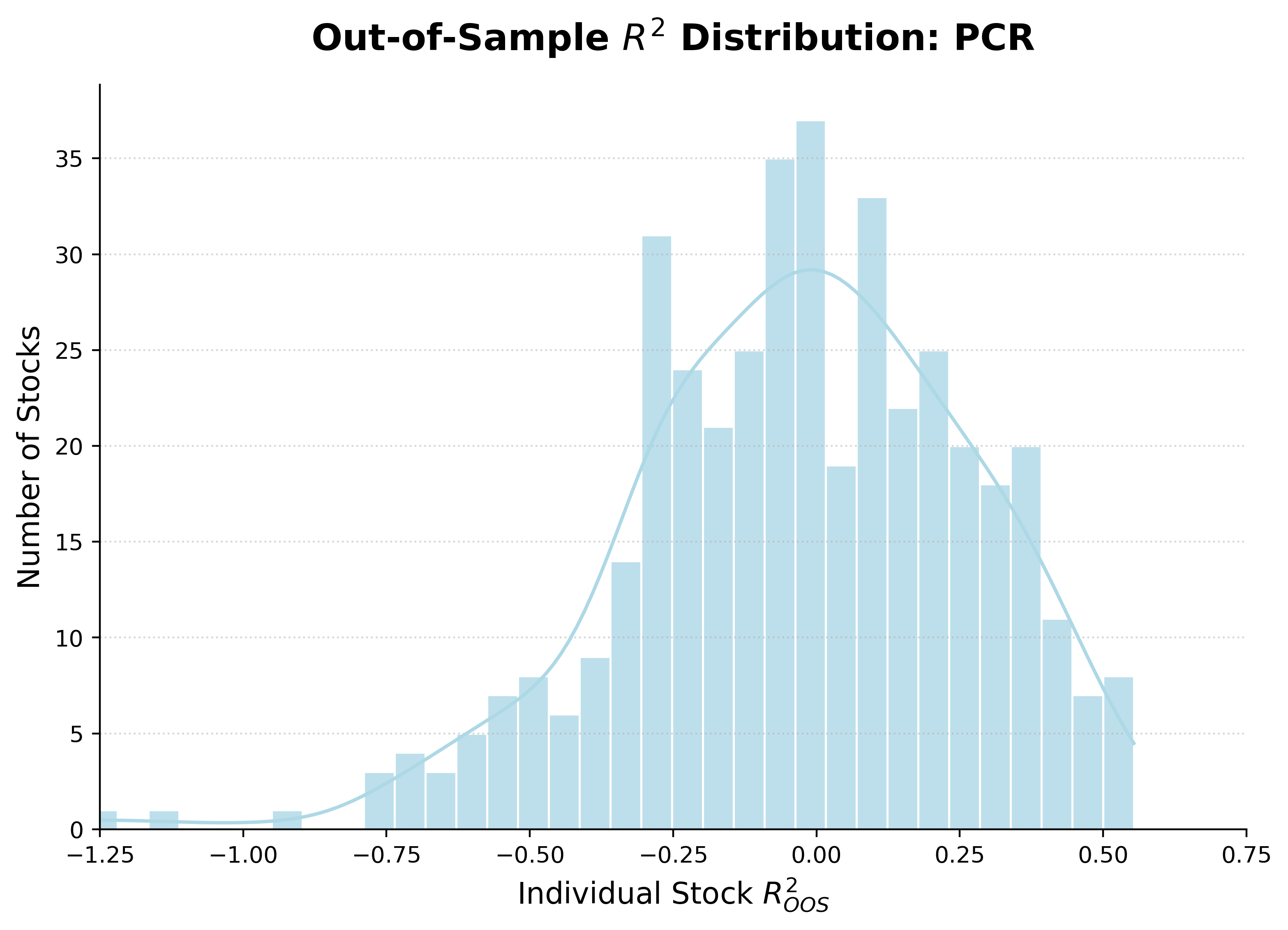}
        \caption{PCR(1911)}
    \end{subfigure}
    \hfill
    \begin{subfigure}{0.48\textwidth}
        \centering
        \includegraphics[width=\linewidth]{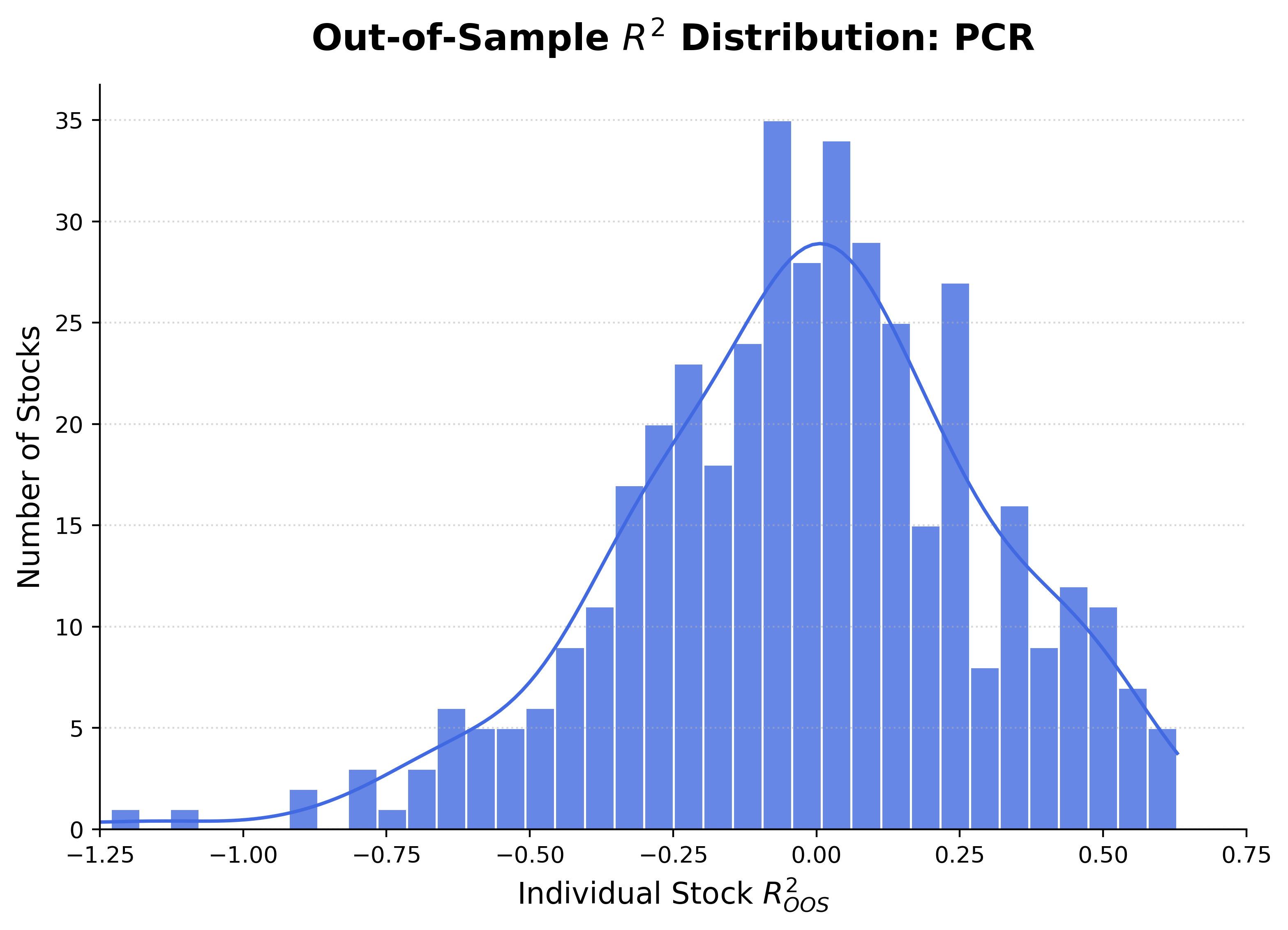}
        \caption{PCR(2112)}
    \end{subfigure}

    \caption[OOS $R^2$ distribution comparison for fw4, fw5 and PCR.]{OOS $R^2$ distribution comparison for fw4, fw5 and PCR.The X-axis exhibits the OOS $R^2$ values, and the Y-axis exhibits the number of stocks.  The ‘$fw(i)(2112)$ and PCR(2112), $i=1,2,\dots,5$’ means the proposed MLP models with the COVID-19 effect, and the '$fw(i)(1911)$' and PCR(1911), $i=1,2,…,5$ means the proposed MLP models without the COVID-19 effect.}
    \label{fig:r2_dis_ch2_fw4_PCR}
\end{figure}

\begin{figure}[htbp]
    \centering
    
    \begin{subfigure}{0.48\textwidth}
        \centering
        \includegraphics[width=\linewidth]{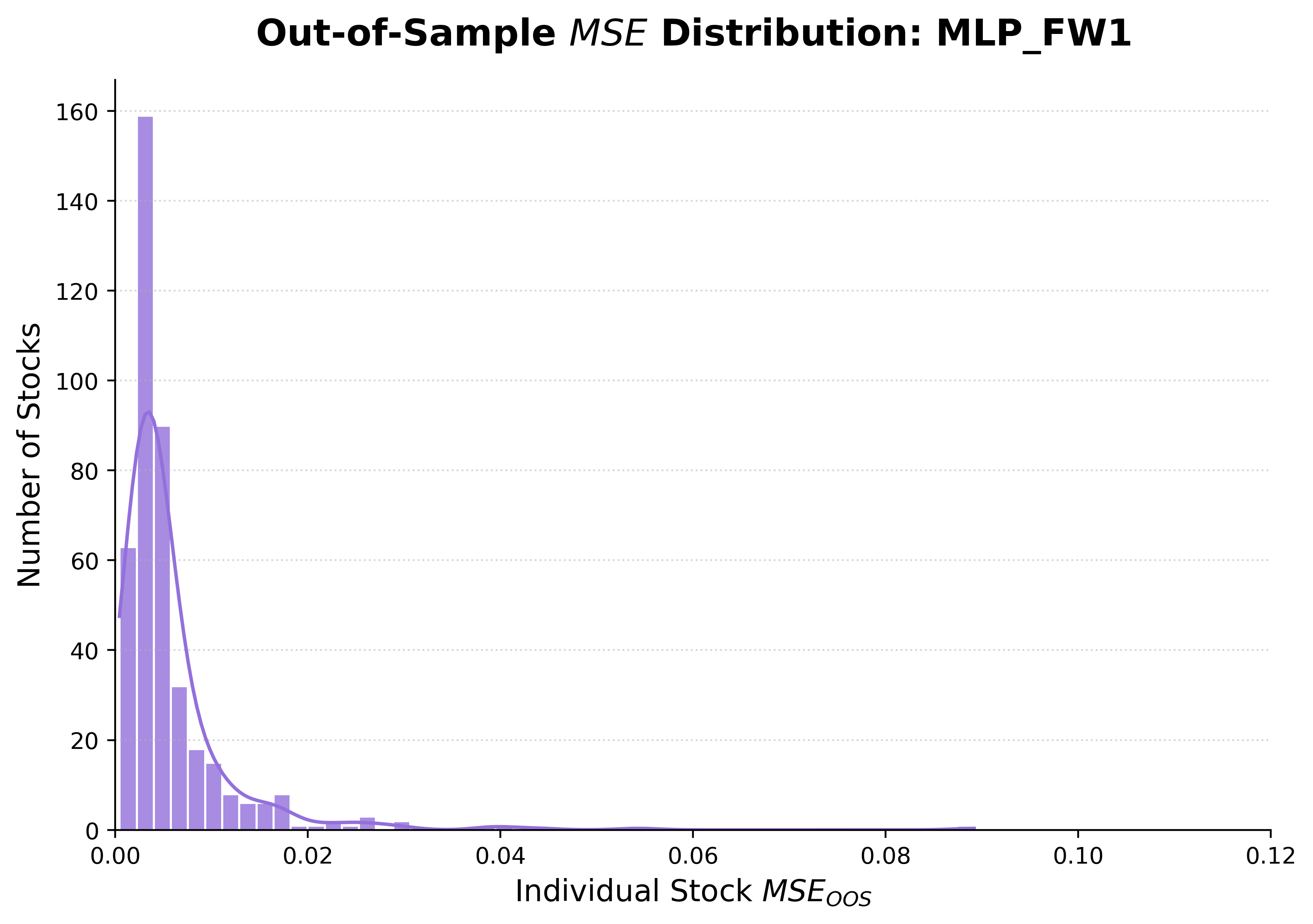}
        \caption{fw1(1911)}
    \end{subfigure}
    \hfill 
    \begin{subfigure}{0.48\textwidth}
        \centering
        \includegraphics[width=\linewidth]{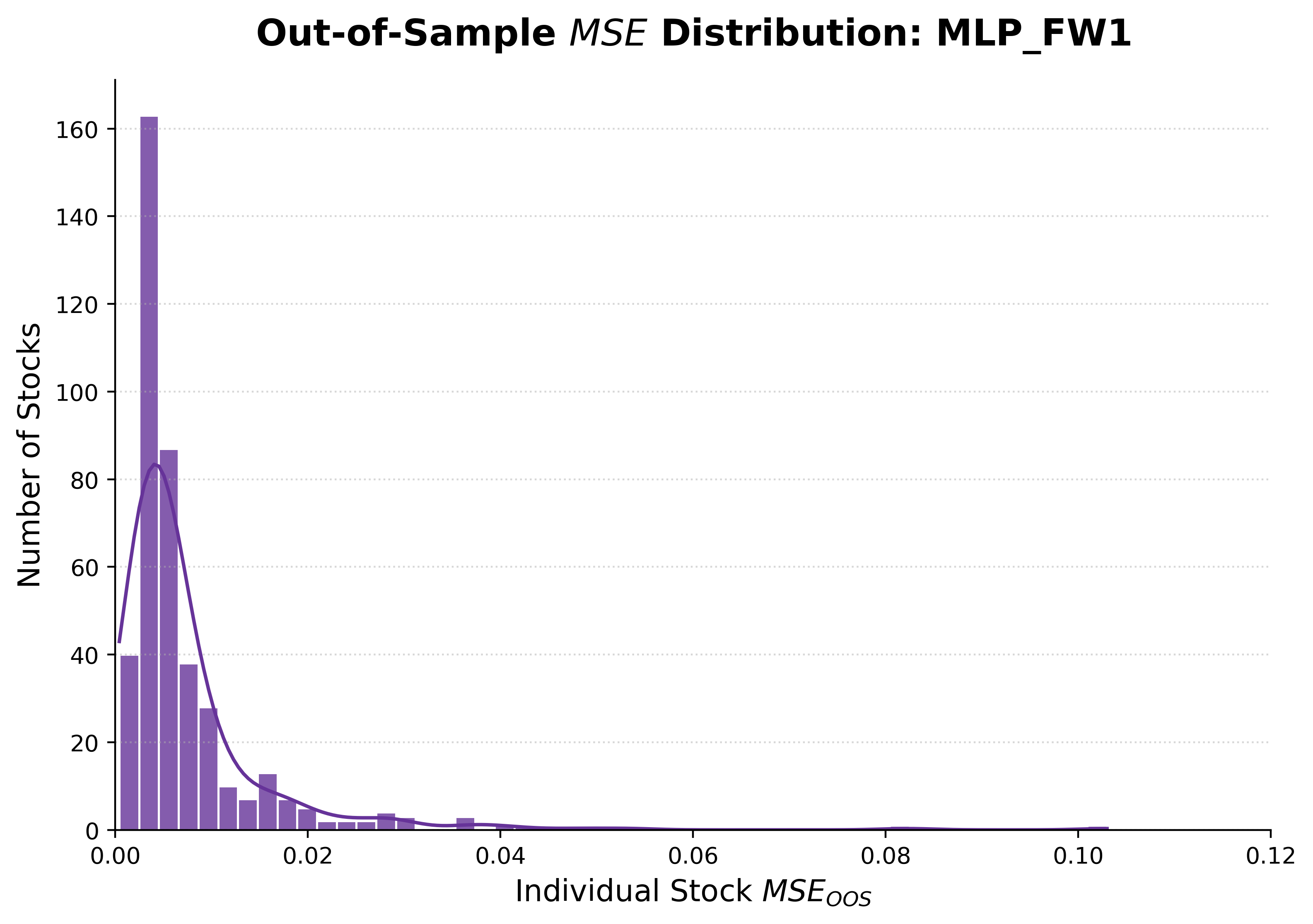}
        \caption{fw1(2112)}
    \end{subfigure}

    \vspace{0.5cm} 

    \begin{subfigure}{0.48\textwidth}
        \centering
        \includegraphics[width=\linewidth]{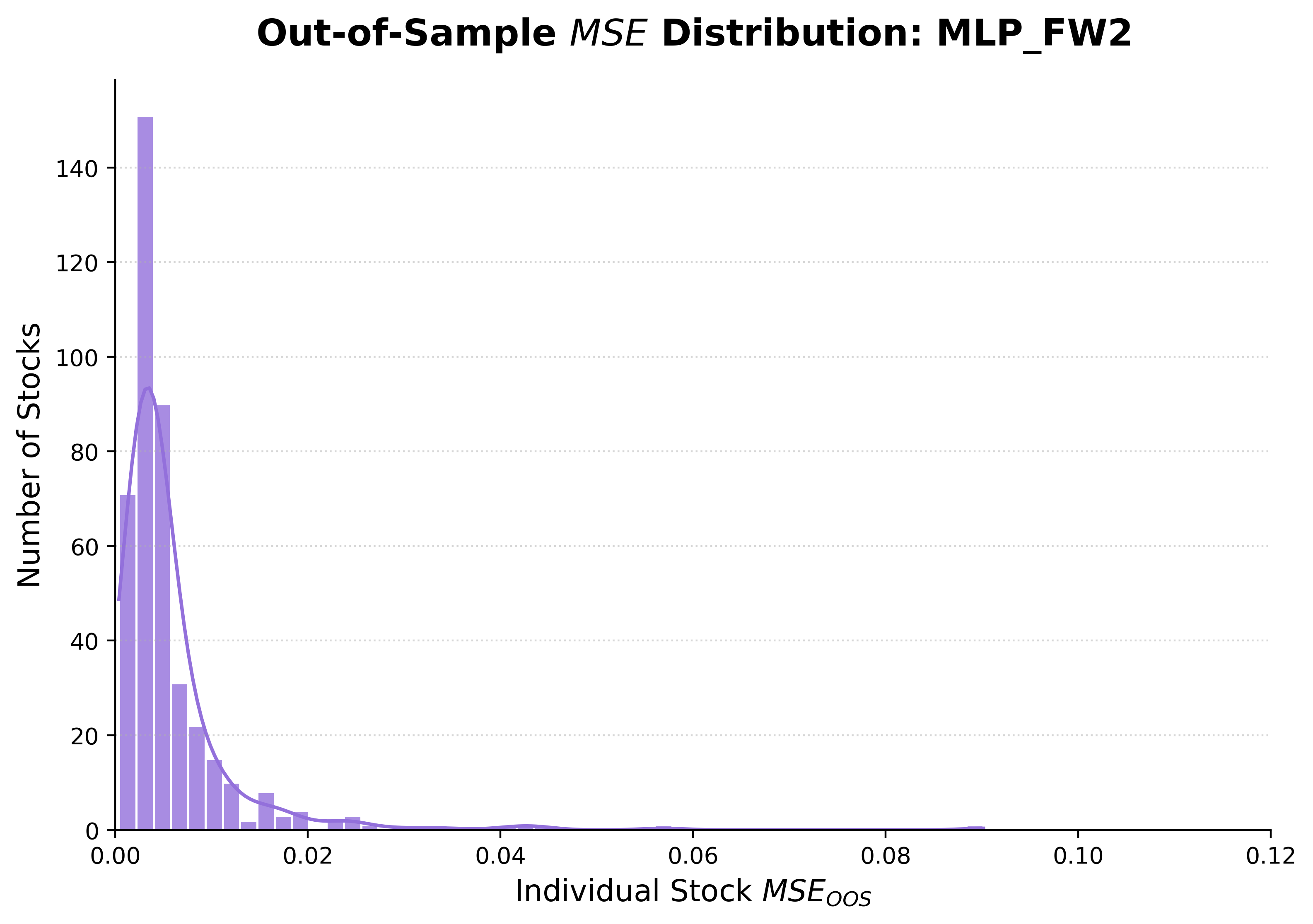}
        \caption{fw2(1911)}
    \end{subfigure}
    \hfill
    \begin{subfigure}{0.48\textwidth}
        \centering
        \includegraphics[width=\linewidth]{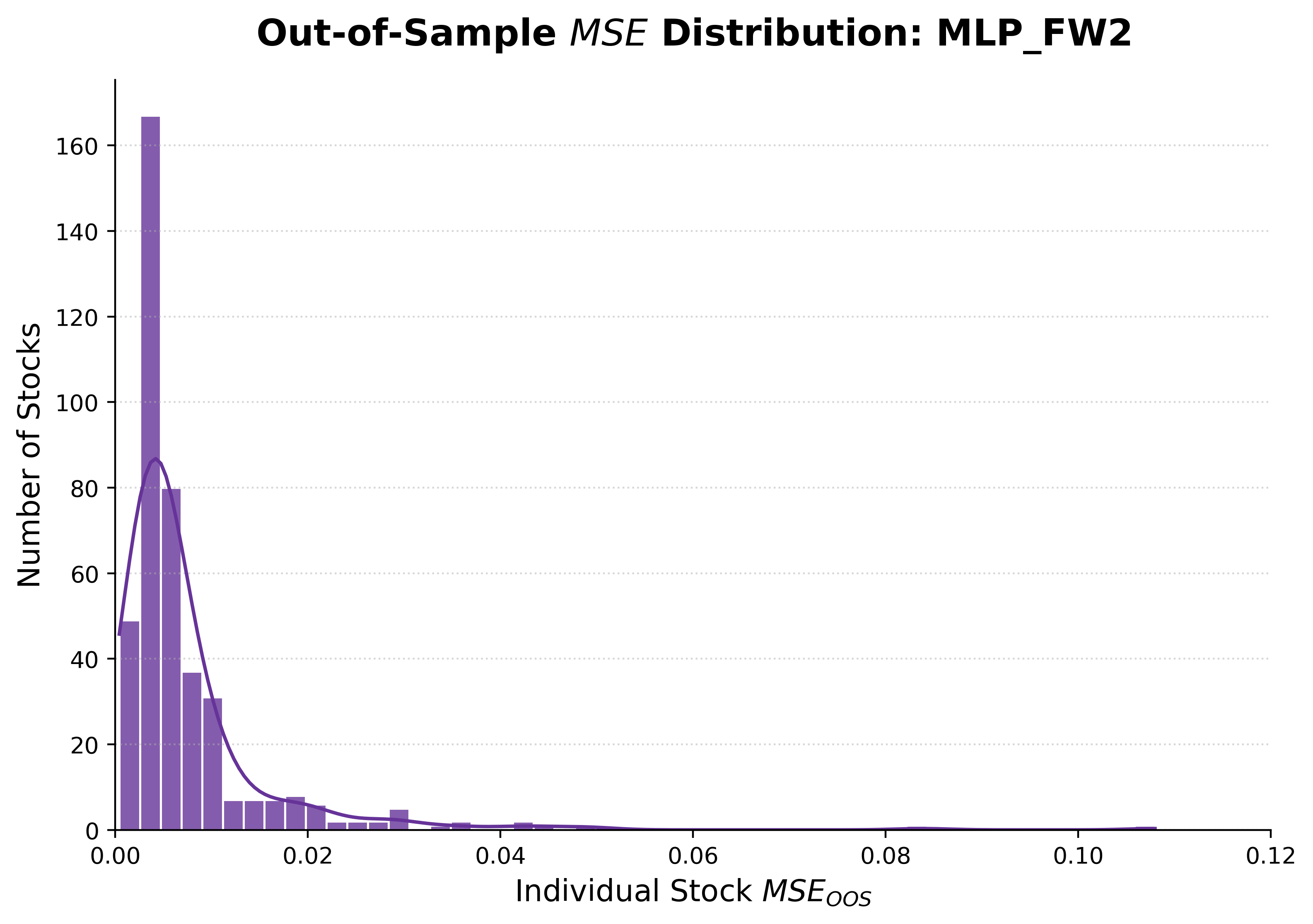}
        \caption{fw2(2112)}
    \end{subfigure}

    \vspace{0.5cm}

    \begin{subfigure}{0.48\textwidth}
        \centering
        \includegraphics[width=\linewidth]{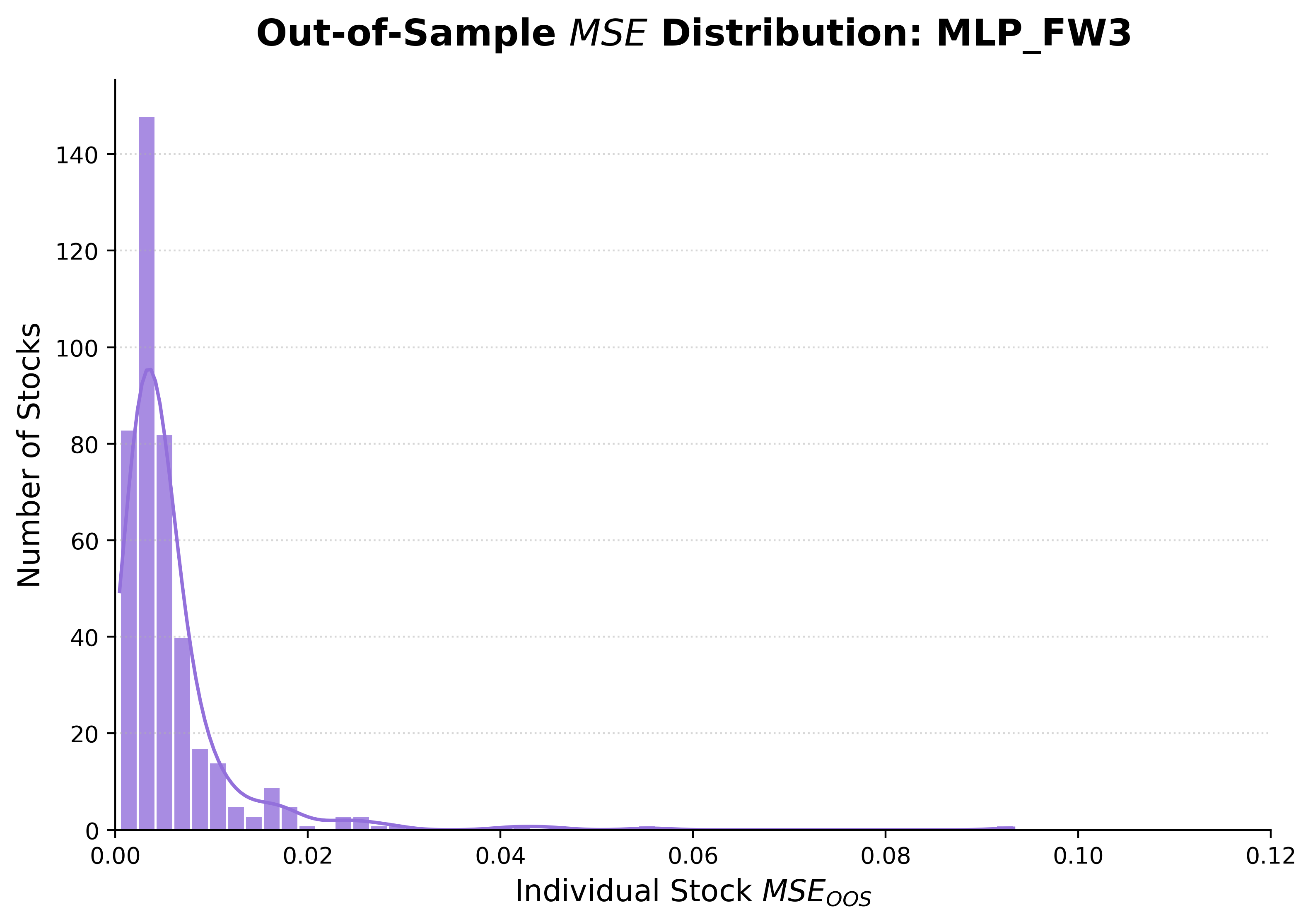}
        \caption{fw3(1911)}
    \end{subfigure}
    \hfill
    \begin{subfigure}{0.48\textwidth}
        \centering
        \includegraphics[width=\linewidth]{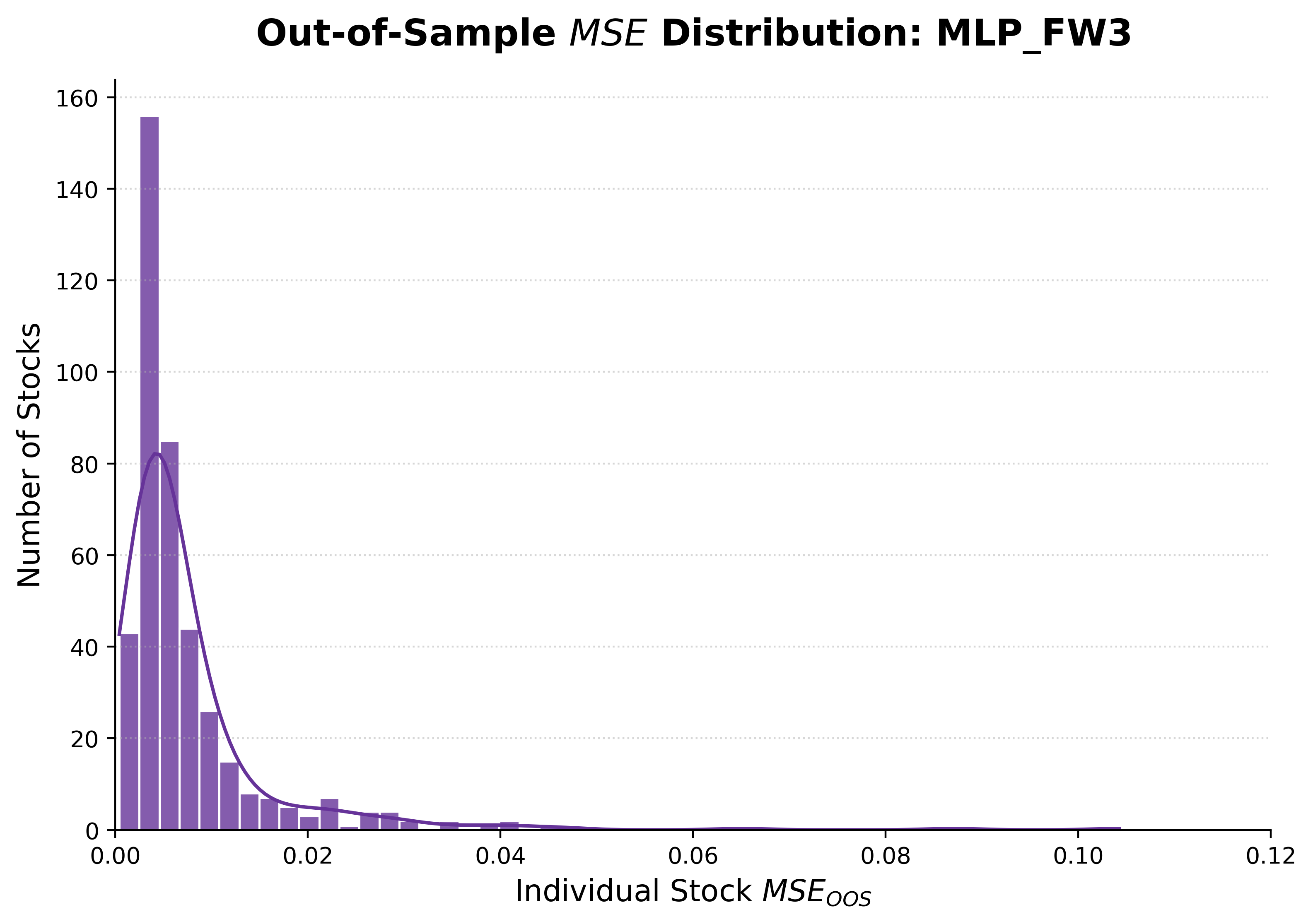}
        \caption{fw3(2112)}
    \end{subfigure}

    \caption[OOS $MSE$ distribution comparison for fw1, fw2 and fw3.]{OOS $MSE$ distribution comparison for fw1, fw2 and fw3.The X-axis exhibits the OOS $MSE$ values, and the Y-axis exhibits the number of stocks. The ‘$fw(i)(2112)$, $i=1,2,\dots,5$’ means the proposed MLP models with the COVID-19 effect, and the '$fw(i)(1911)$', $i=1,2,…,5$ means the proposed MLP models without the COVID-19 effect.}
    \label{fig:mse_dis_ch2_fw1_fw3}
\end{figure}

\begin{figure}[htbp]
    \centering
    
    \begin{subfigure}{0.48\textwidth}
        \centering
        \includegraphics[width=\linewidth]{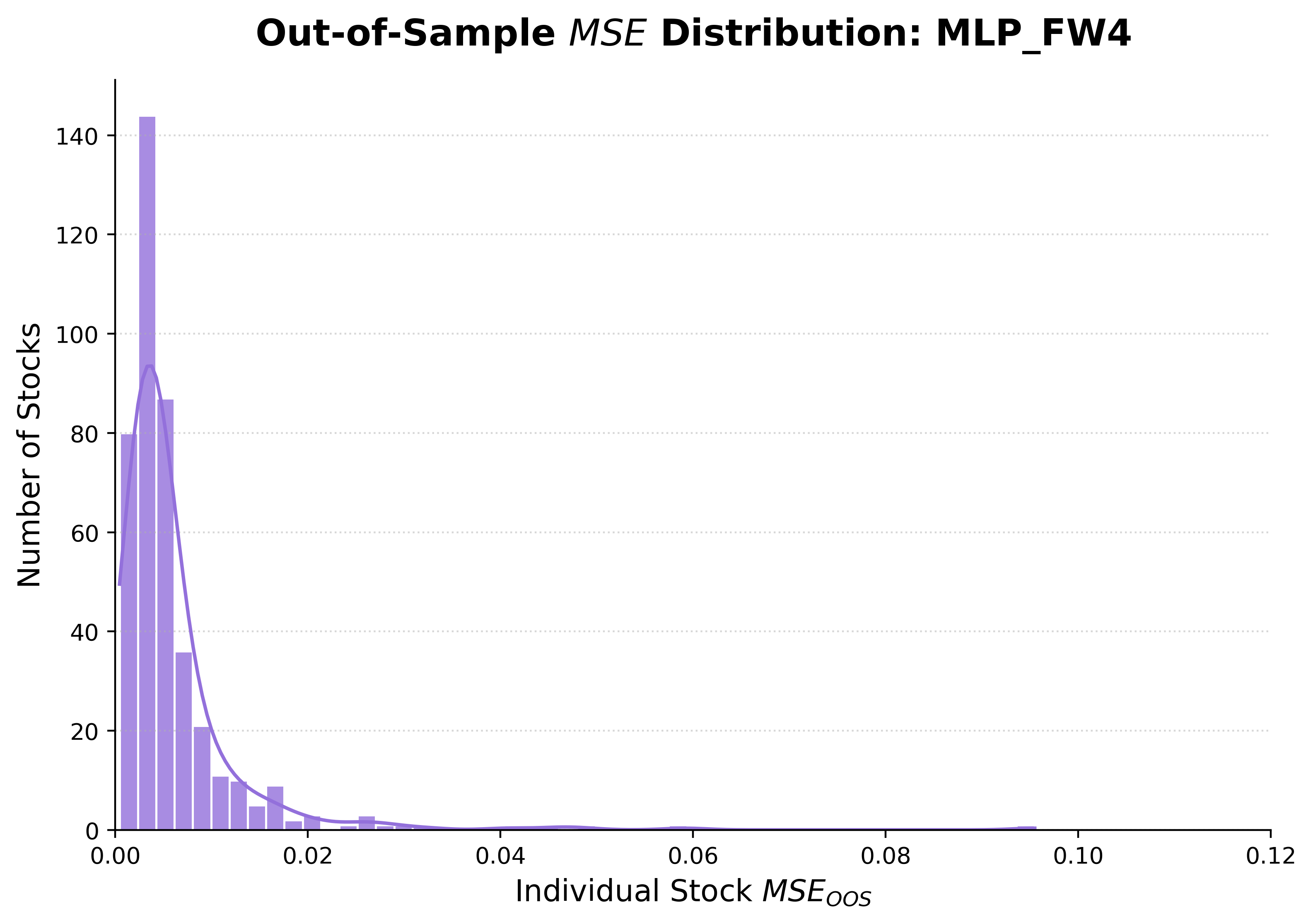}
        \caption{fw4(1911)}
    \end{subfigure}
    \hfill 
    \begin{subfigure}{0.48\textwidth}
        \centering
        \includegraphics[width=\linewidth]{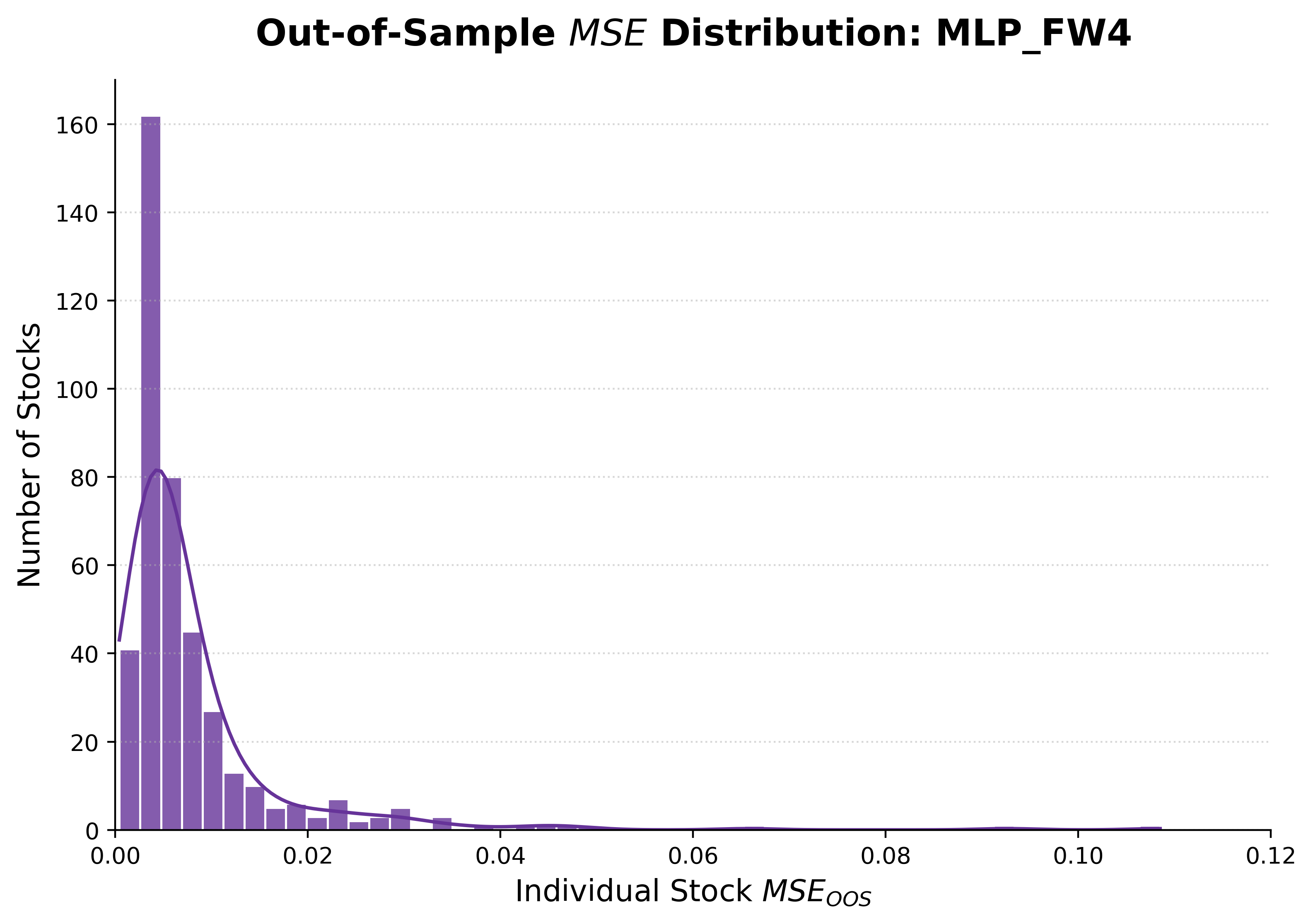}
        \caption{fw4(2112)}
    \end{subfigure}

    \vspace{0.5cm} 

    \begin{subfigure}{0.48\textwidth}
        \centering
        \includegraphics[width=\linewidth]{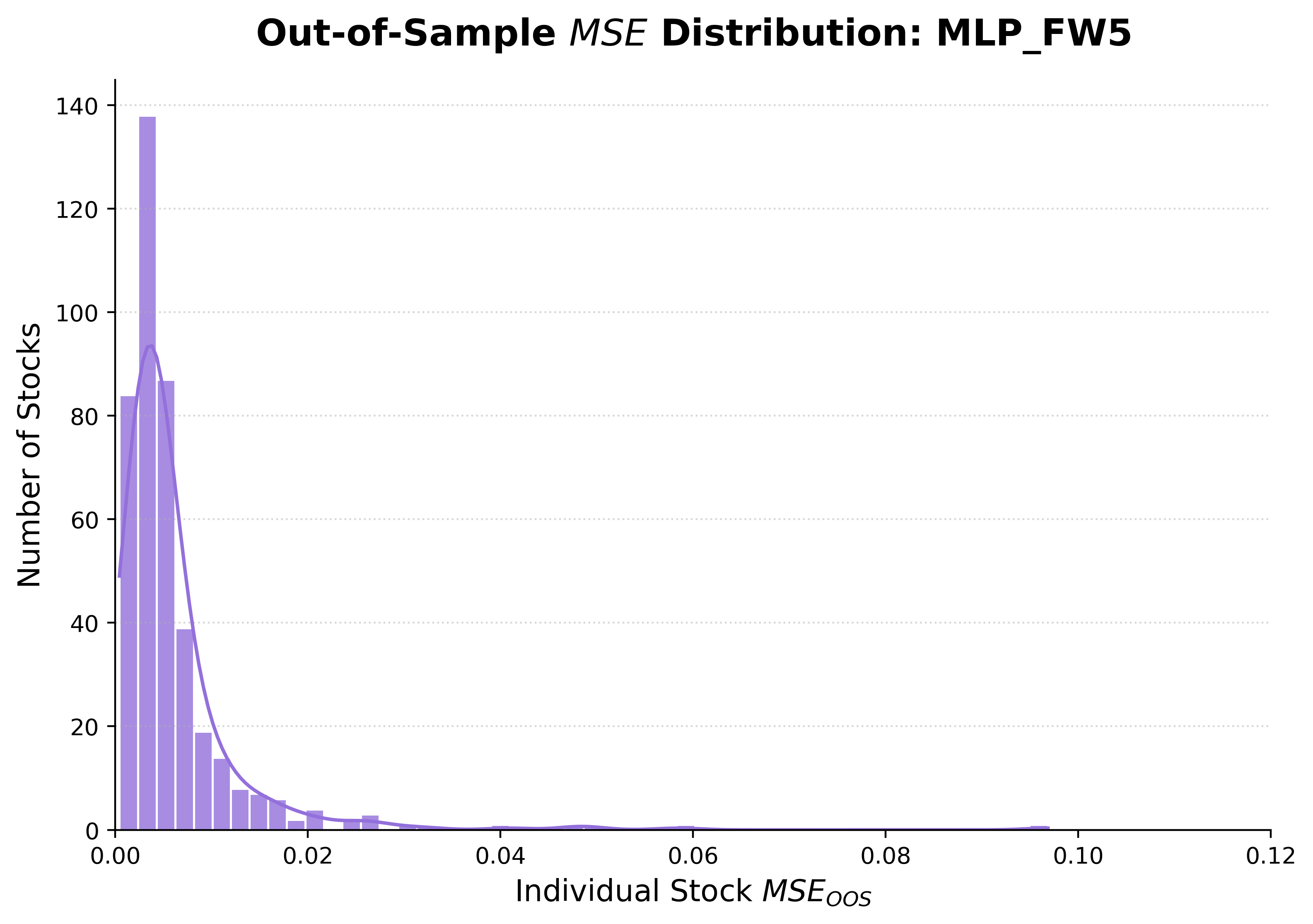}
        \caption{fw5(1911)}
    \end{subfigure}
    \hfill
    \begin{subfigure}{0.48\textwidth}
        \centering
        \includegraphics[width=\linewidth]{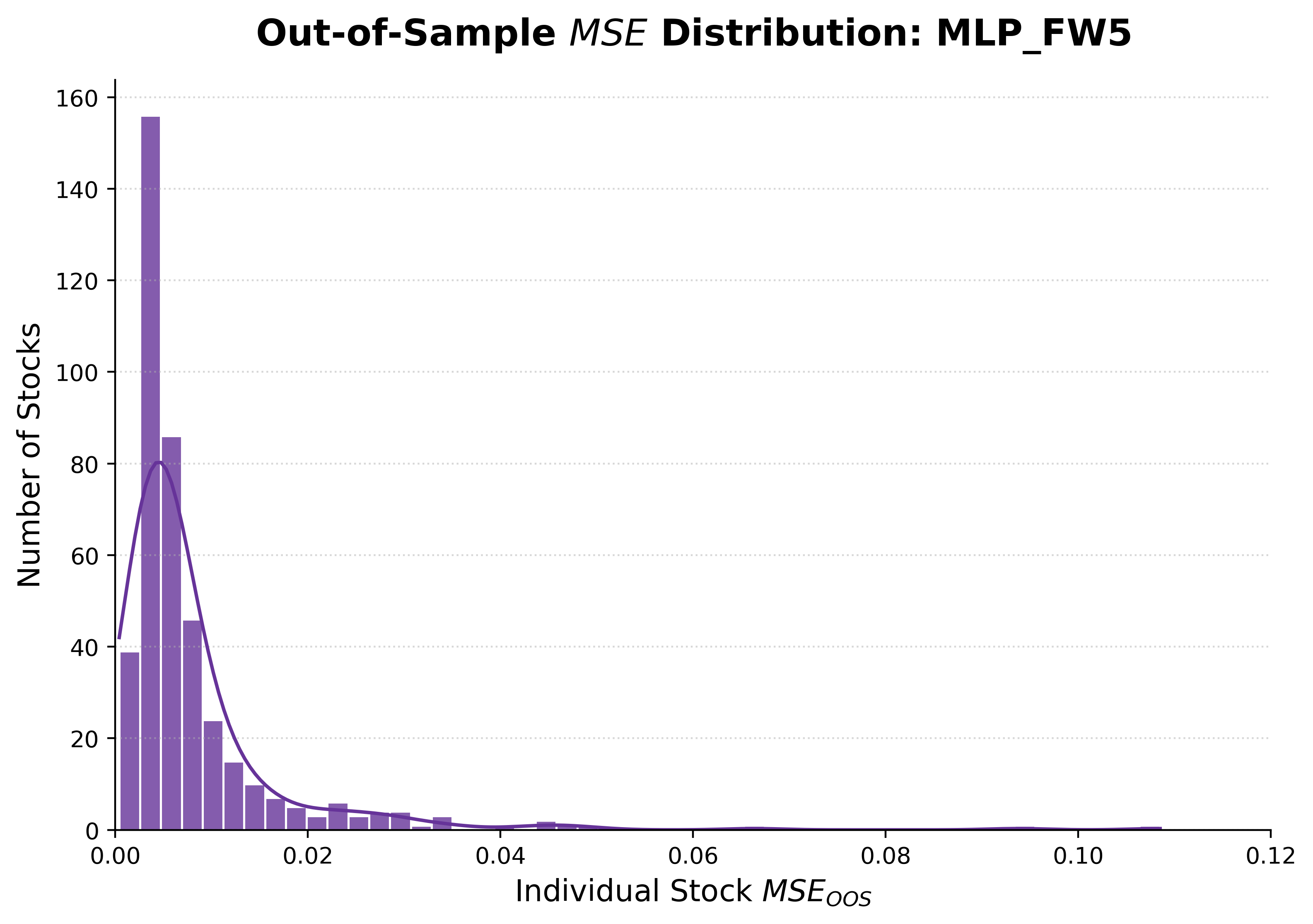}
        \caption{fw5(2112)}
    \end{subfigure}

    \vspace{0.5cm}

    \begin{subfigure}{0.48\textwidth}
        \centering
        \includegraphics[width=\linewidth]{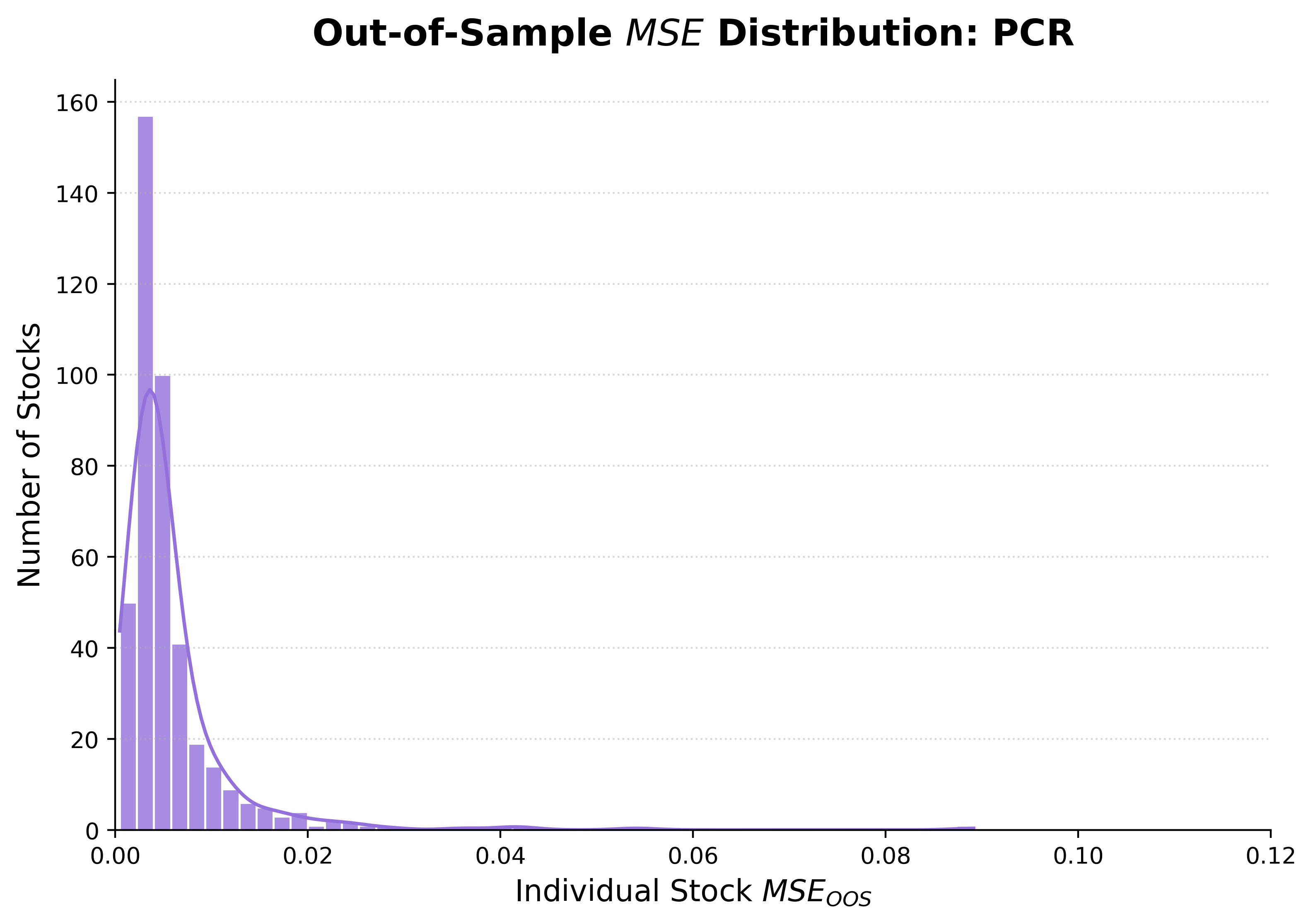}
        \caption{PCR(1911)}
    \end{subfigure}
    \hfill
    \begin{subfigure}{0.48\textwidth}
        \centering
        \includegraphics[width=\linewidth]{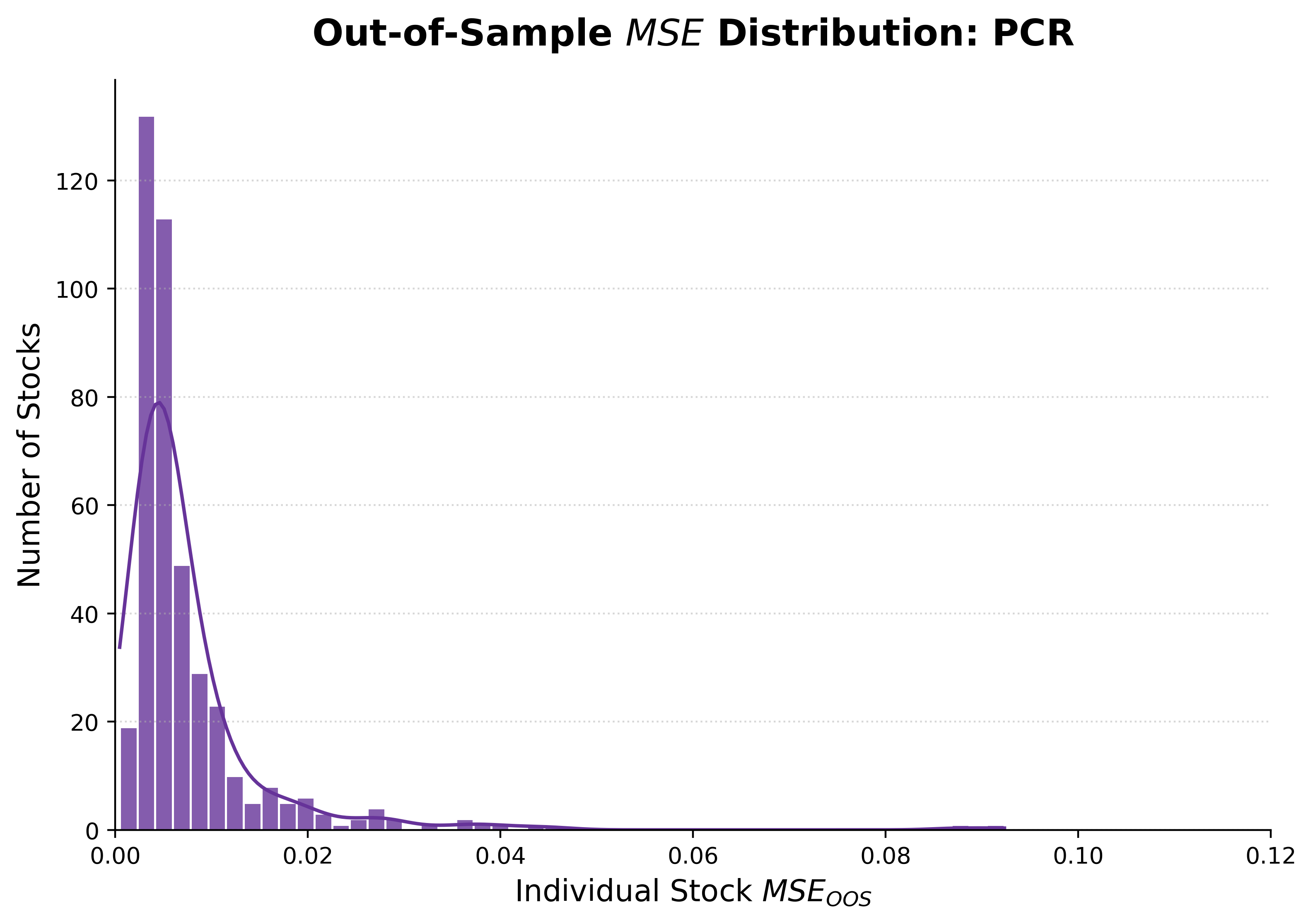}
        \caption{PCR(2112)}
    \end{subfigure}

    \caption[OOS $MSE$ distribution comparison for fw4, fw5 and PCR.]{OOS $MSE$ distribution comparison for fw4, fw5 and PCR.The X-axis exhibits the OOS $MSE$ values, and the Y-axis exhibits the number of stocks.  The ‘$fw(i)(2112)$ and PCR(2112), $i=1,2,\dots,5$’ means the proposed MLP models with the COVID-19 effect, and the '$fw(i)(1911)$' and PCR(1911), $i=1,2,\dots,5$ means the proposed MLP models without the COVID-19 effect.}
    \label{fig:mse_dis_ch2_fw4_PCR}
\end{figure}

Figure ~\ref{fig:r2_dis_ch2_fw1_fw3} and Figure~\ref{fig:r2_dis_ch2_fw4_PCR} exhibit the distribution of OOS $R^2$ of the proposed MLP models and the PCR model in the Pre–COVID-19 Period (1911) and the COVID-19–Inclusive Period (2112). And Table ~\ref{tab:r2_stats_ch2} records the descriptive statistics of the OOS $R^2$. All distributions exhibit a strong right-skewed shape with a sharp peak near zero. The majority of individual stock OOS $R^2$ values fall within the range of -0.3 to +0.3, with very few stocks showing OOS $R^2$ higher than 0.5. This confirms the well-known challenge in cross-sectional asset pricing: stock return predictability is inherently weak and noisy at the individual stock level. The proposed shallow MLP models (fw1 and fw2) produce the widest distributions with the longest left tails. Model fw2, in particular, consistently shows the strongest positive OOS $R^2$ mass and the most heterogeneous predictions across stocks. It generates more stocks with meaningfully positive OOS $R^2$ (especially in the 0.1 to 0.5 range) and exhibits a moderate negative tail. This indicates that shallow networks are better at capturing differences in stock-specific predictability. The medium-depth MLP model (fw3) illustrates the transitional nature of this architecture. The distribution is sharply peaked around zero, with a high concentration of stocks near $R^2 = 0$. Compared to Model fw2, the left tail is noticeably shorter, and extreme positive or negative predictions are reduced. Model fw3 sits between the high-heterogeneity shallow models and the overly conservative deeper models. And the deep MLP models (fw4 and fw5) produce the narrowest and most concentrated distributions, with an extremely sharp peak at zero and very short tails on both sides. Predictions for the vast majority of stocks are tightly clustered around zero, indicating that deeper networks tend to generate overly smooth, conservative forecasts that lose much of the cross-sectional variation in predictability. All the phenomena indicate shallow MLP models (especially fw2) deliver more heterogeneous and informative predictions, while deeper models (fw4 and fw5) tend to “average out” signals, resulting in safer but less discriminative forecasts. Comparing the two periods, in the Pre-COVID-19 period (1911), distributions (particularly for fw1 and fw2) are relatively wider with longer left tails, and in the COVID-19–Inclusive Period (2112), all models show more concentrated distributions around zero. This suggests that the COVID-19 shock increased market noise and triggered a regime shift in pricing dynamics, making it more difficult for models to identify strong, predictable signals at the individual stock level. The PCR, as the best-performing traditional statistical benchmark, produces wider distributions than the deeper MLPs but still lags behind fw2. Shallow MLPs consistently demonstrate superior ability to generate positive predictions compared to this linear benchmark. This aligns with the findings in Table ~\ref{tab:model_perform_ch2}. Overall, the OOS $R^2$ distribution analysis suggests that model depth significantly influences prediction heterogeneity. Shallow MLPs, especially Model fw2, yield more informative forecasts with stronger positive OOS $R^2$ tails, whereas deeper architectures produce overly conservative predictions clustered tightly around zero (this also applies in the GKX2020 MLP benchmarks). The COVID-19–Inclusive Period (2112) further improves predictability across the proposed MLP models, highlighting the regime-dependent nature of the forecasting. These findings reinforce the superiority of shallow MLP in this domain. In addition, the OOS MSE distribution diagrams, Figure ~\ref{fig:mse_dis_ch2_fw1_fw3} and Figure ~\ref{fig:mse_dis_ch2_fw4_PCR}, agree with the findings of OOS $R^2$ distributions. \\

\begin{sidewaystable}[htbp!]
\centering
\small 
\setlength{\tabcolsep}{3pt} 
\begin{tabular}{lccccccccccc}
\toprule
\textbf{Pre-COVID-19 (1911)} & OLS & PCR & PLS & GKX2020\_fw3 & GKX2020\_fw4 & GKX2020\_fw5 & fw1 & fw2 & fw3 & fw4 & fw5 \\
\midrule
Mean & -0.225 & -0.038 & -0.225 & -0.342 & -0.033 & -0.006 & -0.002 & 0.034 & 0.025 & 0.007 & -0.005 \\
Std. Error & 0.022 & 0.015 & 0.022 & 0.104 & 0.027 & 0.001 & 0.016 & 0.013 & 0.006 & 0.003 & 0.002 \\
Median & -0.150 & -0.019 & -0.150 & -0.004 & -0.003 & -0.003 & 0.041 & 0.051 & 0.015 & 0.000 & -0.003 \\
Std. Deviation & 0.455 & 0.317 & 0.454 & 2.123 & 0.558 & 0.019 & 0.327 & 0.259 & 0.130 & 0.061 & 0.037 \\
Variance & 0.207 & 0.100 & 0.207 & 4.506 & 0.311 & 0.000 & 0.107 & 0.067 & 0.017 & 0.004 & 0.001 \\
Kurtosis & 3.994 & 4.860 & 3.996 & 61.843 & 419.005 & 7.207 & 7.515 & 20.904 & 14.305 & 13.458 & 92.607 \\
Skewness & -1.372 & -1.195 & -1.372 & -7.703 & -20.458 & 0.130 & -2.114 & -3.313 & -2.024 & -1.103 & -6.813 \\
Range & 3.536 & 2.686 & 3.536 & 21.545 & 11.532 & 0.204 & 2.344 & 2.656 & 1.435 & 0.748 & 0.660 \\
Minimum & -3.011 & -2.132 & -3.011 & -21.460 & -11.428 & -0.100 & -1.798 & -2.061 & -1.058 & -0.489 & -0.519 \\
Maximum & 0.525 & 0.554 & 0.525 & 0.085 & 0.104 & 0.104 & 0.546 & 0.595 & 0.377 & 0.259 & 0.141 \\
\midrule
\textbf{COVID-19–Inclusive (2112)} & OLS & PCR & PLS & GKX2020\_fw3 & GKX2020\_fw4 & GKX2020\_fw5 & fw1 & fw2 & fw3 & fw4 & fw5 \\
\midrule
Mean & -0.246 & -0.037 & -0.246 & -0.298 & -0.031 & -0.006 & 0.018 & 0.045 & 0.036 & 0.011 & -0.002 \\
Std. Error & 0.024 & 0.017 & 0.024 & 0.097 & 0.021 & 0.001 & 0.017 & 0.013 & 0.006 & 0.004 & 0.002 \\
Median & -0.157 & -0.006 & -0.157 & -0.004 & -0.003 & -0.003 & 0.072 & 0.058 & 0.020 & 0.001 & -0.002 \\
Std. Deviation & 0.490 & 0.349 & 0.490 & 1.993 & 0.439 & 0.014 & 0.339 & 0.261 & 0.129 & 0.075 & 0.035 \\
Variance & 0.240 & 0.122 & 0.240 & 3.974 & 0.193 & 0.000 & 0.115 & 0.068 & 0.017 & 0.006 & 0.001 \\
Kurtosis & 4.781 & 4.716 & 4.780 & 117.900 & 392.707 & 5.764 & 9.909 & 14.585 & 7.469 & 17.430 & 33.292 \\
Skewness & -1.480 & -1.274 & -1.480 & -9.973 & -19.592 & -0.120 & -2.450 & -2.766 & -1.022 & -1.010 & -1.291 \\
Range & 3.700 & 2.584 & 3.700 & 28.826 & 8.929 & 0.137 & 2.711 & 2.347 & 1.290 & 0.941 & 0.560 \\
Minimum & -3.054 & -1.955 & -3.053 & -28.777 & -8.854 & -0.062 & -2.170 & -1.814 & -0.860 & -0.586 & -0.349 \\
Maximum & 0.647 & 0.629 & 0.647 & 0.049 & 0.075 & 0.075 & 0.541 & 0.533 & 0.430 & 0.355 & 0.210 \\
\bottomrule
\end{tabular}
\caption[Descriptive Statistics of $R^2$ for Each Model]{Descriptive Statistics of $R^2$ for Each Model. The upper panel shows the values of the Pre-COVID-19 (1911) testing period, and the lower panel shows the values of the COVID-19-Inclusive (2112) testing period.}
\label{tab:r2_stats_ch2}
\end{sidewaystable}

Furthermore, the DM test is applied to assess the level of the performance difference between the two models. Table~\ref{tab:dm_test_ch2}
shows the DM test statistics and p-values-based statistical significance level. The upper panel of Table~\ref{tab:dm_test_ch2} demonstrates the DM statistics of the Pre–COVID-19 Period (1911), the lower panel reports the results for the COVID-19-inclusive period (2112). In both periods, the OLS model exhibits significantly larger forecast errors than most alternative models, confirming its inferior performance in a high-dimensional factor setting. The PCR model outperforms OLS, but it is consistently dominated by the best proposed MLP model: fw2. The PLS model performs similarly to OLS and remains inferior to most neural network specifications. The proposed MLP models (fw1 to fw5) consistently outperform the GKX2020 benchmark models, especially the proposed MLP models (fw2 to fw4), which are statistically significant. The advantage of the proposed MLP models over the GKX2020 benchmarks remains robust across both the pre-COVID and COVID-inclusive periods. Among the proposed models, fw2 and fw3 generally deliver the strongest performance in both periods, with insignificant differences between them. This pattern suggests that although increasing the depth of MLP models improves predictive accuracy, the marginal advantage diminishes after a certain level. Overall, the DM test results provide strong statistical evidence that the proposed MLP models achieve superior out-of-sample predictive accuracy compared to both traditional linear models and GKX2020 MLP models. The consistency of these findings across the pre-COVID and COVID-19-inclusive periods underscores the robustness of the proposed architecture in capturing complex nonlinear relationships in equity returns, even during periods of significant market disruption.\\

\begin{sidewaystable}[ph!]
\centering
\small
\begin{tabularx}{\textwidth}{l *{10}{X}}
\toprule
& OLS & PCR & PLS & GKX\_fw3 & GKX\_fw4 & GKX\_fw5 & fw1 & fw2 & fw3 & fw4 \\
\midrule
\multicolumn{11}{c}{\textbf{Pre--COVID-19 Period (1911)}} \\
\midrule
OLS &  &  &  &  &  &  &  &  &  &  \\
PCR & -9.30*** &  &  &  &  &  &  &  &  &  \\
PLS & -1.07 & 9.29*** &  &  &  &  &  &  &  &  \\
GKX\_fw3 & 0.06 & 2.66*** & 0.06 &  &  &  &  &  &  &  \\
GKX\_fw4 & -2.61*** & 0.61 & -2.61*** & -2.73*** &  &  &  &  &  &  \\
GKX\_fw5 & -2.65*** & 0.53 & -2.65*** & -2.66*** & -1.27 &  &  &  &  &  \\
fw1 & -8.54*** & -1.49 & -8.55*** & -3.23*** & -1.38 & -1.28 &  &  &  &  \\
fw2 & -8.27*** & -2.30** & -8.27*** & -3.81*** & -2.49** & -2.36** & -2.79*** &  &  &  \\
fw3 & -5.04*** & -0.71 & -5.05*** & -3.58*** & -2.97*** & -2.74*** & 0.01 & 1.64 &  &  \\
fw4 & -3.27*** & 0.25 & -3.27*** & -2.99*** & -2.73*** & -2.21** & 1.05 & 2.33** & 2.84*** &  \\
fw5 & -2.72*** & 0.52 & -2.72*** & -2.69*** & -1.49 & -0.71 & 1.28 & 2.40** & 2.82*** & 2.27** \\
\midrule
\multicolumn{11}{c}{\textbf{COVID-19--Inclusive Period (2112)}} \\
\midrule
OLS &  &  &  &  &  &  &  &  &  &  \\
PCR & -10.71*** &  &  &  &  &  &  &  &  &  \\
PLS & 0.16 & 10.70*** &  &  &  &  &  &  &  &  \\
GKX\_fw3 & -0.52 & 2.09** & -0.52 &  &  &  &  &  &  &  \\
GKX\_fw4 & -2.54** & 0.37 & -2.54** & -3.33*** &  &  &  &  &  &  \\
GKX\_fw5 & -2.64*** & 0.25 & -2.64*** & -3.35*** & -2.79*** &  &  &  &  &  \\
fw1 & -7.81*** & -1.61 & -7.81*** & -3.28*** & -1.33 & -1.16 &  &  &  &  \\
fw2 & -7.06*** & -1.96** & -7.06*** & -4.29*** & -2.47** & -2.23** & -2.02** &  &  &  \\
fw3 & -4.61*** & -0.79 & -4.61*** & -4.48*** & -3.21*** & -2.76*** & 0.04 & 1.51 &  &  \\
fw4 & -3.21*** & -0.04 & -3.21*** & -3.88*** & -3.54*** & -2.47** & 0.89 & 2.12** & 2.77*** &  \\
fw5 & -2.74*** & 0.21 & -2.74*** & -3.44*** & -3.15*** & -1.31 & 1.14 & 2.26** & 2.85*** & 2.63*** \\
\bottomrule
\end{tabularx}
\caption[DM statistics all models during Period 2112 and Period 1911]{DM statistics all models during Period 2112 and Period 1911. The ‘$fw(i)$’, $i=1,2,\cdots,5$ means the proposed MLP models with 1 to 5 hidden layers, while ‘$GKX\_fw(i)$’, $i=3,4,5$, means GKX2020 models with 3 to 5 hidden layers. In ‘DM statistics’ tables, the difference of each model’s OOS absolute error is computed as model $m$ minus $n$ in Equation~\eqref{eq:4.2.6_ch2}, model $n$ is from column labels, while model $m$ is from row labels. For example, in the upper panel, the first number ‘-9.30’ in the second column means `PCR'’s absolute error minus `OLS'’s. This indicates that Model `PCR' performs better than `OLS'.}
\label{tab:dm_test_ch2}
\end{sidewaystable}

\begin{sidewaysfigure}[htbp]
    \centering
    
    \begin{subfigure}{0.32\textwidth}
        \centering
        \includegraphics[height=5.7cm, keepaspectratio]{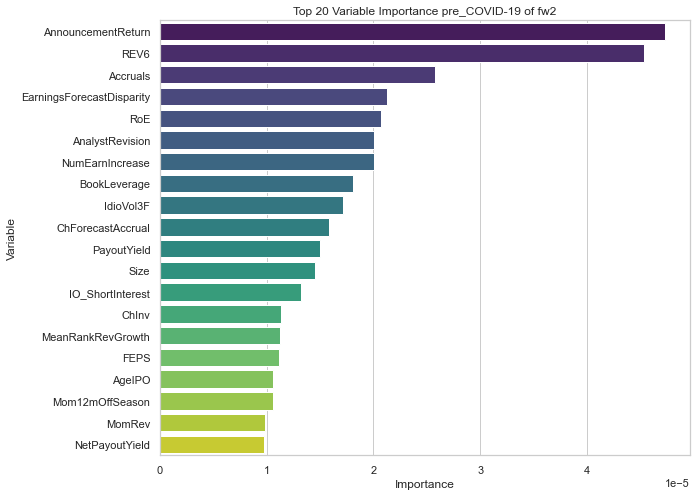}
        \caption{fw2(1911)}
    \end{subfigure}
    \hfill
    \begin{subfigure}{0.32\textwidth}
        \centering
        \includegraphics[height=5.7cm, keepaspectratio]{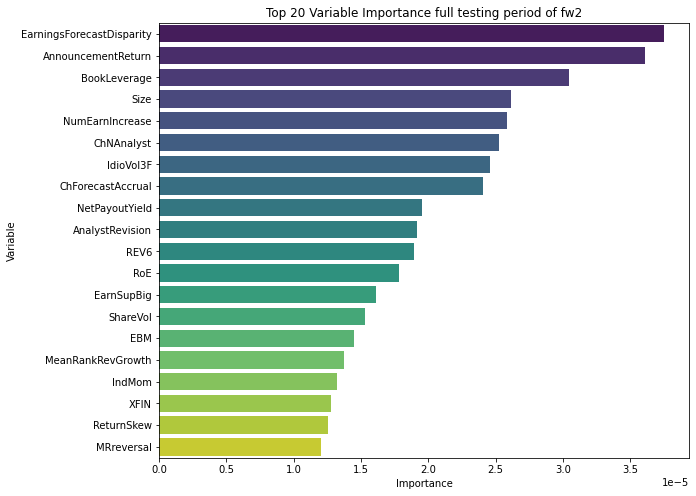}
        \caption{fw2(2112)}
    \end{subfigure}
    \hfill
    \begin{subfigure}{0.32\textwidth}
        \centering
        \includegraphics[height=5.7cm, keepaspectratio]{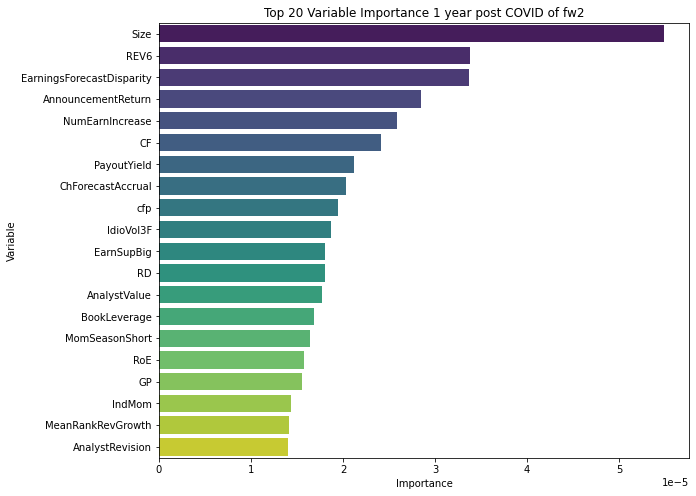}
        \caption{fw2(2212)}
    \end{subfigure}

    \vspace{0.2cm}   

    \begin{subfigure}{0.32\textwidth}
        \centering
        \includegraphics[height=5.7cm, keepaspectratio]{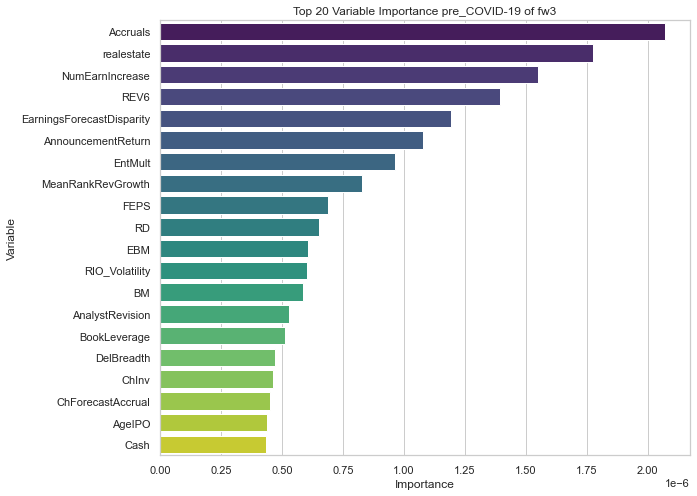}
        \caption{fw3(1911)}
    \end{subfigure}
    \hfill
    \begin{subfigure}{0.32\textwidth}
        \centering
        \includegraphics[height=5.7cm, keepaspectratio]{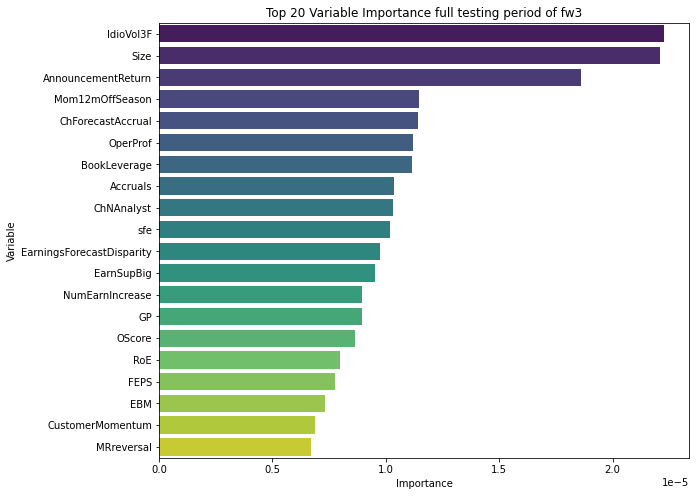}
        \caption{fw3(2112)}
    \end{subfigure}
    \hfill
    \begin{subfigure}{0.32\textwidth}
        \centering
        \includegraphics[height=5.7cm, keepaspectratio]{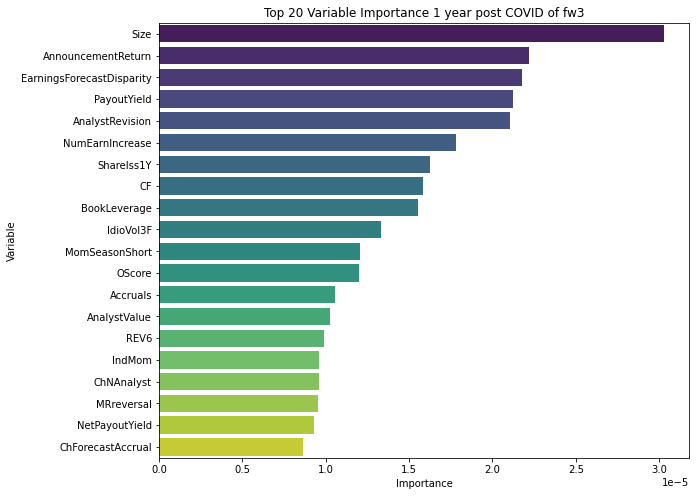}
        \caption{fw3(2212)}
    \end{subfigure}
    \caption[VI plots for the top 20 important factors of the best proposed MLP models for the three market regimes.]{VI plots for the top 20 important factors of the best proposed MLP models for Pre–COVID-19 Period (1911), COVID-19–Inclusive Period (2112) and Period Including COVID-19 and One-Year After (2212). The Y-axis is the factor names, and the X-axis is the value of factor importance.}
    \label{fig:iv_ch2}
\end{sidewaysfigure}

For understanding which features or factors are most important for the proposed models, variable (factor) importance examination is typically employed. As presented in Section~\ref{subsec:perform_evalu_ch2}, the permutation-based variable importance test is applied to the proposed MLP models for measuring the importance of the input factors. It shows that factors have strong time-varying patterns. In the pre-COVID-19 period, the importance patterns reflect relatively normal market conditions with a smooth uptrend. For Model fw2, Announcement Return (AnnouncementReturn) shows as the overwhelmingly dominant predictor, followed by short-term reversal (REV6), Accruals, and Earnings Forecast Disparity. This indicates the central role of information shocks and short-term reversal effects, with earnings quality signals, during Pre-COVID periods. In contrast, Model fw3 assigns the highest importance to Accruals. And factors of real estate (realestate), Number of Earn Increase (NumEarnIncrease), short-term reversal (REV6), and Earnings Forecast Disparity (EarningsForecastDisparity) also rank prominently. The deeper architecture thus places heavier weight on accounting-based signals and fundamental quality metrics, suggesting that additional hidden layers enable the model to capture more complex interactions among earnings management and related characteristics.\\

For COVID-19–Inclusive Period (2112), the factor importance shifts significantly, aligning with increased uncertainty and risk premia during the pandemic. In Model fw2, Earnings Forecast Disparity (EarningsForecastDisparity) rises to the top position, with Announcement Return (AnnouncementReturn), Leverage (BookLeverage), and Size following closely. This indicates that leverage-related risk exposures acquire substantial pricing relevance in response to the shock. For Model fw3, idiosyncratic volatility (IdioVol3F) and Size jointly dominate, accompanied by Announcement Return (AnnouncementReturn) and momentum-off-season effects. These changes highlight how crisis periods amplify the importance of volatility, size, and leverage factors, while information-related variables such as Announcement Return (AnnouncementReturn) remain robust across both architectures. This all aligns with the findings in the work of \citet{AndrewY.Zimmermann2020OpenPricing}.\\

This study also examined the factor importance of Period Including COVID-19 and One-Year After (2212), which includes the period of one year post-pandemic (from January 2022 to December 2022), with a regime recovery pattern shown in Figure~\ref{fig:price_index_ch2}. Size surpasses all other factors by a wide margin in both fw2 and fw3. This suggests a strong resurgence of the size premium in the recovery phase, potentially reflecting market re-pricing of scale as a buffer against residual uncertainty or favoring smaller firms during rebound dynamics. Announcement Return (AnnouncementReturn) and Earnings Forecast Disparity (EarningsForecastDisparity) continue to rank highly in both models, confirming their status as the most persistent and regime-robust predictors throughout the entire sample. Additionally, payout-related variables (such as Payout Yield (PayoutYield) and Net Payout Yield (NetPayoutYield)) have notable prominence in the post-COVID period, especially in fw3, indicating that investor attention shifted toward shareholder-friendly cash distribution policies during economic normalization. Short-term reversal (REV6) also regains strength in fw2, consistent with mean-reversion effects reasserting in a recovering market.\\

The variable (factor) importance results illustrate some crucial insights. Firstly, cross-sectional factor premia are highly time-varying and regime-dependent, with pronounced rotations around the extreme market turbulence, such as the COVID-19 pandemic. Secondly, information-based characteristics, particularly Announcement Return (AnnouncementReturn) and Earnings Forecast Disparity (EarningsForecastDisparity), exhibit remarkable stability across all three periods and both model depths, reinforcing their role as core elements of the stock pricing. Eventually, modest yet systematic differences between fw2 and fw3 (e.g., fw3’s stronger pre-crisis emphasis on accounting signals and post-crisis focus on payout policy) demonstrate that one-layer deeper network depth enables the capture of higher-order nonlinear interactions among firm-characteristics-sorted portfolio factors. In addition, high impact factors are quite consistent in alternative proposed MLP models, and they are fading in fw5, which shows in the factor importance plots in the Appendix.\\

\begin{figure}[htbp!]
\centering
\includegraphics[width=1\columnwidth]{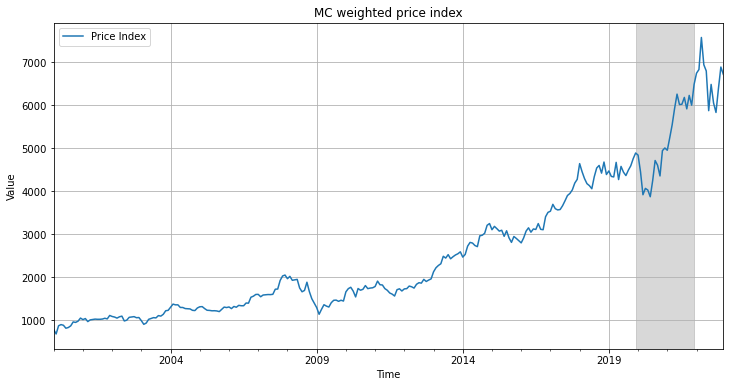}
\caption[Market-cap-weighted (MC) Price Index.]{Market-cap-weighted (MC) Price Index.}
\label{fig:price_index_ch2}
\end{figure}

To summarize, the proposed MLP models with 2 hidden layers outperform all alternative models with the highest OOS $R^2$ value in both periods, which is followed by the one with 3 hidden layers. All proposed MLP models’ performance is significantly improved when the COVID-19 period data is counted in the testing period, but the benchmark models show less sensitivity to the pandemic effect. Additionally, individual stock performance may have a significant impact on overall portfolio performance. Size, as one of the most important factors, has the most crucial impact on the best proposed models when the market was recovering from extreme fluctuations, as well as momentum factors and the factor of ‘earnings surprise of big firms (EarnSupBig)’. The general fading variable importance in fw4 to fw5, together with results from OOS $R^2$, challenges the legend ‘deep, better’ of MLP models. It agrees with the suggestion of \citet{Coqueret2020MachineVersion} that data size should be beyond parameter size to a certain level.\\

\subsection[Empirical Application and Results for Chapter 2]{Evaluation of Back Testing Performance}
\label{subsec:back_testing_eva_ch2}
The portfolio performance is commonly evaluated via indicators such as annualized returns (annual return), Sharpe Ratio (SR), Sortino ratio (SO), maximum drawdown (MDD) and $\alpha$. Table~\ref{tab:EW_ch2} shows the equal-weighted portfolios’ performance of the selected 420 stocks with and without the pandemic period in the static transaction cost (50bps) scenario. In equal-weighted portfolios, the overall average annual return is clearly higher when the pandemic period is included, and the annual returns for individual models show the same trend. Although the overall standard deviation (Std) is higher for the COVID-19–Inclusive Period (2112), it does not affect the annual return much. For both periods tested, the lowest annual returns are from the models of ‘fw1’, ‘OLS’,’PLS’ and ‘PCR’, and the rest of the MLP-based models show a trend of ‘deeper, better’, but interestingly, none of them achieves higher excess returns compared with the ‘buy-and-hold’ strategy. Also, for the proposed models, a deeper structure comes with the higher Std, but the highest Std is shown by the ‘buy-and-hold’ strategy. The 'buy-and-hold' strategy, as a well-applied benchmark in backtesting of trading strategy constructions, is a passive strategy that opens a position from the first trading day, and holds the position until the final trading day. It does not need any market timing or signal generation techniques. Sharpe ratios reflect the capital gain under a certain level of risk which is measured by the standard deviation of excess returns. Period 2112’s performance on SR is better than Period 1911. All models in the Period 2112 have higher Sharpe ratios than or very close to the benchmark strategy of ‘buy-and-hold’. However, during the pre-COVID period (1911), all models were lower than or extremely close to ‘buy-and-hold’. In terms of downside risk performance, an analogous phenomenon appears on Sortino ratios as well, which scales the gain from downside risks (e.g. the standard deviation from the negative excess returns). MDD, which measures the largest downfall from peak to the bottom of portfolio excess returns, shows the opposite trends to SR and SO, all are lower than ‘buy-and-hold’ in the full period but higher in the pre-pandemic period (1911). During the full testing period, our best-performing model, MLP with 2 hidden layers, showed a noticeable advantage which achieving the SR of 0.2651, Sortino Ratio of 0.4154 and the lowest MDD of 39.03\% followed by MLP with 3 hidden layers. This model beat the traditional statistical models and the shallower MLP structure model as well. From the factor investing angle, alpha measures the extra gains from factors; all models have positive alphas, which are significant from zero for both periods. The proposed deep MLP models have higher alphas in the full testing period (2112), but no higher than GKX2020’s MLP models. From the empirical asset pricing perspective, the insignificant $\alpha$ indicates a better interpretability of these factors. The positive residual OOS $\alpha$ of the best proposed MLP models with patterns (lower in Period 1911, a mild uptrend, and higher in Period 2112, a sharp uptrend) implies the existence of 'omitted factors' or the arbitrage opportunities of the market. From a factor investing perspective, practitioners prefer a higher positive $\alpha$, which implies their trading strategy derives higher abnormal returns beyond factors. Noticeably, the performance of GKX2020 models on the proposed dataset exhibits the similarity and approach to the ‘buy-and-hold’ strategy, which is detected as the trading signal capture failing. \\

\begin{sidewaysfigure}[htbp!]
\begin{center}
\includegraphics[width=0.9\columnwidth]{"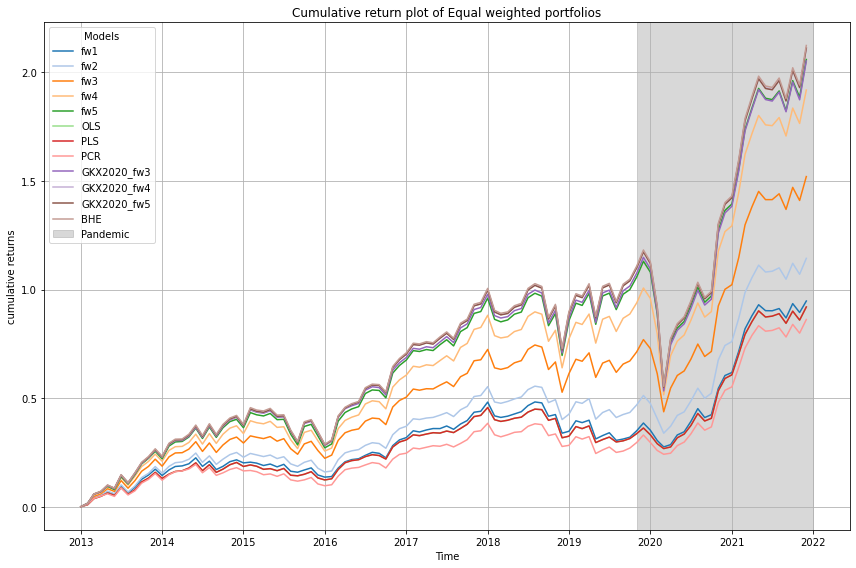"}
\end{center}
\caption[Cumulative excess return plot for equal-weighted portfolio considering the static transaction cost.]{Cumulative excess return plot for equal-weighted portfolio considering the static transaction cost. The $fw(i)$, $i=1,2,\dots,5$, are denoted as the proposed MLP models with i hidden layers. 'BHE' means buy-and-hold benchmark for the equal-weighted portfolio. }
\label{fig:EW_cumulative_ch2}
\end{sidewaysfigure}

As shown in the pre-COVID testing period (1911), the results have a mild upward trend. The overall annualized return is around 2\% lower than the full testing period. The Std is around 0.01 lower as well. Shape Ratio and Sortino Ratio present that the proposed MLP models have a ‘deep, better’ trend, better than the ‘OLS’, ‘PLS’ and ‘PCR’ models, but the difference between SR and SO is less significant than the ones in the full testing period. It implies that models suffered higher downside risks in the pre-COVID period than in the pandemic period. The lowest MDD shows in the MLP with 3 hidden layers, but the best SR and SO appear in the MLP with 5 hidden layers, which also has the lowest alpha value. However, the ‘fw5’ and GKX2020 models’ performance approaches the ‘buy-and-hold’ benchmark, which implies the failure to detect sign-based trading signals. The ‘OLS’, ‘PLS’ and ‘PCR’ models achieve similar but better $\alpha$s in the pre-COVID period, which implies that practitioners have extra gain from the factor investing. Their alphas are even higher than the best alpha in the full testing period, which implies that, without high market fluctuation, the simplest models may lead to a greater gain from factors. But anyway, by comparing the full testing period and pre-COVID testing period results, it shows that the best proposed MLP models have higher profitability with relatively low risks. Figure~\ref{fig:EW_cumulative_ch2} illustrates the cumulative excess returns of all models for the full testing period. It clearly shows the best proposed MLP model well manages the downside risks and achieves a relatively high excess return in the equal-weighted portfolio, no matter the testing period includes or excludes the pandemic period although the absolute returns are less than those in the simplest ‘buy-and-hold’ strategy.\\
\begin{sidewaystable}[htbp!]
\centering
\small
\begin{tabular}{lcccccccccc}
\hline
\multicolumn{11}{c}{\textbf{Equal-weighted portfolio with period of 1911}} \\
\hline
\textbf{model} & \textbf{AR} & \textbf{Std} & \textbf{SR} & \textbf{Ann.SR} & \textbf{SO} & \textbf{Ann.SO} & \textbf{MDD} & \textbf{$\alpha$} & \textbf{Ann.$\alpha$} & \textbf{t\_stat\_$\alpha$} \\
\hline
fw5 & 0.10997 & 0.0345 & 0.2574 & 0.8917 & 0.3611 & 1.2509 & -0.3745 & 0.0039 & 0.0465 & 10.3241 \\
fw4 & 0.10045 & 0.0319 & 0.2527 & 0.8752 & 0.3463 & 1.1997 & -0.3527 & 0.0040 & 0.0478 & 10.2502 \\
fw3 & 0.08095 & 0.0273 & 0.2351 & 0.8144 & 0.3042 & 1.0539 & -0.3166 & 0.0043 & 0.0521 & 11.3607 \\
fw2 & 0.05719 & 0.0221 & 0.1996 & 0.6913 & 0.2528 & 0.8756 & -0.3598 & 0.0045 & 0.0542 & 10.4123 \\
fw1 & 0.04432 & 0.0206 & 0.1625 & 0.5630 & 0.1945 & 0.6736 & -0.3952 & 0.0041 & 0.0487 & 8.6578 \\
OLS & 0.04286 & 0.0193 & 0.1661 & 0.5755 & 0.1961 & 0.6792 & -0.3958 & 0.0064 & 0.0768 & 15.1934 \\
PLS & 0.04286 & 0.0193 & 0.1661 & 0.5755 & 0.1961 & 0.6792 & -0.3951 & 0.0064 & 0.0767 & 15.1780 \\
PCR & 0.03819 & 0.0187 & 0.1504 & 0.5211 & 0.1915 & 0.6635 & -0.5044 & 0.0062 & 0.0746 & 15.1854 \\
GKX2020\_fw5 & 0.11322 & 0.0354 & 0.2585 & 0.8955 & 0.3609 & 1.2502 & -0.3680 & 0.0043 & 0.0511 & 11.5900 \\
GKX2020\_fw4 & 0.11346 & 0.0353 & 0.2594 & 0.8986 & 0.3622 & 1.2549 & -0.3682 & 0.0042 & 0.0504 & 11.0508 \\
GKX2020\_fw3 & 0.11135 & 0.0348 & 0.2584 & 0.8950 & 0.3613 & 1.2517 & -0.3684 & 0.0043 & 0.0512 & 4.9815 \\
buy-and-hold (E) & 0.11386 & 0.0356 & 0.2588 & 0.8967 & 0.3631 & 1.2577 & -0.3701 & & & \\
\hline
\multicolumn{11}{c}{\textbf{Equal-weighted portfolio with period of 2112}} \\
\hline
\textbf{model} & \textbf{AR} & \textbf{Std} & \textbf{SR} & \textbf{Ann.SR} & \textbf{SO} & \textbf{Ann.SO} & \textbf{MDD} & \textbf{$\alpha$} & \textbf{Ann.$\alpha$} & \textbf{t\_stat\_$\alpha$} \\
\hline
fw5 & 0.13228 & 0.0434 & 0.2512 & 0.8703 & 0.3098 & 1.0731 & -0.5109 & 0.0055 & 0.0664 & 17.0422 \\
fw4 & 0.12637 & 0.0402 & 0.2565 & 0.8885 & 0.3257 & 1.1281 & -0.4884 & 0.0053 & 0.0634 & 15.7385 \\
fw3 & 0.10816 & 0.0333 & 0.2608 & 0.9036 & 0.3517 & 1.2184 & -0.4319 & 0.0048 & 0.0579 & 14.0525 \\
fw2 & 0.08841 & 0.0262 & 0.2651 & 0.9184 & 0.4154 & 1.4391 & -0.3903 & 0.0039 & 0.0468 & 9.8960 \\
fw1 & 0.07684 & 0.0239 & 0.2512 & 0.8703 & 0.3816 & 1.3218 & -0.4292 & 0.0030 & 0.0358 & 6.7956 \\
OLS & 0.07514 & 0.0235 & 0.2573 & 0.8914 & 0.3938 & 1.3642 & -0.4136 & 0.0034 & 0.0406 & 8.1689 \\
PLS & 0.07516 & 0.0227 & 0.2574 & 0.8917 & 0.3954 & 1.3698 & -0.4139 & 0.0034 & 0.0405 & 8.1319 \\
PCR & 0.07151 & 0.0220 & 0.2513 & 0.8704 & 0.4155 & 1.4392 & -0.5044 & 0.0050 & 0.0598 & 12.4912 \\
GKX2020\_fw5 & 0.13442 & 0.0451 & 0.2471 & 0.8558 & 0.2980 & 1.0324 & -0.5315 & 0.0060 & 0.0718 & 18.4553 \\
GKX2020\_fw4 & 0.13441 & 0.0451 & 0.2470 & 0.8555 & 0.2977 & 1.0314 & -0.5321 & 0.0061 & 0.0731 & 15.6569 \\
GKX2020\_fw3 & 0.13199 & 0.0446 & 0.2453 & 0.8499 & 0.2950 & 1.0221 & -0.5356 & 0.0060 & 0.0715 & 7.5860 \\
buy-and-hold (E) & 0.13491 & 0.0452 & 0.2472 & 0.8564 & 0.2988 & 1.0349 & -0.5310 & & & \\
\hline
\end{tabular}%
\caption[Equal-weighted portfolio performance with (2112) and without (1911) the COVID-19 period considering the static transaction cost.]{Equal-weighted portfolio performance with (2112) and without (1911) the COVID-19 period considering the static transaction cost. Report for equal-weighted portfolio excess returns. The lower panel is the report for the full testing period, while the upper panel is for the testing period without the pandemic. Both reports include the indicators of annualized returns (AR), standard deviations (Std), Sharpe Ratio (SR), Sortino Ratio (SO), Maximum drawdown (MDD), $\alpha$ and the $t$ statistic of the alpha which is listed in the last column. The ‘buy-and-hold’(E for equal-weighted and V for value-weighted) applies as a benchmark strategy for all models without any prediction technique, hence no $\alpha$ is presented. The notation 'Ann' means annualized.}
\label{tab:EW_ch2}
\end{sidewaystable}

\begin{sidewaysfigure}[htbp!]
\begin{center}
\includegraphics[width=0.9\columnwidth]{"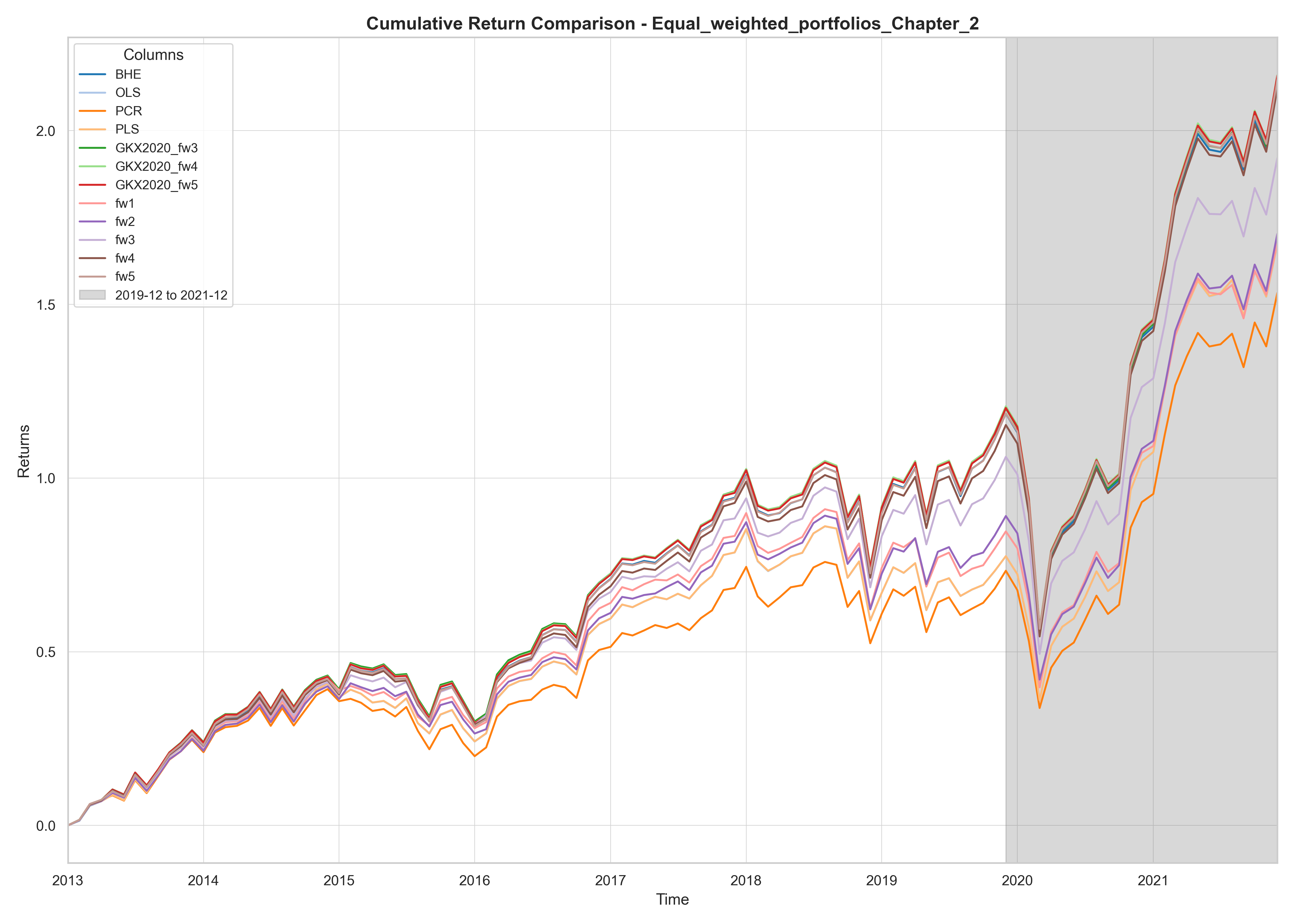"}
\end{center}
\caption[Cumulative excess return plot for equal-weighted portfolio considering the dynamic transaction cost.]{Cumulative excess return plot for equal-weighted portfolio considering the dynamic transaction cost. The $fw(i)$, $i=1,2,\dots,5$, are denoted as the proposed MLP models with i hidden layers. 'BHE' represents the buy-and-hold benchmark strategy in the equal-weighted method.}
\label{fig:EW_cumulative_ch2_turnover}
\end{sidewaysfigure}

For investigating how the turnover rates affect the profitability of the models, the turnover rate-based dynamic transaction cost is applied for a robust examination. Table~\ref{tab:EW_ch2_turnover} shows the robust backtesting results for equal-weighted portfolios, and Figure~\ref{fig:EW_cumulative_ch2_turnover} shows the robust cumulative return of the equal-weighted portfolios. The dynamic transaction cost robust examination in equal-weighted portfolios reveals a different picture from that in static transaction cost. The annualized returns generally improved in the dynamic transaction cost framework. This indicates that static transaction cost with an average rate may cause the over-penalization for the models' profitability in the highly variable turnover rate scene. Conretely, in the Pre-COVID-19 Period (1911), all models' annualized returns do not exceed the buy-and-hold benchmark (12.52\%), and traditional statistical models (OLS, PCR, PLS) are significantly lower than the MLP models. GKX2020 MLP models slightly outperform the proposed deep MLP models (fw4 to fw5), but they significantly outperform the proposed shallow MLP models. However, when introducing the volatility as an indicator for risks (Sharpe ratio and Sortino ratio), Model fw5 and GKX2020 MLP models perform insignificantly better than the buy-and-hold benchmark. And the proposed MLP models show a tendency of `deeper, better'. The maximum dwarndown (MDD) between models is insignificant (approximately -14.8\%). The turnover rates are negatively correlated with the annualized returns, where the traditional statistical models show the highest turnover rates. In the COVID-19-Inclusive Period (2112), deep MLP models slightly outperform the buy-and-hold benchmark in annualized return, and MLP models show superiority to the traditional statistical models. The proposed MLP models show a `deeper, better' pattern as well. This is aligned with the Sharpe Ratio. Nonetheless, Sortino ratios present a different picture. Different from the Sharpe Ratio, which considers the entire volatility as risk, the Sortino ratios only introduce the downside risk. The Sotinuo ratios of traditional statistical models surprisingly outperform the MLP models and the buy-and-hold benchmark with the highest turnover rates. The proposed MLP models exhibit superiority over the GKX2020 models, except Model fw3, in the Sortino ratio. This provides evidence that high turnover rates are not the primary reason for the deterioration of profitability in specific conditions. In addition, the OOS $\alpha$ for all models in both periods is positive at a 1\% significance level. The traditional statistical models outperform MLP models in the Pre-COVID-19 Period, but lag behind during the COVID-19-Inclusive Period. The OOS $\alpha$ of the proposed models exhibits the 'deep, better' pattern, but none of them exceeds the GKX2020 models during the COVID-19-Inclusive Period. The cumulative return plot Figure~\ref{fig:EW_cumulative_ch2_turnover} also supports these findings.\\

\begin{sidewaystable}[ph]
    \centering
    \footnotesize
    \begin{tabular}{lcccccccccc}
        \toprule
        \multicolumn{11}{c}{\textbf{Equal-weighted portfolio with Pre-COVID-19 Period (1911)}} \\
        \midrule
        \textbf{Model} & \textbf{AR} & \textbf{Ann.SR} & \textbf{SR} & \textbf{Ann.SO} & \textbf{SO} & \textbf{MDD} & \textbf{Turnover} & \textbf{$\alpha$} & \textbf{Ann. $\alpha$} & \textbf{t\_stats} \\
        \midrule
        buy-and-hold(E) & 0.1252 & 0.8892 & 0.2567 & 1.2303 & 0.3552 & -0.1483 & 0.0120 & -- & -- & -- \\
        OLS             & 0.0827 & 0.7214 & 0.2083 & 0.9451 & 0.2728 & -0.1456 & 0.7576 & 0.0064 & 0.0768 & 15.193*** \\
        PCR             & 0.0780 & 0.6847 & 0.1977 & 0.9508 & 0.2745 & -0.1386 & 0.7587 & 0.0062 & 0.0746 & 15.185*** \\
        PLS             & 0.0827 & 0.7219 & 0.2084 & 0.9455 & 0.2730 & -0.1456 & 0.7573 & 0.0064 & 0.0767 & 15.178*** \\
        GKX2020\_fw3    & 0.1152 & 0.8986 & 0.2594 & 1.2437 & 0.3590 & -0.1483 & 0.0215 & 0.0043 & 0.0512 & 4.9815*** \\
        GKX2020\_fw4    & 0.1156 & 0.8994 & 0.2596 & 1.2455 & 0.3595 & -0.1482 & 0.0155 & 0.0042 & 0.0504 & 11.051*** \\
        GKX2020\_fw5    & 0.1153 & 0.8967 & 0.2589 & 1.2416 & 0.3584 & -0.1483 & 0.0151 & 0.0043 & 0.0511 & 11.590*** \\
        fw1             & 0.0884 & 0.7553 & 0.2180 & 0.9820 & 0.2835 & -0.1494 & 0.6699 & 0.0041 & 0.0487 & 8.6578*** \\
        fw2             & 0.0917 & 0.8035 & 0.2320 & 1.0618 & 0.3065 & -0.1426 & 0.4951 & 0.0045 & 0.0542 & 10.412*** \\
        fw3             & 0.1049 & 0.8643 & 0.2495 & 1.1526 & 0.3327 & -0.1458 & 0.2611 & 0.0043 & 0.0521 & 11.361*** \\
        fw4             & 0.1115 & 0.8846 & 0.2554 & 1.2164 & 0.3511 & -0.1472 & 0.1018 & 0.0040 & 0.0478 & 10.250*** \\
        fw5             & 0.1140 & 0.8892 & 0.2567 & 1.2315 & 0.3555 & -0.1499 & 0.0357 & 0.0039 & 0.0465 & 10.324*** \\
        \bottomrule
    \end{tabular}
    
    \vspace{0.5em}

    \begin{tabular}{lcccccccccc}
        \toprule
        \multicolumn{11}{c}{\textbf{Equal-weighted portfolio with COVID-19-Inclusive Period (2112)}} \\
        \midrule
        \textbf{Model} & \textbf{AR} & \textbf{Ann.SR} & \textbf{SR} & \textbf{Ann.SO} & \textbf{SO} & \textbf{MDD} & \textbf{Turnover} & \textbf{$\alpha$} & \textbf{Ann. $\alpha$} & \textbf{t\_stats} \\
        \midrule
        buy-and-hold(E) & 0.1353 & 0.8512 & 0.2457 & 1.0181 & 0.2939 & -0.2890 & 0.0093 & -- & -- & -- \\
        OLS             & 0.1154 & 0.8117 & 0.2343 & 1.0934 & 0.3157 & -0.2693 & 0.7280 & 0.0034 & 0.0406 & 8.1689*** \\
        PCR             & 0.1088 & 0.7940 & 0.2292 & 1.0981 & 0.3170 & -0.2391 & 0.7280 & 0.0050 & 0.0598 & 12.491*** \\
        PLS             & 0.1155 & 0.8122 & 0.2344 & 1.0938 & 0.3157 & -0.2693 & 0.7278 & 0.0034 & 0.0405 & 8.1319*** \\
        GKX2020\_fw3    & 0.1357 & 0.8519 & 0.2459 & 1.0147 & 0.2929 & -0.2911 & 0.0175 & 0.0060 & 0.0715 & 7.5860*** \\
        GKX2020\_fw4    & 0.1364 & 0.8567 & 0.2473 & 1.0260 & 0.2962 & -0.2890 & 0.0120 & 0.0061 & 0.0731 & 15.657*** \\
        GKX2020\_fw5    & 0.1363 & 0.8564 & 0.2472 & 1.0251 & 0.2959 & -0.2889 & 0.0116 & 0.0060 & 0.0718 & 18.455*** \\
        fw1             & 0.1158 & 0.8129 & 0.2347 & 1.0282 & 0.2968 & -0.2684 & 0.6401 & 0.0030 & 0.0358 & 6.7956*** \\
        fw2             & 0.1168 & 0.8315 & 0.2400 & 1.0399 & 0.3002 & -0.2495 & 0.4714 & 0.0039 & 0.0468 & 9.8960*** \\
        fw3             & 0.1265 & 0.8365 & 0.2415 & 1.0003 & 0.2888 & -0.2752 & 0.2467 & 0.0048 & 0.0579 & 14.053*** \\
        fw4             & 0.1349 & 0.8576 & 0.2476 & 1.0386 & 0.2998 & -0.2828 & 0.0948 & 0.0053 & 0.0634 & 15.739*** \\
        fw5             & 0.1358 & 0.8591 & 0.2480 & 1.0376 & 0.2995 & -0.2837 & 0.0320 & 0.0055 & 0.0664 & 17.042*** \\
        \bottomrule
    \end{tabular}
    
    \medskip
    \raggedright
\caption[Equal-weighted portfolio performance with (2112) and without (1911) the COVID-19 period considering the dynamic transaction cost.]{Equal-weighted portfolio performance with (2112) and without (1911) the COVID-19 period considering the dynamic transaction cost. Report for equal-weighted portfolio excess returns. The lower panel is the report for the full testing period, while the upper panel is for the testing period without the pandemic. Both reports include the indicators of annualized returns (AR), standard deviations (Std), Sharpe Ratio (SR), Sortino Ratio (SO), Maximum drawdown (MDD), $\alpha$ and the $t$ statistic of the alpha which is listed in the last column. The ‘buy-and-hold’(E for equal-weighted and V for value-weighted) applies as a benchmark strategy for all models without any prediction technique, hence no $\alpha$ is presented. The notation 'Ann' means annualized. ***, **, and * denote statistical significance at the 1\%, 5\%, and 10\% levels, respectively.}
\label{tab:EW_ch2_turnover}
\end{sidewaystable}

\begin{sidewaysfigure}[htbp!]
\begin{center}
\includegraphics[width=0.9\columnwidth]{"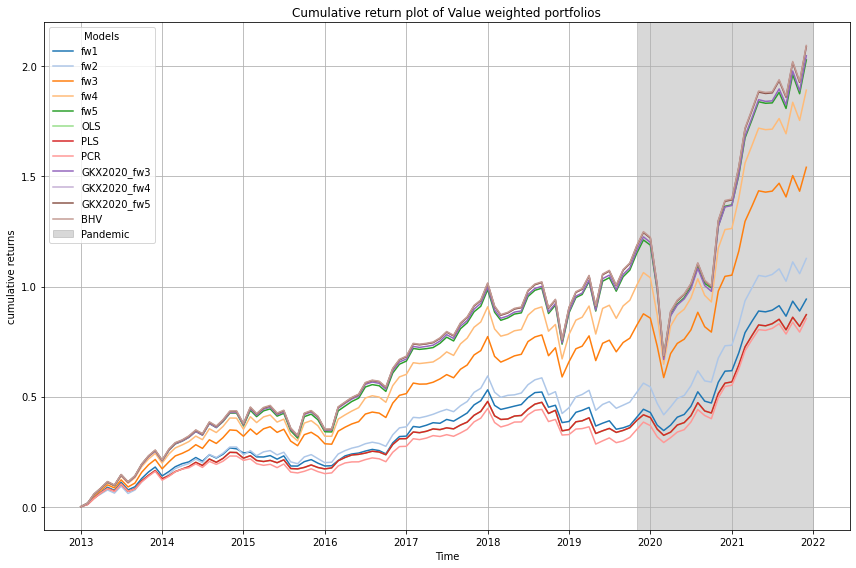"}
\end{center}
\caption[Cumulative excess return plot for value-weighted portfolios considering the static transaction cost.]{Cumulative excess return plot for value-weighted portfolios considering the static transaction cost. The $fw(i)$, $i=1,2,\cdots,5$ denoted for the proposed MLP models with i hidden layers. The ‘PLS’ and ‘PCR’ mean PLS and PCR with 95\% variance coverage, which is equal to the notation of ‘PLS’ and ‘PCR’ in previous tables and diagrams. ‘OLS’ means OLS linear regression model.}
\label{fig:VW_cumulative_ch2}
\end{sidewaysfigure}

Apart from the equal-weighted portfolios, the value-weighted method is widely used as the asset allocation strategy. As early as \citet{Fama1993CommonBonds} to later GKX2020 \citep{Gu2020EmpiricalLearning}, all factor models have employed the value-weighted method. Specifically, value weights are generated from the ratio of an individual stock’s market capitalization to the total market capitalization of all selected stocks. The portfolio returns are generated by the weighted average of the individual stocks' returns. Table~\ref{tab:VW_ch2} is the back-testing results of the value-weighted portfolios of the selected stocks. Compared with the equal-weighted method, the value-weighted method distributes more weight on larger-cap stocks, which increases the impact of the larger-cap stocks and decreases the impact of smaller-cap stocks. By comparing the performance of the equal-weighted portfolios and the value-weighted portfolios, we can evaluate the sensitivity of the models to the stock's market capitalization or namely, the volatility of the stock returns. With the full testing period, the conclusions drawn from the equal-weighted portfolio still hold for the value-weighted portfolio. From the table, the annualized returns of all models are quite similar to the equal-weighted portfolio’s annualized returns, but standard deviations are lower than the equal-weighted portfolio’s. The Sharpe Ratios are relatively higher, but the Sortino Ratios vary among models. The differences in Sortino Ratio between individual models are less pronounced when compared with the equal-weighted portfolio’s Sortino Ratios. The overall maximum drawdowns are dropped considerably, and the highest Sortino Ratio value shows up on the proposed MLP model with 3 hidden layers, which comes with the second lowest MDD (33.04\%). Compared to the lowest MDD in the equal-weighted portfolio with the full testing period, it dropped 15.37\%. This implies that the value-weighted portfolio moderates the downside risks with a reasonable gain in the full testing period. Figure~\ref{fig:VW_cumulative_ch2} also shows how the value-weighted portfolio moderates the downside risks. Figure~\ref{fig:VW_cumulative_ch2} presents the full testing period cumulative returns of all models. It exhibits clearly that MLP with 3 hidden layers (green line) well reduced the downside fluctuation with a decent gain. Signal-detecting failure still exists in the deep MLP model and GKX2020 models. \\

In the pre-COVID testing period, compared with the equal-weighted portfolio, the annual returns for all models are slightly higher than annualized returns in the equal-weighted portfolio and the Std is noticeably decreased for MLP-based models including GKX2020, but no great change in ‘OLS’, ‘PLS’ and ‘PCR’ models. The lowest MDD took place at MLP with 3 hidden layers which is 26.50\%, but the proposed MLP model with the highest SR and SO cannot surpass the ‘buy-and-hold’ strategy. Whereas the overall Sortino ratios are higher than the ones in the equal-weighted portfolio at the same period settings. The SR and SO of proposed MLP models show a ‘deeper, higher’ trend. The best proposed MLP model, with 2 hidden layers shows lower SR and SO in Period 2112 of the value-weighted portfolio compared with the ones in the equal-weighted portfolio, which indicates that shallow MLP models are more flexible and have slightly better capability in capturing the patterns in high fluctuation scenarios. Since portfolio weighting has no impact on signal detecting, the GKX2020 and deep MLP model signal detecting failure still exists under the value weighting mode in both periods. Figure~\ref{fig:VW_cumulative_ch2} shows the cumulative return plot, which exhibits that the meaning of MLP models lies upon hedging downside risks instead of deriving the absolute returns.\\
\begin{table}[htbp!]
\centering
\resizebox{\textwidth}{!}{%
\begin{tabular}{lccccccc}
\hline
\multicolumn{8}{c}{\textbf{Value-weighted portfolio with period of 2112}} \\
\hline
\textbf{Model} & \textbf{AR} & \textbf{Std} & \textbf{SR} & \textbf{Ann.SR} & \textbf{SO} & \textbf{Ann.SO} & \textbf{MDD} \\
\hline
fw5 & 0.1310 & 0.0393 & 0.2708 & 0.9382 & 0.3327 & 1.1524 & -0.4345 \\
fw4 & 0.1252 & 0.0359 & 0.2801 & 0.9704 & 0.3677 & 1.2737 & -0.3913 \\
fw3 & 0.1092 & 0.0299 & 0.2893 & 1.0023 & 0.3938 & 1.3642 & -0.3304 \\
fw2 & 0.0875 & 0.0241 & 0.2831 & 0.9808 & 0.3984 & 1.3800 & -0.3564 \\
fw1 & 0.0766 & 0.0223 & 0.2666 & 0.9236 & 0.3706 & 1.2838 & -0.3488 \\
OLS & 0.0721 & 0.0209 & 0.2661 & 0.9219 & 0.3546 & 1.2284 & -0.3221 \\
PLS & 0.0722 & 0.0209 & 0.2662 & 0.9223 & 0.3548 & 1.2290 & -0.3219 \\
PCR & 0.0711 & 0.0211 & 0.2600 & 0.9006 & 0.3744 & 1.2970 & -0.3627 \\
GKX2020\_fw5 & 0.1335 & 0.0407 & 0.2671 & 0.9252 & 0.3257 & 1.1283 & -0.4518 \\
GKX2020\_fw4 & 0.1335 & 0.0407 & 0.2672 & 0.9255 & 0.3257 & 1.1284 & -0.4518 \\
GKX2020\_fw3 & 0.1318 & 0.0405 & 0.2653 & 0.9190 & 0.3231 & 1.1194 & -0.4555 \\
buy-and-hold (V) & 0.1337 & 0.0408 & 0.2673 & 0.9259 & 0.3263 & 1.1302 & -0.4513 \\
\hline
\multicolumn{8}{c}{\textbf{Value-weighted portfolio with period of 1911}} \\
\hline
\textbf{Model} & \textbf{AR} & \textbf{Std} & \textbf{SR} & \textbf{Ann.SR} & \textbf{SO} & \textbf{Ann.SO} & \textbf{MDD} \\
\hline
fw5 & 0.1171 & 0.0316 & 0.2948 & 1.0213 & 0.4136 & 1.4328 & -0.2982 \\
fw4 & 0.1059 & 0.0293 & 0.2864 & 0.9922 & 0.3839 & 1.3298 & -0.2827 \\
fw3 & 0.0908 & 0.0253 & 0.2821 & 0.9773 & 0.3617 & 1.2531 & -0.2650 \\
fw2 & 0.0623 & 0.0214 & 0.2244 & 0.7774 & 0.2692 & 0.9327 & -0.3564 \\
fw1 & 0.0505 & 0.0201 & 0.1910 & 0.6617 & 0.2238 & 0.7754 & -0.3374 \\
OLS & 0.0489 & 0.0187 & 0.1971 & 0.6827 & 0.2183 & 0.7563 & -0.3209 \\
PLS & 0.0490 & 0.0187 & 0.1972 & 0.6833 & 0.2186 & 0.7572 & -0.3208 \\
PCR & 0.0444 & 0.0186 & 0.1782 & 0.6174 & 0.2100 & 0.7276 & -0.3627 \\
GKX2020\_fw5 & 0.1194 & 0.0322 & 0.2954 & 1.0233 & 0.4231 & 1.4656 & -0.2968 \\
GKX2020\_fw4 & 0.1195 & 0.0322 & 0.2959 & 1.0250 & 0.4237 & 1.4677 & -0.2946 \\
GKX2020\_fw3 & 0.1180 & 0.0319 & 0.2943 & 1.0194 & 0.4224 & 1.4633 & -0.2946 \\
buy-and-hold (V) & 0.1197 & 0.0323 & 0.2956 & 1.0241 & 0.4243 & 1.4698 & -0.2963 \\
\hline
\end{tabular}%
}
\caption[Value-weighted portfolio performance with and without COVID-19 period considering the static transaction cost.]{Value-weighted portfolio performance with and without COVID-19 period considering the static transaction cost. Report for value-weighted portfolio excess returns. The upper panel is the report for the full testing period, while the lower panel is for the testing period without the pandemic. Both reports include the indicators of annualised returns (AR), standard deviations (Std), Sharpe Ratio (SR), Sortino Ratio (SO), Maximum drawdown (MDD). The ‘buy-and-hold (V)’ applies as a benchmark strategy for all models without any prediction technique, hence no $\alpha$ is presented. The notation of 'Ann' means annualized.}
\label{tab:VW_ch2}
\end{table}

\begin{sidewaysfigure}[htbp!]
\begin{center}
\includegraphics[width=0.9\columnwidth]{"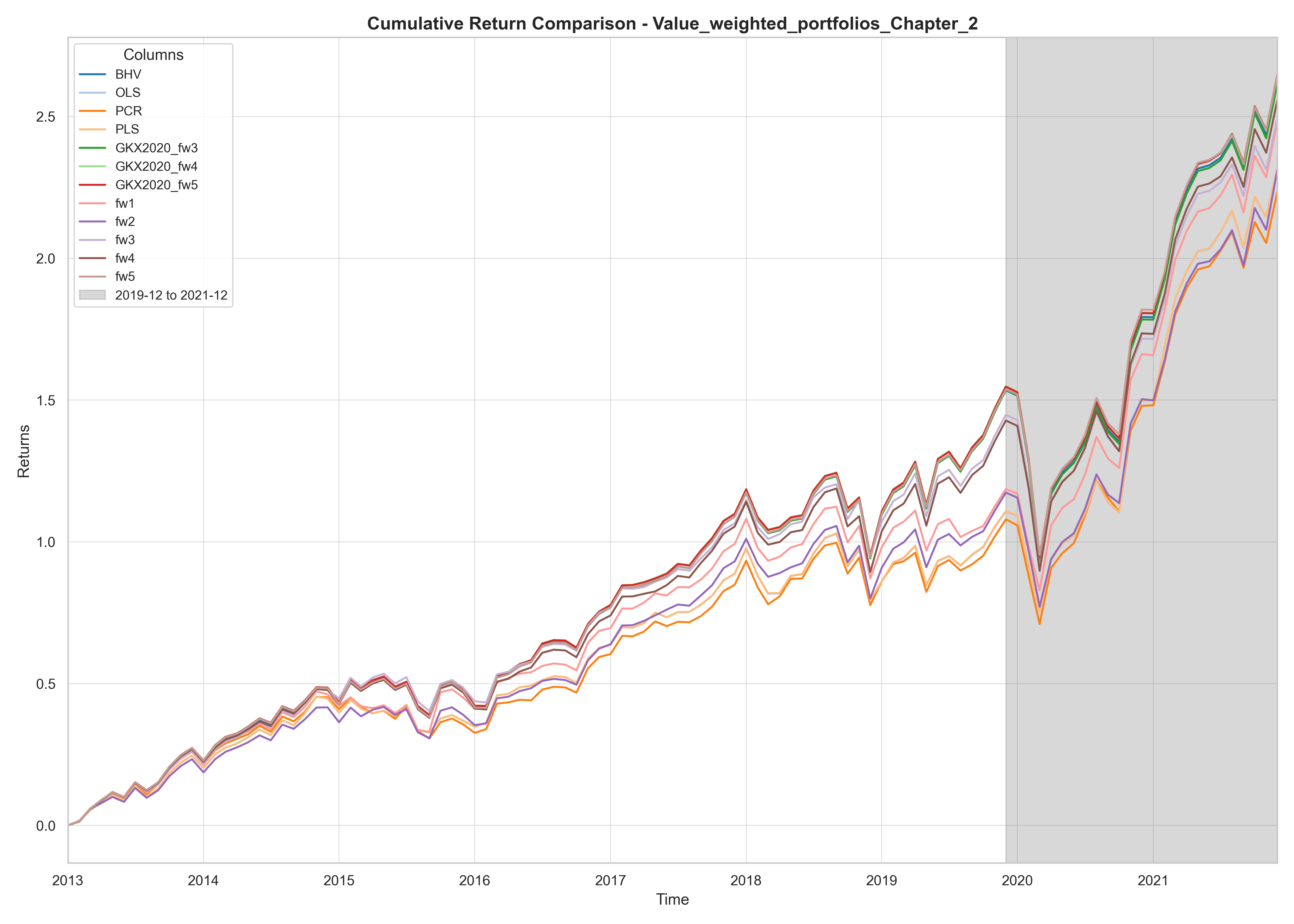"}
\end{center}
\caption[Cumulative excess return plot for value-weighted portfolio considering the dynamic transaction cost.]{Cumulative excess return plot for value-weighted portfolio considering the dynamic transaction cost. The $fw(i)$, $i=1,2,\cdots,5$ denoted for the proposed MLP models with i hidden layers. 'BHV' represents the buy-and-hold strategy in the value-weighted method.}
\label{fig:VW_cumulative_ch2_turnover}
\end{sidewaysfigure}

The transaction cost robustness examination is conducted on value-weighted method as well. Table~\ref{tab:VW_ch2_turnover} is the backtesting results of the value-weighted portfolios considering the dynamic transaction cost. The conclusions drawn from the equal-weighted portfolio are consistent with the value-weighted method. From the table, the annualized returns of all models are generally higher than the equal-weighted portfolio’s annualized returns, as well as the Sharpe ratio and Sortino ratio. Noticeably, the maximum drawdown is significantly moderated in the value-weighted method, particularly in the COVID-19-Inclusive Period. In Pre-COVID-19 Period, deep MLP models' annulized returns are superior to the buy-and-hold benchmark, as well as the Sharpe ratio and Sortino ratio. In the COVID-19-Inclusive Period, deep MLP models still show excellence in annulized returns, but it becomes complicated when considering turnover rate, volatility (SR) and downside risks (SO). Market capitalization has a higher impact on Model PCR than alternatives. Model fw1 surprisingly achieves the highest Sharpe ratio and Sortino ratio with a relatively high turnover rate. It proves again that the high turnover rate could affect the absolute capital returns, but not the profitability concerning the volatility and downside risks in the large-cap scenario of stocks. However, in conjunction with the report on equal-weighted portfolios, the turnover rates for the proposed deep MLP models and the GKX2020 models are approaching the buy-and-hold benchmark, suggesting a possible failure in trading signal detection. The cumulative return plot ~\ref{fig:VW_cumulative_ch2_turnover} supports the findings in the value-weighted portfolio backtesting report.\\ 

\begin{table}[htbp]
\centering
\small 
\tabcolsep=0.18cm
\begin{tabular}{lccccccc}
\toprule
\multicolumn{8}{c}{\textbf{Value-weighted portfolio with Pre-COVID-19 Period (1911)}} \\ 
\midrule
\textbf{Model} & \textbf{AR} & \textbf{Ann.SR} & \textbf{SR} & \textbf{Ann.SO} & \textbf{SO} & \textbf{MDD} & \textbf{Turnover} \\ 
\midrule
buy-and-hold (V) & 0.1388 & 1.1795 & 0.3405 & 1.6054 & 0.4634 & -0.1297 & 0.0120 \\
OLS              & 0.1095 & 1.0190 & 0.2941 & 1.2788 & 0.3692 & -0.1160 & 0.7300 \\
PCR              & 0.1069 & 0.9953 & 0.2873 & 1.2989 & 0.3750 & -0.1101 & 0.7430 \\
PLS              & 0.1096 & 1.0199 & 0.2944 & 1.2799 & 0.3695 & -0.1160 & 0.7303 \\
GKX2020\_fw3       & 0.1388 & 1.1802 & 0.3407 & 1.6059 & 0.4636 & -0.1297 & 0.0193 \\
GKX2020\_fw4       & 0.1398 & 1.1863 & 0.3425 & 1.6155 & 0.4664 & -0.1297 & 0.0152 \\
GKX2020\_fw5       & 0.1396 & 1.1851 & 0.3421 & 1.6139 & 0.4659 & -0.1297 & 0.0149 \\
fw1              & 0.1151 & 1.0189 & 0.2941 & 1.3395 & 0.3867 & -0.1190 & 0.6672 \\
fw2              & 0.1139 & 1.0483 & 0.3026 & 1.3350 & 0.3854 & -0.1244 & 0.4941 \\
fw3              & 0.1331 & 1.1890 & 0.3432 & 1.5782 & 0.4556 & -0.1153 & 0.2557 \\
fw4              & 0.1318 & 1.1353 & 0.3277 & 1.5084 & 0.4355 & -0.1348 & 0.1067 \\
fw5              & 0.1391 & 1.1832 & 0.3416 & 1.6125 & 0.4655 & -0.1294 & 0.0344 \\

\midrule \addlinespace[0.5em] 

\multicolumn{8}{c}{\textbf{Value-weighted portfolio with COVID-19-Inclusive Period (2112)}} \\ 
\midrule
\textbf{Model} & \textbf{AR} & \textbf{Ann.SR} & \textbf{SR} & \textbf{Ann.SO} & \textbf{SO} & \textbf{MDD} & \textbf{Turnover} \\ 
\midrule
buy-and-hold (V) & 0.1539 & 1.0688 & 0.3085 & 1.3057 & 0.3769 & -0.2352 & 0.0093 \\
OLS              & 0.1424 & 1.0854 & 0.3133 & 1.5324 & 0.4424 & -0.1678 & 0.7110 \\
PCR              & 0.1394 & 1.0629 & 0.3068 & 1.5111 & 0.4362 & -0.1773 & 0.7083 \\
PLS              & 0.1425 & 1.0861 & 0.3135 & 1.5331 & 0.4426 & -0.1678 & 0.7112 \\
GKX2020\_fw3       & 0.1535 & 1.0635 & 0.3070 & 1.2913 & 0.3728 & -0.2381 & 0.0155 \\
GKX2020\_fw4       & 0.1546 & 1.0725 & 0.3096 & 1.3110 & 0.3785 & -0.2352 & 0.0117 \\
GKX2020\_fw5       & 0.1546 & 1.0724 & 0.3096 & 1.3106 & 0.3783 & -0.2352 & 0.0114 \\
fw1              & 0.1487 & 1.1022 & 0.3182 & 1.6028 & 0.4627 & -0.1631 & 0.6387 \\
fw2              & 0.1423 & 1.0834 & 0.3128 & 1.4456 & 0.4173 & -0.1853 & 0.4814 \\
fw3              & 0.1491 & 1.0786 & 0.3114 & 1.3434 & 0.3878 & -0.2219 & 0.2501 \\
fw4              & 0.1515 & 1.0713 & 0.3092 & 1.3659 & 0.3943 & -0.2184 & 0.1031 \\
fw5              & 0.1545 & 1.0810 & 0.3120 & 1.3311 & 0.3842 & -0.2300 & 0.0324 \\
\bottomrule
\end{tabular}

\caption[Value-weighted portfolio performance with and without COVID-19 period considering the dynamic transaction cost.]{Value-weighted portfolio performance with and without COVID-19 period considering the dynamic transaction cost. Report for value-weighted portfolio excess returns. The upper panel is the report for the full testing period, while the lower panel is for the testing period without the pandemic. Both reports include the indicators of annualised returns (annual return), standard deviations (Std), Sharpe Ratio (SR), Sortino Ratio (SO), Maximum drawdown (MDD). The ‘buy-and-hold (V)’ applies as a benchmark strategy for all models without any prediction technique, hence no $\alpha$ is presented. The notation of 'Ann' means annualized.}
\label{tab:VW_ch2_turnover}
\end{table}

In conclusion, whether the equal-weighted portfolio or the value-weighted portfolio, the overall annualized returns from the COVID-19–Inclusive Period (2112) perform better than the Pre–COVID-19 Period (1911) due to the strong rebound trend during the pandemic in the static transaction scene. The strong upward trend with deep drawdown caused by the pandemic increases the SR and SO as well. The standard deviations for all models are smaller in the pre-pandemic period, while the MDDs are generally smaller in the pandemic as well. Between the equal-weighted portfolio and the value-weighted portfolio, the value-weighted portfolio for the COVID-19–Inclusive Period (2112) performs well for moderating the downside risks, while for the Pre–COVID-19 Period (1911), it increases the capital gain under certain risks for all models. Among the main and benchmark models, all proposed MLP models outperform the traditional statistical models, such as OLS, PLS and PCR in the scenario of static transaction cost. By considering risks and returns, the proposed MLP model with 3 hidden layers outperforms in the value-weighted portfolio for both periods and the equal-weighted portfolio for the Pre–COVID-19 Period (1911). It manages the balance of downside risk and relatively large capital gain, but for the equal-weighted portfolio COVID-19–Inclusive Period (2112), the MLP with 2 hidden layers overtakes the alternative models, achieving a better gain in terms of its risks. The robustness examination results show that high turnover rates significantly affect the models' profitability, but they have a lower impact than the transaction deduction rates (50bps for static and 20bps for dynamic). All models in all scenarios exhibit significant positive OOS $\alpha$s, indicating additional gains from factors. However, in all scenarios, the GKX2020 models are suspected of failing to detect trading signals. Their performance approaches the performance of the ‘buy-and-hold’ strategy. The turnover rate and transaction deduction rate have a high impact on the profitability of the models. Moreover, the transaction deduction rate has a higher impact than the turnover rate in this chapter. The main findings in this section are presented as: firstly, the model overfitting issue traps the sign-based trading signal detection; secondly, when the market experiences a strong upward trend, the fundamental trading strategy such as ‘buy-and-hold’ could be the most efficient trading strategy without considering the risks. In this sense, MLP models’ meaning for factor investing is more about managing downside risks and deriving a relatively high return instead of ‘beating the market’. Finally, the value-weighted portfolio reduces downside risks compared to the equal-weighted portfolio. Finally, trading signal filtering techniques can be applied to the system to further decrease the turnover rate and optimize the profitability. \\

\section[Conclusion for Chapter 2]{Conclusion and Discussion}\label{sec:conclusion_ch2}
This chapter revisits and develops the GKX2020’s MLP models via employing a dynamic MLP structure introduced by \citet{Coqueret2020MachineVersion} and features as firm characteristics sorted factors provided by \citet{AndrewY.Zimmermann2020OpenPricing}, re-evaluating the model and portfolio performance for the testing period of COVID-19. It considers the more realistic U.S. stock trading environment and practical conditions, such as high liquidity, ‘going concern’, variable borrowing rates and available borrowing shares from the brokers. Specifically, in the real trading environment, low liquidity brings systematic risks, such as failing to close the position in time to stop loss or confirm the trading profit. The `going concern' is due to the listed firms' unexpected insolvency or delisting, which may cause severe loss to the investors. For the short positions, which borrow the securities from either brokers or companies, investors face variable borrowing rates according to the demand or estimated market risks, and sometimes the rates can be high enough to make the short trading unavailable. Moreover, brokers usually provide limited borrowing shares each day.\\

The variable (factor) importance analysis reveals that the predictive power of firm-characteristics-sorted portfolio factors of the best-performing MLP models (fw2 and fw3) is highly time-varying and regime-dependent. In the pre-COVID period, AnnouncementReturn is the dominant predictor in the shallower model (fw2), whereas the deeper architecture (fw3) places greater emphasis on accounting-based signals such as Accruals. During the COVID-19-Inclusive Period, risk-related factors, including Earnings Forecast Disparity (EarningsForecastDisparity), leverage (BookLeverage), size, and idiosyncratic volatility (IdioVol3F), stably ranked in the top 20, reflecting heightened uncertainty and risk premia. In the post-COVID recovery phase (up to December 2022), Size becomes overwhelmingly dominant in both models, with the persistent strength of Announcement Return (AnnouncementReturn) and Earnings Forecast Disparity (EarningsForecastDisparity), as well as the renewed relevance of payout-related variables. These findings highlight three key insights: cross-sectional factor premia exhibit pronounced regime shifts during the extreme market turbulence, such as the one caused by COVID-19, capture the patterns that linear models such as Fama-MacBeth regressions would struggle to capture; information-based factors, particularly Announcement Return and Earnings Forecast Disparity, demonstrate considerable robustness of both models in all periods, confirming these factors as core drivers of the expected returns of the examined stocks; the modest but systematic differences between fw2 and fw3 suggest that deeper neural networks can capture higher-order nonlinear interactions among the factors that shallower models cannot capture. To conclude, variable importance analysis under the permutation method improves the economic and financial interpretability of the ML models in asset pricing.\\

This work also shows that a model's strong predictive power does not guarantee the profitability of a strategy built on it. The meaning of MLP models focuses on moderating the downside risks rather than achieving the absolute capital gain exceeding the buy-and-hold benchmarks, which verifies the results from \citet{Avramov2021MachinePredictability}. They achieve absolute returns during strong uptrends and deep downtrends. As the directional trading signal approach, which is commonly used in the trend-following strategy, is applied in this study, future researchers can explore alternative signal methods for improving profitability by improving the trading signal quality, such as deploying a signal filtering mechanism or configuring a restricted threshold for the correct market timing. Additionally, it is also worth investigating the generalization of the proposed MLP models, for example, examining them in alternative markets or assets, or on small or medium-cap stocks. Also, all models examined have significant positive $\alpha$,  which indicates the abnormal returns for the investors, but with these frames and factor configurations, stocks are not fully explained. The pattern of $\alpha$ shows no sign of model mis-specification. \\ 

Overall, MLP models bring promising performance on downside risk hedging and relatively high absolute return after transaction cost as the factor investing strategy in this case, especially in the COVID-19-Inclusive Period and the value-weighted method. This means models have a higher preference for stocks with larger market capitalization. 
The design of the factor investing strategy with the proposed MLP models highly approaches the real stock market practical conditions, which makes it more applicable to practitioners. It highlights the new inspirations on neural network models in the empirical asset pricing and factor investing. Whereas, the existing challenges of limited data length, trading signal filtering, model hyperparameters tuning and stock selection highly influence the model and portfolio performance. Topics such as how to moderate the restriction of data length, improve the trading signal quality and tune the model hyperparameter effectively are still worth further exploration. In addition, the MLP models pre-assume the sequential configuration of the data, with no mechanism embedded for processing data sequentially. More temporally specific neural network models can be employed for the asset pricing tasks. In addition, to further improve the financial and economic interpretability of these neural network models, the DM test can be applied to individual stocks to investigate the model's preference for an industry or sector, while investigating further on MSE and OOS $R^2$ for the contribution of characters, for example, which industry sectors have higher predictive power in a certain framework, and what is the impact of market capitalization size. These structures can also be applied to the volatility modelling for risk control. \\

\bmhead{Acknowledgements}
I would like to thank my supervisor, Professor Peter N Smith (University of York), for his support and guidance throughout this project. I am also grateful to Dr Penn Rainford, Dr Mark Stevenson, Dr Mark Hallam, and Professor Laura Coroneo for their comments and suggestions. I appreciate the technical support from Qiran Lai and Jingbo Yang. I thank Dr Grega Smrkolj for his advice on LaTeX formatting and diagrams, and my family and friends for their encouragement during this work. This research was supported by the Viking Computing Center (University of York) through access to its GPU-based high-performance computing (HPC) cluster.\\

Constructive feedback is greatly appreciated.





\begin{appendices}

\section{Algorithm Pseudocode}\label{sec:Algorithm Pseudocode}

\subsection{Adam optimizer}\label{subsec:Adam optimizer}
Adam updates parameters by maintaining the first and second moments of the gradients.
\begin{algorithm}[H]
\caption{Adam Optimization}
\begin{algorithmic}[1]
\State \textbf{Initialize:} $m_0 = 0$, $v_0 = 0$, $l = 0$
\State \textbf{Set hyperparameters:} learning rate $\eta = 0.001$, $\beta_1 = 0.9$, $\beta_2 = 0.999$, $\epsilon = 10^{-8}$
\While{$\theta_l$ not converged}
    \State $l \gets l + 1$
    \State Compute gradient: $g_l = \nabla_\theta L(\theta_{l-1})$
    \State $m_l \gets \beta_1 m_{l-1} + (1 - \beta_1) g_l$
    \State $v_l \gets \beta_2 v_{l-1} + (1 - \beta_2) g_l \odot g_l$ \Comment{$\odot$: element-wise multiplication}
    \State $\hat{m}_l \gets \frac{m_l}{1 - \beta_1^l}$ \Comment{Bias correction}
    \State $\hat{v}_l \gets \frac{v_l}{1 - \beta_2^l}$
    \State $\theta_l \gets \theta_{l-1} - \eta \cdot \frac{\hat{m}_l}{\sqrt{\hat{v}_l} + \epsilon}$
\EndWhile
\State \textbf{Return:} $\theta_l$
\end{algorithmic}
\begin{flushleft}
\textit{Source: Adapted from \citet{Gu2020EmpiricalLearning}}
\end{flushleft}
\end{algorithm}
where $g_t$ and $\theta_t$ represent the gradients and parameters computed from previous sections respectively.
\subsection{Early stopping}\label{subsec:Early stopping}
\begin{algorithm}[H]
\caption{Early Stopping}
\begin{algorithmic}[1]
\State \textbf{Initialize:} $j = 0$, $\epsilon = \infty$, select patience parameter $p$.
\While{$j < p$}
    \State Update $\theta$ using the training algorithm (e.g., for $h$ steps).
    \State Calculate the prediction error from the validation sample, denoted as $\epsilon'$.
    \If{$\epsilon' < \epsilon$}
        \State $j \gets 0$
        \State $\epsilon \gets \epsilon'$
        \State $\theta' \gets \theta$
    \Else
        \State $j \gets j + 1$
    \EndIf
\EndWhile
\State \textbf{Return:} $\theta'$
\end{algorithmic}
\begin{flushleft}
\textit{Source: Adapted from \citet{Gu2020EmpiricalLearning}}
\end{flushleft}
\end{algorithm}

\section{Mechanism of Backpropagation}
The context shows the backpropagation mechanism of a simple MLP. Concretely, when the results are calculated from Equation~\eqref{eq:4.2.5_ch2} and Equation~\eqref{eq:4.2.6_ch2}, we derive the loss from the forward round. Here, $W^{(l)}$ is the weight matrix for the $l$-th layer, $b^{(l)}$ is the bias, $\sigma(\cdot)$ is the activation function, and $a^{(l-1)}$ is the activation from the previous layer.
\begin{align}
z^{(l)} = W^{(l)} a^{(l-1)} + b^{(l)}\label{eq:4.2.5_ch2}\\
a^{(l)} = \sigma\left( z^{(l)} \right)\label{eq:4.2.6_ch2}
\end{align}
After loss from the output layer $\sigma^{(L)}$ has been computed via Equation~\eqref{eq:4.2.7_ch2}, backpropagation can take place. Here $L$ is the loss function.
\begin{equation}
\delta^{(L)} = \frac{\partial L}{\partial a^{(L)}} \cdot \sigma'\left( z^{(L)} \right)
\label{eq:4.2.7_ch2}
\end{equation}
Concretely,
\begin{equation}
\delta^{(l)} = \left( \delta^{(l+1)} W^{(l+1)T} \right) \cdot \sigma'\left( z^{(l)} \right)
\label{eq:4.2.8_ch2}
\end{equation}
Where $W^{(l+1)T}$ is the transpose of the weight matrix of the $(l+1)$-th layer, $\sigma'\left( z^{(l)} \right)$ is the derivative of the activation function from the $l$-th layer. Then, the new gradients can be computed via:
\begin{align}
\frac{\partial L}{\partial W^{(l)}} = a^{(l-1)T} \delta^{(l)}\\
\frac{\partial L}{\partial b^{(l)}} = \delta^{(l)}
\end{align}
Eventually, the weights are updated via the GD process:
\begin{align}
W^{(l)} = W^{(l)} - \eta \frac{\partial L}{\partial W^{(l)}}\\
b^{(l)} = b^{(l)} - \eta \frac{\partial L}{\partial b^{(l)}}
\end{align}

\begin{landscape}
\section{Definitions of Financial Factors}

\thispagestyle{plain} 

\setlength{\tabcolsep}{6pt} 
\small 

\begin{longtable}{p{0.15\linewidth} p{0.1\linewidth} p{0.08\linewidth} p{0.15\linewidth} p{0.45\linewidth}}
\caption{Detailed Definitions and Sources of Factor Variables} \label{tab:factor_defs} \\
\toprule
\textbf{Acronym} & \textbf{Authors} & \textbf{Year} & \textbf{Short Description} & \textbf{Detailed Definition} \\
\midrule
\endfirsthead

\multicolumn{5}{c}{{\bfseries \tablename\ \thetable{} -- Continued from previous page}} \\
\toprule
\textbf{Acronym} & \textbf{Authors} & \textbf{Year} & \textbf{Short Description} & \textbf{Detailed Definition} \\
\midrule
\endhead

\midrule
\multicolumn{5}{r}{{Continued on next page}} \\
\bottomrule
\endfoot

\bottomrule
\endlastfoot

AbnormalAccruals & Xie & 2001 & Abnormal Accruals & Define Accruals as net income (ib) minus operating cash flow (oancf), divided by average total assets (at) for years $t-1$ and $t$. If oancf is missing, replace operating cash flow with funds from operations (fopt) minus the annual change in total current assets (act) plus the annual change in cash and short-term investments (che) plus the annual change in current liabilities (lct) minus the annual change in debt in current liabilities (dlc). For each year $t$ and 2-digit sic code, regress Accruals on: the inverse of average total assets for year $t-1$, the change in revenue (sale) from year $t-1$ to $t$ divided by total assets for $t-1$, propery plant and equipment (ppegt) divided by total assets for $t-1$.  AbnormalAccrual is the residual from this cross-sectional regression. \\
\hline
Accruals & Sloan & 1996 & Accruals & Annual change in current total assets (act) minus annual change in cash and short-term investements (che) minus annual change in current liabilities (lct) minus annual change in debt in current liabilities (dlc) minus change in income taxes (txp). All divided by average total assets (at) over this year and last year. Exclude if $abs(prc) < 5$. \\
\hline
AccrualsBM & Bartov and Kim & 2004 & Book-to-market and accruals & Binary variable equal to 1 if stock is in the highest Accrual quintile and the lowest BM quintile, and equal to 0 if stock is in the lowest Accrual quintile and the highest BM quintile. Exclude if book equity (ceq) is negative. \\
\hline
AM & Fama and French & 1992 & Total assets to market & Total assets (at) divided by market value of equity. \\
\hline
AnalystRevision & Hawkins, Chamberlin, Daniel & 1984 & EPS forecast revision & keep $fpi=="1"$, last obs each month.  Signal is meanest / last month's meanest. \\
\hline
Announcement\\Return & Chan et al. & 1996 & Earnings announcement return & Get announcement date for quarterly earnings from IBES ($fpi = 6$). AnnouncementReturn is the sum of $(ret - mktrf + rf)$ from one day before an earnings announcement to 2 days after the announcement. \\
\hline
AssetGrowth & Cooper et al. & 2008 & Asset growth & Annual growth rate of total assets (at). \\
\hline
BetaLiquidityPS & Pastor and Stambaugh & 2003 & Stambaugh liquidity beta & Monthly excess return (ret -rf) regressed on innovations in liquidity from Pastor's website (\url{https://faculty.chicagobooth.edu/lubos.pastor/research/liq_data_1962_2018.txt}). Use 60 month rolling window regression, and require at least 36 non-missing observations. \\
\hline
BetaTailRisk & Kelly and Jiang & 2014 & Tail risk beta & Each month, compute the 5th percentile over daily returns over all firms. For all daily return observations with return below that 5th percentile, compute the average of (log(ret/5th percentile of cross-sectional return distribution). Call that average tailEX. BetaTailRisk is the coefficient of a 120-month rolling regression of a firm's stock return on tailEX. Exclude if price less than 5 or share code greater than 11. \\
\hline
BM & Stattman & 1980 & Book to market & Log of tangible book equity over market equity matched at fiscal year end. \\
\hline
BMdec & Fama and French & 1992 & BM using Dec ME & Book-to-market using December market equity values. \\
\hline
BookLeverage & Fama and French & 1992 & Book leverage (annual) & Total assets (at) divided by book value of equity plus deferred taxes (txditc) and preferred stock. Equity is shareholder equity (seq) if available, or book equity (ceq) plus preferred stock (pstk, if missing pstkrv, if missing pstkl), or total assets minus total liabilities (lt). \\
\hline
BPEBM & Penman et al. & 2007 & Leverage component of BM & $BP - EBM$, where $BP = (ceq + tstkp - dvpa)/(shrout*abs(prc))$, and $EBM$ is defined above. Exclude if price less than 5. \\
\hline
Cash & Palazzo & 2012 & Cash to assets & Ratio of quarterly cash and short-term investments (cheq) and total assets (atq). \\
\hline
CashProd & Chandrashekar and Rao & 2009	& Cash Productivity & Calculate market value of equity (mve\_c) as absolute price (prc) times number of shares outstanding (shrout). Cash productivity is equal to the difference between mve\_c and total assets (at) divided by cash and short-term investments (che). \\
\hline
CBOperProf & Ball et al. & 2016	& Cash-based operating profitability &	Revenue (revt) minus cost (cogs) - (administrative expenses (xsga) - R\&D expenses (xrd))  minus annual change in receivables (rect), annual change in investment (invt) and annual change in prepaid expenses, plus annual change in current deferred revenue (drc), long-term deferred revenue (drlt), accounts payable (ap) and accrued expenses (xacc), all divided by total assets (at) in year t. Replace all variables in the numerator with 0 if they are missing. Exclude if share code is greater than 11, market value of equity, BM or total assets are missing, or if SIC code is between 6000 and 6999.\\
\hline
CF & Lakonishok et al. & 1994 & Cash flow to market & Net income (ib) plus depreciation (dp) divided by market equity. Exclude NASDAQ stocks.\\
\hline
cfp & Desai et al. & 2004 &	Operating Cash flows to price &	Operating cash-flow (oancf) divided by market value of equity. If operating cash-flow is missing, replace by difference betwee net income (ib) and level of accruals, where the latter is the annual change in current assets (act) minus the annual change in cash and short-term investments (che), minus the annual change in current liabilities (lct) plus the annual change in debt in current liabilities (dlc) plus the annual change in payable income taxes (txp) plus depreciation (dp).\\
\hline
ChAssetTurnover & Soliman &	2008 & Change in Asset Turnover & Annual change in AssetTurnover (defined above). Exclude if price less than 5.\\
\hline
ChEQ & Lockwood and Prombutr & 2010	& Growth in book equity & Ratio of book equity (ceq) to book equity in the previous year. Include only if book equity is positive this year and last year.\\
\hline
ChForecastAccrual &	Barth and Hutton &	2004 &	Change in Forecast and Accrual &	Within upper half of Accruals distribution, equal to 1 if mean earnings estimate increased relative to the previous month. 0 if it decreased.\\
\hline
ChInv &	Thomas and Zhang &	2002 &	Inventory Growth &	12-month change in inventory (invt) divided by average total assets.\\
\hline
ChInvIA & Abarbanell and Bushee & 1998 & Change in capital inv (ind adj) &	Growth in capital expenditure (capx) minus average growth in capital expenditure in the same industry (two-digit SIC). If capx is missing, capital expenditure is defined as the annual change in property, plant and equipment (ppent). Capital expenditure growth is defined as the percentage growth of capx today relative to the average capx over the previous two years (.5*(capx$_{t-1}$ + capx$_{t-2}$), or as percentage growth relative to the previous year only if $t-2$ is missing.\\
\hline
ChNAnalyst & Scherbina & 2008 &	Decline in Analyst Coverage & Using IBES forecasts, keep if $fpi = 1$, fpedats not missing, and $fpedats > statpers + 30$.  Binary variable equal to 1 if the number of analysts (numest) for next quarter's EPS estimate decreased relative to three months ago, and 0 if it increased. Keep if in bottom two size quintiles among all firms to match OP's table.\\
\hline
ChNNCOA	& Soliman &	2008 & Change in Net Noncurrent Op Assets & Twelve-month change in noncurrent operating assets. Noncurrent operating assets is $( (at - act - ivao) - (lt - dlc - dltt) )/at$.\\
\hline
ChNWC &	Soliman & 2008 & Change in Net Working Capital & 	Twelve-month change in net working capital. Net working capital is $( (act - che) - (lct - dlc) )/at$\\
\hline
ChTax &	Thomas and Zhang & 2011 & Change in Taxes &	4-quarter change in quarterly total taxes (txtq), scaled by lagged total assets (at).\\
\hline
CompEquIss & Daniel and Titman & 2006 &	Composite equity issuance & 5-year growth rate of market value of equity minus 5-year stock return. \\
\hline
CompositeDebtIssuance &	Lyandres et al. & 2008 &	Composite debt issuance & Log of long-term debt (dltt) plus debt in current liabilities (dlc) minus log of the same variable 5 years ago.\\
\hline
ConvDebt &	Valta &	2016 & Convertible debt indicator &	Binary variable equal to 1 if deferred charges (dc) greater than 0 or common shares reserved for convertible debt (cshrc) greater than 0.\\
\hline
CoskewACX &	Ang et al. & 2006 &	Coskewness using daily returns &	Signal is the sample counterpart of $E[\tilde{r}_{it} \tilde{r}_{mt}^2]/(SD[\tilde{r}_{it} ] SD[\tilde{r}_{mt}]^2$ where $\tilde{r}_{it}$ is the de-meaned stock return and $\tilde{r}_{mt}$ is the de-meaned market excess return.  Signal is computed using the past year of daily data, and using the NYSE CRSP VW index for the market (dsia), with returns continuously compounded. \\
\hline
CustomerMomentum & Cohen and Frazzini &	2008 & Customer momentum &	Based on firms' principals customers from Compustat Segment data as in Cohen and Frazzini.\\
\hline
DelBreadth & Chen et al. & 2002 & Breadth of ownership &	Quarterly change in the number of institutional owners (numinstowners) from 13F data. Exclude if in the lowest quintile of stocks by market value of equity (based on NYSE stocks only).\\
\hline
DelCOA & Richardson et al. & 2005 &	Change in current operating assets & Difference in current operating assets (total current assets (act) minus cash and short-term investments (che)) between years t-1 and t, scaled by average total assets (at) in years t-1 and t.\\
\hline
DelCOL & Richardson et al. & 2005 & Change in current operating liabilities & Difference in current operating liabilities (total current liabilities (lct) minus debt in current liabilities (dlc)) between years t-1 and t, scaled by average total assets (at) in years t-1 and t.\\
\hline
DelEqu & Richardson et al. & 2005 &	Change in equity to assets & Difference in book equity (ceq) between years t-1 and t, scaled by average total assets (at) in years t-1 and t.\\
\hline
DelFINL & Richardson et al. & 2005 & Change in financial liabilities & Difference in financial liabilities (sum of long-term debt (dltt), current liabilitites (dlc) and preferred stock (pstk)) between years t-1 and t, scaled by average total assets (at) in years t-1 and t.\\
\hline
DelLTI & Richardson et al. & 2005 &	Change in long-term investment & Difference in investment and advances (ivao) between years t-1 and t, scaled by average total assets (at) in years t-1 and t.\\
\hline
DelNetFin &	Richardson et al. &	2005 & Change in net financial assets & Compute the sum of short-term investments (ivst) and investments and advances (ivao) minus the sum of long-term debt (dltt), debt in current liabilities (dlc) and preferred stock capital. Divide the difference between the current and one-year lagged sum by total assets (at) averaged over the current and previous fiscal years.\\
\hline
DivInit & Michaely et al. & 1995 & Dividend Initiation &	Keep only $distcd$ 2nd d$igit = 2 or 3$.  Define dividend initiation as having paid a dividend in month $t (divamt > 0)$, and not having paid a dividend in the last 24 months.  DivInit is equal to 1 if a dividend was initiated in the past 6 months, and 0 for all other stocks.\\
\hline
DivOmit & Michaely et al. & 1995 & Dividend Omission & Keep only $distcd$ 2nd $digit = 2 or 3$.  Define firms as quarterly, semi-annual, or annual payers based on payment history.  Define a consistent payer as a firm that paid quarterly or semi-annual dividends for the past 18 months, or annual dividends for the past 2 years.  An omission is the first month where a consistent payer failed to pay a dividend over the past quarter/6-months/year. Finally, DivOmit = 1 if there is an omission in the past 2 months and 0 otherwise.\\
\hline
DivSeason &	Hartzmark and Salomon &	2013 &	Dividend seasonality & Drop if 3rd digit of $distcd = 2 or >= 6$.  Assign DivSeason = 0 if there was a dividend paid in the last 12 months.  Replace DivSeason = 1 if the third digit of $disctcd$ is 3, 0, or 1, and a positive dividend was paid 2, 5, 8, or 11 months ago, if the third digit is 4 and a dividend was paid 5 or 11 months ago, or if the third digit is 5 and a dividiend was paid 11 months ago.\\
\hline
DivYieldST & Litzenberger and Ramaswamy & 1979 &	Predicted div yield next month & Using CRSP distributions, keep only $distcd$ beginning in 12 and if the third digit of $distcd$ is 3, 4, or 5.  Also keep only if stock paid a dividend in the last 12 months.  Define $Ediv1 = div$ 2 months ago if $distcd$'s 3rd digit is 0, 1, or 3.  Define $Ediv1 = div$ 5 months ago if the $distcd$ 3rd digit is 4.  Define $Ediv = div$ 11 months ago if the $distcd$ 3rd digit is 5.  Define $Edy1$ as $Ediv1/abs(prc)$.  Finally, discretize $Edy1$ to smooth around the huge mass at 0 as follows: $DivYieldST = 0$ if $Edy1 = 0$, $1$ if $Edy$ is between 0 and 0.005, 2 if $Edy1$ is between 0.005 and 0.010, and 3 if $Edy > 0.010$.\\
\hline
dNoa &	Hirshleifer et al. & 2004 &	change in net operating assets & 12-month growth in Net Operating Assets scaled by lagged total assets (at). Net Operating assets are operating assets minus operating liabilities. Operating assets are total assets (at) minus cash and short-term investments (che), operating liabilities are total assets minus long-term debt (dltt), minority interest (mib), deferred charges (dlc), book equity (ceq) and preferred stock (pstk), all items (except at and ceq) replaced with 0 if missing.\\
\hline
DolVol & Brennan et al. &	1998 &	Past trading volume &	Log of two-month lagged trading volume (vol) times two-month lagged price (prc).\\
\hline
Earnings\\Consistency & Alwathainani &	2009 &	Earnings consistency &	Average earnings growth over previous 48 months. Earnings growth is defined as EPS (epspx) minus EPS 12 months ago divided by average EPS 12 and 24 months ago. Exclude if price less than 5, absolute value of 12 month earnings growth greater 600\%, or earnings growth and earnings growth 12 months ago have different signs.\\
\hline
Earnings\\ForecastDisparity	& Da and Warachka &	2011 &	Long-vs-short EPS forecasts &	Analyst forecasted 5-year earnings growth (fgr5yr) minus 100 times the difference between mean earnings forecast (meanest) and fiscal year earnings expectations (fy0a) scaled by the absolute value of fy0a.  Drop if fpedats is missing or $fpedats - statpers < 30$\\
\hline
EarningsSurprise &	Foster et al. &	1984 &	Earnings Surprise &	EPS (epspxq) minus EPS twelve months ago - Drift, scaled by standard deviation of that expression. Drift is the average earnings growth (EPS - EPS twelve months ago) over the past two years. Exclude if price less than 5.\\
\hline
EarnSupBig &	Hou&	2007&	Earnings surprise of big firms&	Average monthly value of EarningsSurprise (defined above) of the 30\% largest companies by market value of equity in the same Fama-French 48 industry. Exclude the largest 30\% of companies for EarnSupBig (not to compute the anomaly).\\
\hline
EBM & Penman et al. & 2007&	Enterprise component of BM&$	(ceq + che - dltt - dlc - dc - dvpa+ tstkp ) / (mve\_c + che - dltt - dlc - dc - dvpa+ tstkp)$. Exclude if price less than 5.\\
\hline
EntMult & Loughran and Wellman & 2011 &	Enterprise Multiple & Market value of $equity + long-term debt (dltt) + debt in current liabilities (dlc) + deferred charges (dc) - cash$ and short-term investments (che) , divided by operating income (oibdp). Exclude if missing book equity or negative operating income.\\
\hline
EP &	Basu&	1977&	Earnings-to-Price &	$ib / lag(market value of equity$, 6 months).  NYSE stocks only.  Exclude if $EP < 0$.  Lag simulates the Dec 31 market equity used in original paper.\\
\hline
EquityDuration& 	Dechow et al.&	2004 &	Equity Duration&	see code of \citet{AndrewY.Zimmermann2020OpenPricing}.\\
\hline
ExchSwitch & 	Dharan and Ikenberry&	1995&	Exchange Switch&	Binary variable equal to 1 if a firm switched from AMEX or NASDAQ to NYSE within the past year, or from NASDAQ to AMEX within the past year.\\
\hline
FEPS&	Cen et al. &	2006&	Analyst earnings per share&	Using IBES unadjusted forecasts, keep fpi == 1, signal is meanest.\\
\hline
fgr5yrLag&	La Porta&	1996&	Long-term EPS forecast&	Lag long-term earnings forecast (fgr5yr) by 6 months.   Then keep only June observations, and fill in missing with most recent obs.  Exclude if book equity (ceq), net income (ib), deferred taxes (txdi), dividends (dvp), revenue (sale) or depreciation (dp) is missing. \\
\hline
FirmAgeMom&	Zhang&	2006&	Firm Age - Momentum&	6 month return, restricted to the bottom quintile of the cross-sectional firm age distribution. Exclude if price less than 5 or firm younger than 12 months.\\
\hline
ForecastDispersion&	Diether et al. &	2002&	EPS Forecast Dispersion&	Keep fpi = 1 and $fpedats > statpers + 30$.  Standard deviation of earnings estimates $(stdev\_est)$ scaled by mean earnings estimate.\\
\hline
Frontier&	Nguyen and Swanson&	2009&	Efficient frontier index&	Frontier is the residual of a regression of log(BM) on log(book equity (ceq)), long-term debt (dltt) to assets (at), capital expenditures (capx) to revenue (sale), R\&D expense (xrd) to revenue, advertising expense (xad) to revenue, property plant and equipment (ppent) to assets, EBIT (ebitda) to assets, and dummies for Fama-French's 48 industry definitions. Regression is updated each month with a rolling window of 60 months.\\
\hline
GP&	Novy-Marx&	2013&	gross profits / total assets&	Revenue (sale) - cost of goods solds (cogs), divided by total assets (at).  Drop if financial.\\
\hline
GrAdExp&	Lou&	2014&	Growth in advertising expenses&	Log of advertising expense (xad) minus log of advertising expense last year. Exclude if price less than 5, xad less than .1 or stock in the lowest decile of market value of equity.\\
\hline
grcapx&	Anderson and Garcia-Feijoo&	2006&	Change in capex (two years)&	Growth rate of capital expenditures (capx) relative to two years ago.   If capx is missing, replace with annual change in property, plant and equipment (ppent).\\
\hline
grcapx3y&	Anderson and Garcia-Feijoo&	2006	&Change in capex (three years)&	Capital expenditures (capx) divided by the sum of capital expenditures from year - 1, year -2, and year -3.   If capx is missing, replace with annual change in property, plant and equipment (ppent).\\
\hline
Herf&	Hou and Robinson&	2006&	Industry concentration (sales)&	Three-year rolling average of the three digit industry Herfindahl index based on firm revenue (sale). Exclude regulated industries (4011, 4210, 4213 \& year $\leq$ 1980; 4512 \& year $\leq$ 1978, 4812, 4813 \& year $\leq$ 1982, 4900-4999 in any year).\\
\hline
HerfBE&	Hou and Robinson&	2006&	Industry concentration (equity)&	Three-year rolling average of the three digit industry Herfindahl index based on firm book equity. Exclude regulated industries (4011, 4210, 4213 \& year $\leq$ 1980; 4512 \& year $\leq$ 1978, 4812, 4813 \& year $\leq$ 1982, 4900-4999 in any year).\\
\hline
hire&	Bazdresch et al.&	2014&	Employment growth&	Change in number of employees (emp) between t-1 and t, scaled by average number of employees in t-1 and t. Replace hire with 0 if emp or lagged emp is missing.\\
\hline
IdioVol3F&	Ang et al.&	2006&	Idiosyncratic risk (3 factor)&	Standard deviation of residuals from Fama-French three factor regressions using the past month of daily data.  Value weighted.\\
\hline
IdioVolAHT&	Ali et al.&	2003&	Idiosyncratic risk (AHT)&	Standard deviation of residuals from CAPM regressions using the past year of daily data. Require at least 100 non-missing observations.\\
\hline
Illiquidity& Amihud& 2002&	Amihud's illiquidity& Past twelve-month average of: daily return (abs(ret)) divided by turnover($(abs(prc)*vol$).\\
\hline
IndIPO&	Ritter&	1991&	Initial Public Offerings&	1 if IPO in the past 6-36 months.  0 otherwise.  IPO dates are taken from Jay Ritter's IPO data available at: http://bear.warrington.ufl.edu/ritter/ipodata.htm.  Missing IPO dates imply IndIPO = 0.\\
\hline
IndMom&	Grinblatt and Moskowitz&	1999&	Industry Momentum&	Weighted average of firm-level 6 month buy-and-hold return. Average is taken over two digit industries each month and weights are based on market value of equity.\\
\hline
IndRetBig&	Hou&	2007&	Industry return of big firms&	Average monthly return (ret) of the 30\% largest companies by market value of equity in the same Fama-French 48 industry. Exclude the largest 30\% of companies for IndRetBig (not to compute the anomaly!).\\
\hline
IntanBM&	Daniel and Titman&	2006&	Intangible return using BM&	In each month, run a cross-sectional regression of a firm's five-year stock return on 5-year lagged BM (defined above) and a constructed regressor that is the change in BM from 5 years ago to today plus the five-year stock return. The residual from that regression is IntanBM.\\
\hline
IntanCFP&	Daniel and Titman&	2006&	Intangible return using CFtoP&	In each month, run a cross-sectional regression of a firm's five-year stock return on the 5-year lagged  CFP = (net income (ni) plus depreciation (dp))/market value of equity and a constructed regressor that is the change in CFP from 5 years ago to today plus the five-year stock return. The residual from that regression is IntanCFP.\\
\hline
IntanEP&	Daniel and Titman&	2006&	Intangible return using EP&	In each month, run a cross-sectional regression of a firm's five-year stock return on the 5 year lagged  EP = net income (ni)/market value of equity and a constructed regressor that is the change in EP from 5 years ago to today plus the five-year stock return. The residual from that regression is IntanEP.\\
\hline
IntanSP&	Daniel and Titman&	2006&	Intangible return using Sale2P&	In each month, run a cross-sectional regression of a firm's five-year stock return on 5-year lagged SP (defined above) and a constructed regressor that is the change in SP from 5 years ago to today plus the five-year stock return. The residual from that regression is IntanSP.\\
\hline
IntMom&	Novy-Marx&	2012&	Intermediate Momentum&	Stock return between months t-12 and t-6.\\
\hline
Investment&	Titman et al.&	2004&	Investment to revenue&	Ratio of capital investment (capx) to revenue (revt) divided by the firm-specific 36-month rolling mean of that ratio. Exclude if revenue less than \$10m.\\
\hline
InvestPPEInv&	Lyandres et al.&	2008&	change in ppe and inv/assets&	One-year change in property, plants and equipment (ppegt) plus one year change in inventory (invt), scaled by one-year lagged assets (at).\\
\hline
InvGrowth&	Belo and Lin&	2012&	Inventory Growth&	Defate $invt$ growth using $gnp$ deflator.  Signal is deflated $invt$ growth rate from fiscal year t to fiscal year t-1.  Drop if 1 digit sic code is 4 or 6, or if at or $ppent <= 0$.\\
\hline
Leverage&	Bhandari&	1988&	Market leverage&	Total liabilities (lt) divided by market value of equity.\\
\hline
LRreversal&	De Bondt and Thaler&	1985&	Long-run reversal&	Stock return between months t-36 and t-13.\\
\hline
MaxRet&	Bali et al.&	2011&	Maximum return over month&	Maximum of daily returns (ret) over the previous month.\\
\hline
MeanRank\\RevGrowth&	Lakonishok, Shleifer, Vishny&	1994&	Revenue Growth Rank&	Rank firms by their annual revenue growth each year over the past 5 years. MeanRankRevGrowth is the weighted average of ranks over the past 5 years, that is, MeanRankRevGrowth = (5*Rank$_{t-1}$ + 4*Rank$_{t-2}$ + 3*Rank$_{t-3}$ + 2*Rank$_{t-4}$ + 1*Rank$_{t-5}$)/15. Exclude NASDAQ stocks.\\
\hline
Mom12m&	Jegadeesh and Titman&	1993&	Momentum (12-month)&	Stock return between months t-12 and t-1.\\
\hline
Mom12mOff\\Season&	Heston and Sadka&	2008&	Momentum without the seasonal part&	Average return in other months over the previous year.\\
\hline
Mom6m&	Jegadeesh and Titman&	1993&	Momentum (6 month)&	Stock return between months t-6 and t-1.\\
\hline
MomOffSeason&	Heston and Sadka&	2008&	Off season long-term reversal&	Average return in other months over the preceding 2-5 years.\\
\hline
MomOffSeason\\06YrPlus&	Heston and Sadka&	2008&	Off season reversal years 6 to 10&	Average return in other months over the preceding 6-10 years.\\
\hline
MomOffSeason\\16YrPlus&	Heston and Sadka&	2008&	Off season reversal years 16 to 20&	Average return in other months over the preceding 16-20 years.\\
\hline
MomRev&	Chan and Ko&	2006&	Momentum and LT Reversal&	Binary variable equal to 1 if firm is in the highest Mom6m quintile and the lowest Mom36m quintile, and equal to 0 if firm is in the lowest Mom6m quintile and the highest Mom36m quintile. Exclude if price less than 5.\\
\hline
MomSeason&	Heston and Sadka&	2008&	Return seasonality years 2 to 5&	Average return in the same month over the preceding 2-5 years.\\
\hline
MomSeason\\06YrPlus&	Heston and Sadka&	2008&	Return seasonality years 6 to 10&	Average return in the same month over the preceding 6-10 years.\\
\hline
MomSeason\\11YrPlus	&Heston and Sadka&	2008&	Return seasonality years 11 to 15&	Average return in the same month over the preceding 11-15 years.\\
\hline
MomSeason\\16YrPlus	&Heston and Sadka&	2008&	Return seasonality years 16 to 20&	Average return in the same month over the preceding 16-20 years.\\
\hline
MomSeason\\Short	&Heston and Sadka&	2008&	Return seasonality last year&	Average return in the same month in the previous year.\\
\hline
MomVol&	Lee and Swaminathan&	2000&	Momentum in high volume stocks&	Define momentum as Mom6m, and volume as the rolling average of the past 6 months of monthly turnover (minimum 5 months).  Independent sort stocks into 10 momentum ports and 3 volume ports. Keep if volume is in the top port, and assign signal = momentum port.  Drop if less than 2 years on CRSP.\\
\hline
MS&	Mohanram&	2005&	Mohanram G-score&	See code of \citet{AndrewY.Zimmermann2020OpenPricing} for details.  MS is only evaluated for low BM firms and comes from combining three signals related to profitability and cash flow, two signals related to income volatility, and three signals related to investment.\\
\hline
NetDebtFinance&	Bradshaw et al.&	2006&	Net debt financing&	Long-term debt issuance (dltis) minus long-term debt reduction (dltr) minus current debt changes (dlcch), scaled by average total assets (at) in years t-1 and t. Replace missing values of $dlcch$ with 0. Exclude if ratio is greater than 1.\\
\hline
NetDebtPrice&	Penman et al.&	2007&	Net debt to price&	Long-term debt (dltt) plus debt in current liabilities (dlc) plus preferred stock (pstk) plus preferred dividends in arrears (dvpa) minus treasury stock (tstkp) minus cash and short-term investments (che), scaled by market value of equity. Exclude if SIC between 6000 and 6999, or if missing value for total assets (at), net income (ib), common shares outstanding (csho), book value of equity (ceq) or price close fiscal year (prcc\_f).  Keep only 3rd B/M Quintile.\\
\hline
NetEquityFinance&	Bradshaw et al.&	2006&	Net equity financing&	Sale of common stock (sstk) minus  purchase of common stock (prstkc), scaled by average total assets (at) from years t and t-1. Exclude if absolute value of ratio is greater than 1.\\
\hline
NetPayoutYield&	Boudoukh et al.&	2007&	Net Payout Yield&	Dividends (dvc) plus purchase of common and preferred stock (prstkc) minus sale of common and preferred stock (sstk), divided by market value of equity lagged 6 months.  Exclude if NetPayoutYield is 0, financial firm based on SIC code, $ceq <= 0$, or less than 2 years in CRSP.\\
\hline
NOA&	Hirshleifer et al.&	2004&	Net Operating Assets&	Difference between operating assets and operating liabilities, scaled by lagged total assets. Operating assets are total assets (at) minus cash and short-term investments (che), operating liabilities are total assets minus long-term debt (dltt), minority interest (mib), deferred charges (dc) and book equity (ceq).\\
\hline
OPLeverage&	Novy-Marx&	2011&	Operating leverage&	Sum of administrative expenses (xsga) and cost of goods sold (cogs), scaled by total assets (at).  Use xsga = 0 if xsga is missing.\\
\hline
OrderBacklog&	Rajgopal et al.&	2003&	Order backlog&	Order backlog (ob) divided by average total assets (at) in years t-1 and t. Exclude if order backlog is 0.\\
\hline
OrderBacklog\\Chg&	Baik and Ahn&	2007&	Change in order backlog&	Define normalized order backlog as order backlog (ob) divided by average total assets (at) in years t-1 and t. Exclude if order backlog is 0.  Signal is normalized order backlog minus normalized order backlog one year ago.\\
\hline
OrgCap&	Eisfeldt and Papanikolaou&	2013&	Organizational capital	&Defined recursively. Set xsga to 0 if missing.  Initialize with OrgCap = 4*general expenses (xsga) in the first year, and calculate as .85*OrgCap previous year + xsga/gnpdeflator current year thereafter. Scale by total assets (at).  Keep only december year ends, and non-financials, and non-missing sic code.  Then winsoriz at the 1\% level and standardized by subtracting the same month within-industry mean (Fama French 17 industries) and dividing by the same month within industry standard deviation of OrgCap.\\
\hline
OScore&	Dichev&	1998&	O Score&	OScore = -1.32 - .407*log(at/GNP deflator) + 6.03*(lt/at) - 1.43*( (act - lct)/at) + .076*(lct/act) - 1.72*I($lt > at$) - 2.37*(ib/at) - 1.83*(fopt/lt) + .285*(ib + ib$_{t-12}$ + ib$_{t-24}$ $< 0$) -  .521*( (ib - ib$_{t-12}$)/(abs(ib) + .abs(ib$_{t-12}$)) ). $fopt = oancf$ if fopt is missing.  Exclude if SIC code between 3999 and 4999, or greater than 5999. Exclude if price less than 5.  Then exclude if OScore is in bottom quintile of OScore (original paper shows non-monotonic returns, as does our replication).\\
\hline
PayoutYield&	Boudoukh et al.&2007&	Payout Yield&	Sum of dividends (dvc), purchase of common and preferred stock (prstkc) and max(preferred stock redemption value (pstkrv), 0) over market value of equity lagged 6 months.  Exclude if PayoutYield $\leq$ 0, financial firm based on SIC code, $ceq <= 0$, or less than 2 years in CRSP.\\
\hline
PctAcc&	Hafzalla et al.&	2011&	Percent Operating Accruals&	Income before extraordinary items (ib) minus net cash flow (oancf) divided by absolute value of ib. If oancf is missing, PctAcc is defined as ((act - act$_{t-12}$) - (che - che$_{t-12}$) - ((lct - lct$_{t-12}$) - (dlc - dlc$_{t-12}$) - (txp - txp$_{t-12}$) - dp))/abs(ib). In either case, if ib is equal to 0, divide by .01 instead. Exclude if price less than 5.\\
\hline
Price&	Blume and Husic&	1973&	Price&	Log of absolute value of price (prc).\\
\hline
PriceDelayRsq&	Hou and Moskowitz&	2005&	Price delay r square&	Each July regress stock return on day $t$ on market return in $t, t-1, \ldots, t-4$ using observations from July 1 of the previous year to June 30 of the current year.  Then regress again with no market return lags.  PriceDelay Rsq = 1 - [Rsq from second regression]/[Rsq from first regression].\\
\hline
PS&	Piotroski&	2000&	Piotroski F-score&	Sum of nine indicator variables which are: 1 if net income (ib) greater 0; 1 if net cash flow (oancf) greater 0; 1 if return on assets (ib/at) increased relative to previous year; 1 if net cash flow greater net income; 1 if long-term debt to assets (dltt/at) declined over the previous year;  if current assets to current liabilities (act/lct) increased over the previous year; 1 if ebit/sale (ebit = ib + txt + xint) increased over the previous year; 1 if revenue to assets increased over the previous year; 1 if shrout $\leq$ shrout last year. Include highest quintile of book-to-market only.  Exclude if missing any of the input variables.\\
\hline
RD&	Chan et al.&	2001&	R\&D over market cap&	R\&D expense (xrd) over market value of equity.\\
\hline
RDAbility&	Cohen et al.&	2013&	R\&D ability& Regress log of sales growth (sale over sale in previous year) on log of (1+ xrd/sale) in 5 bivariate regressions with (1+xrd/sale) lagged by $1, \ldots, 5$ years. Run regressions over previous 8 years and require at least 6 non-missing observations. Also require at least half of past research and development observations to be non-zero. RDAbility is the mean of the coefficients on the five lags of log(1+xrd/sale). Set to missing if firm is not in the highest tercile of xrd/sale in a year or if reseach and development expenses are non-positive.\\
\hline
RDcap&	Li&	2011&	R\&D capital-to-assets&	R\&D capital to assets is the weighted sum of lagged R\&D expenditures scaled by the assets (at). I replace xrd with 0 if xrd is missing and compute the numerator as $xrd_t + .8 xrd_{t-1}+ .6xrd_{t-2} + .4 xrd_{t-3} + .2 xrd_{t-4}$. Replace RDcap with missing before 1980 or if firm is in upper two thirds of market cap (shrout*abs(prc)) distribution in a month.\\
\hline
RDIPO&	Gou, Lev and Shi&	2006&	IPO and no R\&D spending&	Binary variable equal to 1 if R\&D expense (xrd) = 0 and IndIPO = 1. 0 otherwise.\\
\hline
RDS&	Landsman et al.&	2011&	Real dirty surplus	&Define Dirty Surplus as annual change in marketable securities adjustment $msa$ plus annual change in retained earnings adjustment (recta) + .65 times the annual change in min(Unrecognized prior service cost ($pcupsu$) - Pension additional minimum liability ($paddml$),0). Real dirty surplus is the annual change in book equity ($ceq$) minus dirty surplus minus (net income ($ni$) minus dividends preferred ($dvp$)) + dividends ($divamt$) - end-of-fiscal-year-stock-price $(prcc\_f)$*annual change in common shares outstanding ($csho$).\\
\hline
RealizedVol&	Ang et al.&	2006& Realized (Total) Volatility&	Standard deviation of residuals from CAPM regressions using the past month of daily data. Value weighted.\\
\hline
Residual\\Momentum&	Blitz et al.&	2011&	Momentum based on FF3 residuals&	Run a rolling regression over 36 months of excess return (retrf) on excess market return (mktrf), size and value factors (smb, hml) and compute idiosyncratic returns as the one-month lagged residual. ResidualMomentum is the rolling mean of the residual divided by the rolling standard deviation of the residual, both computed over the past 11 months.\\
\hline
retConglomerate&	Cohen and Lou&	2012&	Conglomerate return&	Identify conglomerate firms as those with multiple OPSEG or BUSSEG entries in the Compustat segment data (and require that at least 80\% of firm's total assets are covered by segment data). Compute monthly stock return at the 2-digit SIC level for stand-alone (non-conglomerate) firms only, and match those returns to conglomerates' segments. Compute weighted conglomerate return as the industry return of stand-alone companies, weighted with a conglomerate's total sales in each industry.\\
\hline
ReturnSkew&	Bali et al.&	2015&	Return skewness&	Skewness of daily returns (ret) over previous month.\\
\hline
ReturnSkew3F& Bali et al.&	2015&	Idiosyncratic skewness (3F model)&	Skewness of idiosyncratic returns computed as residuals from regression of daily excess returns (ret - rf) on Fama-French factors (mktrf, smb, hml) over the previous month. I require at least 15 non-missing observations.\\
\hline
REV6&	Chan et al.&	1996&	Earnings forecast revisions&	Define revisions as the change in the mean earnings estimate (meanest) for the next quarter from month t-1 to t, scaled by stock price in month t-1. REV6 is the sum of that variable from months t-6 to t.\\
\hline
Revenue\\Surprise&	Jegadeesh and Livnat&	2006&	Revenue Surprise&	Define revenue per share as quarterly revenue (revtq) divided by quarterly common shares outstanding (cshprq). RevenueSurprise is the 4-quarter change in revenue per share minus the average 4-quarter change in revenue per share over the previous 2 years. RevenueSurprise is scaled by its standard deviation over the previous 2 years. Exclude if price less than 5.\\
\hline
RIO\_Disp&	Nagel&	2005&	Inst Own and Forecast& Dispersion	Follow RIO\_MB, except define Disp as the stdev of IBES forecasts where fpi == 1 divided by at, and sort on Disp instead of MB.  Finally, let RIO\_Disp = lagged RIO quntile if the Disp quintile $>= 4$.\\
\hline
RIO\_MB&	Nagel&	2005&	Inst Own and Market to Book&	Residual institutional ownership (RIO) is defined as log(institutional ownership (instown\_perc)/(1-institutional ownership)) + 23.6 - 2.89*log(market value of equity) +.08*log(market value of equity)$^2$. Replace instown\_perc with 0 if it is missing, with .9999 if it's above .9999, and with .0001 if it's below .0001. Then lag RIO by 6 months.  Keep only stocks with size $> 20th$ percentile across exchcd == 1 or 2.  Define MB as market cap / (ceq + txditc), where txditc = 0 if missing, and MB is replaced with missing if ceq $< 0$.  Then sort into quintiles based on the lagged RIO, and sort into quintiles based on MB.  Finally, let RIO\_MB = lagged RIO quintile if the MB quintile = 5.\\
\hline
RIO\_Turnover&	Nagel&	2005&	Inst Own and Turnover&	Follow RIO\_MB, except define turnover as vol/shrout. Let RIO\_Turnover = lagged RIO quintile if the Volatility quintile == 5.\\
\hline
RIO\_Volatility&	Nagel&	2005&	Inst Own and Idio Vol&	Follow RIO\_MB, except define volatility as the rolling standard deviation of the last 12 months of the stock return. Let RIO\_Disp = lagged RIO quintile if the Volatility quintile == 5.\\
\hline
roaq&	Balakrishnan et al.&	2010&	Return on assets (qtrly)&	Quarterly net income (ibq) divided by lagged total assets (atq).\\
\hline
sfe&	Elgers et al.&	2001&	Earnings Forecast to price&	Median estimate for next year eps (fpi = 1) in March, divided by stock price from December.  Dec fiscal year ends only, keep only forecasts more than 90 days out.  Keep only below median analyst coverage each month.\\
\hline
ShareIss1Y&	Pontiff and Woodgate&	2008&	Share issuance (1 year)&	Growth in number of shares between t-18 and t-6. Number of shares is calculated as shrout/cfacshr to adjust for splits.\\
\hline
ShareIss5Y&	Daniel and Titman&	2006&	Share issuance (5 year)&	5-year growth in number of shares. Number of shares is calculated as shrout/cfacshr to adjust for splits.\\
\hline
ShareVol&	Datar et al.&	1998&	Share Volume&	Let tempvol = sum of monthly share trading volume (vol) over the previous three months, scaled by 3 times common shares outstanding (shrout).  Let ShareVol = 1 if tempvol > 10\%, and ShareVol = 0 if tempvol $< 5\%$.\\
\hline
Size&	Banz&	1981&	Size&	Log of monthly market value of equity (abs(prc)*shrout)).\\
\hline
std\_turn&	Chordia et al. &	2001&	Share turnover volatility& Standard deviation of turnover (vol/shrout) over the past 36 months.\\
\hline
STreversal&	Jegadeesh&	1990&	Short term reversal&	Stock return (ret) over the previous month.\\
\hline
SurpriseRD&	Eberhart et al.&	2004&	Unexpected R\&D increase&	Binary variable equal to 1 if: R\&D (xrd) scaled by revenue (revt) is positive, R\&D scaled by total assets (at) is positive, annual R\&D growth is greater than 5\%, annual growth in R\&D over total assets is greater than 5\%. SurpriseRD is 0 otherwise.\\
\hline
tang&	Hahn and Lee&	2009&	Tangibility&	Cash and short-term investments (che) plus .715*receivables (rect) + .547*inventory (invt) + .535* property, plant and equipment (ppent), scaled by total assets (at). Only defined for manufacturing firms (SIC $\geq$ 2000 and SIC $<4000$). Exclude the lowest tercile of manufacturing firms by total assets.\\
\hline
Tax&	Lev and Nissim&	2004&	Taxable income to income&	Ratio of Taxes paid and tax share of net income. Numerator is defined as the sum of foreign (txfo) and federal (txfed) income taxes. If either one is missing, numerator is defined as total taxes (txt) minus deferred taxes (txdi). Denominator is the product of the prevailing tax rate and net income (ib). Tax rate is .48 before 1979, .46 from 1979 to 1986, .4 in 1987, .34 between 1988 and 1992 and .35 from 1993 onwards. If net income is negative, and the numerator is positive, tax is defined as 1. Exclude if price less than 5.\\
\hline
TotalAccruals&	Richardson et al.&	2005&	Total accruals&	Before 1988: Change in net working capital ((act - che) - (lct - dlc)) plus change in net noncurrent assets ( (at - act - ivao) - (lt - dlc - dltt)) plus change in net financial assets ( (ivst + ivao - (dltt + dlc + pstk)). Starting in 1988: net income (ni) minus total, operating and investment cashflows (oancf, ivncf, fincf) plus stock sales minus repurchases and dividends (sstk, prstkc, dv)). Scaled by lagged total assets (at). Replace missings in ivao, ivst, dltt, dlc, pstk sstk, prstkc, dv with 0.\\
\hline
TrendFactor&	Han et al.&	2016&	Trend Factor&	See \citet{AndrewY.Zimmermann2020OpenPricing} section 2.1 and 2.2.\\
\hline
VolSD&	Chordia et al.&	2001&	Volume Variance&	Rolling standard deviation of monthly trading volume (vol) over the past 36 months (require at least 24 observations). Include only NYSE stocks.\\
\hline
XFIN&	Bradshaw, Richardson, Sloan&	2006&	Net external financing&	Sale of common stock (sstk) minus dividends (dv) minus purchase of common stock (prstkc) plus long-term debt issuance (dltis) minus long-term debt reductions (dltr). Scaled by total assets (at).\\
\hline
zerotrade&	Liu&	2006&	Days with zero trades 6&	In each month, count the number of days with no trades. Define zerotrade as the number of days without trades plus (the sum of monthly turnover (vol/shrout) divided by 48*10$^5$), multiplied by 21/number of trading days per month. Zerotrade is the 6-month average of that variable.\\
\hline
zerotradeAlt1&	Liu&	2006&	Days with zero trades 1&	In each month, count the number of days with no trades. Define zerotrade as the number of days without trades plus (the sum of monthly turnover (vol/shrout) divided by 48*10$^5$), multiplied by 21/number of trading days per month. Take 1-month average.\\
\hline
zerotradeAlt12&	Liu	&2006&	Days with zero trades 12&	In each month, count the number of days with no trades. Define zerotrade as the number of days without trades plus (the sum of monthly turnover (vol/shrout) divided by 48*10$^5$), multiplied by 21/number of trading days per month. Take 12-month average.\\
\hline
AdExp&	Chan et al.&	2001&	Advertising Expense&	Advertising expense (xad) over market value of equity (shrout*abs(prc)).\\
\hline
AgeIPO&	Ritter&	1991&	IPO and age	Age& is (current year - founding year from Jay Ritter's dataset). Exclude if IndIPO == 0.\\
\hline
AnalystValue&	Frankel and Lee&	1998&	Analyst Value&	Value based on a three-stage dividend discount model and analyst forecasts, scaled by market value.  See code of \citet{AndrewY.Zimmermann2020OpenPricing} for details.\\
\hline
AOP&	Frankel and Lee&	1998&	Analyst Optimism&	AnalystValue (defined above) minus IntrinsicValue (defined above), divided by abs(IntrinsicValue).\\
\hline
Beta&	Fama and MacBeth&	1973&	CAPM beta&	Coefficient of a 60-month rolling window regression of monthly stock returns minus the riskfree rate on market return minus the risk free rate (ewretd - rf). Exclude if estimate based on less than 20 months of returns.\\
\hline
BetaFP&	Frazzini and Pedersen&	2014&	Frazzini-Pedersen Beta&	Using daily data, call tempRi the sum of today's return and the past 2 days, call tempRm the sum of today's mkt return and the past 2 days mkt return.  Regress return on tempRi and tempRm using past 5 years of data, with a minimum of 3 years.  BetaFP is the square root of the r square from this regression times the ratio of the idiosyncratic stock vol and the market vol.\\
\hline
BidAskSpread&	Amihud and Mendelsohn&	1986&	Bid-ask spread&	Effective bid ask spread based on Corwin-Schulz scaled by stock price.\\
\hline
BrandInvest&	Belo, Lin and Vitorino&	2014&	Brand capital investment&	Advertising expenses (xad) divided by BrandCapital. Brand capital is computed by the perpetual inventory method. In the first year, brand capital is advertising expense divided by (.5 + .1). In subsequent years, we let brand capital depreciate with a rate of .5 and add current advertising expenses. Brand capital is scaled by total assets (at). Set to missing if advertising expense is missing.  Drop if 1st digit of sic = 4 or 6, and keep only Dec fiscal year ends.\\
\hline
Coskewness&	Harvey and Siddique&	2000&	Coskewness&	Signal is the sample counterpart of $E[\tilde{r}_{it} \tilde{r}_{mt}^2]/( SD[\tilde{r}_{it} ] SD[\tilde{r}_{mt}]^2$ where $\tilde{r}_{it}$ is the de-meaned stock excess return and $\tilde{r}_{mt}$ is the de-meaned market excess return.  Signal is computed using the past 60 months of monthly data, and using the NYSE/AMEX CRSP VW index for the market (msic).  See \citet{AndrewY.Zimmermann2020OpenPricing} code for details.\\
\hline
DebtIssuance&	Spiess and Affleck-Graves&	1999&	Debt Issuance&	Equal to 1 if debt issuance (dltis) greater 0 and 0 otherwise. Exclude if share code $> 11$ or missing book-to-market.\\
\hline
FirmAge&	Barry and Brown&	1984&	Firm age based on CRSP&	Months since start of CRSP coverage.\\
\hline
FR&	Franzoni and Marin&	2006&	Pension Funding Status&	FR = (FVPA - PBO), scaled by market value of equity. FVPA is pbnaa from 1980 to 1986, pplao + pplao from 1987 to 1997, and pplao after 1997. PBO is pbnvv from 1980 to 1986, pbpro + pbpru from 1987 to 1997, and pbpro after 1997. Exclude if price less than 5 or shrcd $> 11$.\\
\hline
GrLTNOA&	Fairfield, Whisenant and Yohn&	2003&	Growth in long term operating assets&	Annual growth in net operating assets, minus accruals. Net operating assets are (rect + invt + ppent + aco + intan + ao- ap- lco- lo) / at. Accruals are ( rect-l12.rect + invt - l12.invt + aco - l12.aco - (ap - l12.ap + lco - l12.lco) - dp ) / ((at + l12.at)/2).\\
\hline
GrSaleToGrInv&	Abarbanell and Bushee&	1998&	Sales growth over inventory growth&	Percentage growth in sales (sale) relative to average sales of t-1 and t-2, minus percentage growth in inventory (invt) relative to average inventory of t-1 and t-2. Both growth terms are calculated relative to t-1 only if t-2 is missing.\\
\hline
GrSaleTo\\GrOverhead&	Abarbanell and Bushee&	1998&	Sales growth over overhead growth&	GrSaleToGrOverHead = Percentage growth in sales (sale) relative to average sales of t-1 and t-2, minus percentage growth in administrative expenses (xsga) relative to average administrative expenses of t-1 and t-2. Both growth terms are calculated relative to t-1 only if t-2 is missing.\\
\hline
HerfAsset&	Hou and Robinson&	2006&	Industry concentration (assets)&	Three-year rolling average of the three digit industry Herfindahl index based on firm assets (at). Exclude regulated industries (4011, 4210, 4213 \& year $\leq$ 1980; 4512 \& year $\leq$ 1978, 4812, 4813 \& year $\leq$ 1982, 4900-4999 in any year).\\
\hline
High52&	George and Hwang&	2004&	52-week high&	Price (prc/cfacshr) divided by the maximum price over the previous 12 months.\\
\hline
IO\_ShortInterest&	Asquith Pathak and Ritter&	2005&	Inst own among high short interest&	Exclude all stocks with short interest (ShortInterest) below 99th percentile. IO\_ShortInterest is institutional ownership (instown\_perc).  Keep NYSE Only.\\
\hline
MomOffSeason\\11YrPlus&	Heston and Sadka&	2008&	Off season reversal years 11 to 15&	Average return in other months over the preceding 11-15 years.\\
\hline
MRreversal&	De Bondt and Thaler&	1985&	Medium-run reversal&	Stock return between months t-18 and t-13.\\
\hline
NumEarnIncrease&	Loh and Warachka&	2012&	Earnings streak length&	Number if consecutive 4-quarter increases in ibq, up to 8.\\
\hline
OperProf&	Fama and French&	2006&	operating profits / book equity&	Revenue (revt) minus cost (cogs) - administrative expenses (xsga) - interest expenses (xint), scaled by book value of equity (ceq). Exclude smallest size tercile.\\
\hline
OperProfRD&	Ball et al.&	2016&	Operating profitability R\&D adjusted&	Revenue (revt) minus cost (cogs) - (administrative expenses (xsga) - R\&D expenses (xrd)), all divided by total assets (at) in year t. Replace all variables in the numerator with 0 if they are missing. Exclude if share code is greater 11, market value of equity, BM or total assets are missing, or if SIC code between 6000 and 6999.\\
\hline
PriceDelaySlope&	Hou and Moskowitz&	2005&	Price delay coeff&	Each July regress daily stock return (ret) on market return (mktrf) in $t, t-1, \ldots, t-4$ with observations over the previous year. Define PriceDelaySlope as the ratio of ]1*beta on mktrf$_t-1$ + 2*beta on mktrf$_t-2$ + 3*beta on mktrf$_t-3$ + 4*beta on mktrf$_t-4$], and [beta on mktrf$_t$ + beta on mktrf$_t-1$ + beta on mktrf$_t-2$ + beta on mktrf$_t-3$ + beta on mktrf$_t-4$].\\
\hline
PriceDelayTstat&	Hou and Moskowitz&	2005&	Price delay SE adjusted& Each July regress daily stock return (ret) on market return (mktrf) in $t, t-1, \ldots, t-4$ using observations from July 1 of the previous year to June 30 of the current year.  Trim the highest and lowest 1\% of estimated coefficients, standard errors, and t-stats. Define PriceDelayTstat as the ratio of [1*tstat on mktrf$_t-1$ + 2*tstat on mktrf$_t-2$ + 3*tstat on mktrf$_t-3$ + 4*tstat on mktrf$_t-4$], and [tstat on mktrf$_t$ + tstat on mktrf$_t-1$ + tstat on mktrf$_t-2$ + tstat on mktrf$_t-3$ + tstat on mktrf$_t-4$]. Finally winsor the extreme 10 and 90 percentiles each month.\\
\hline
realestate&	Tuzel&	2010&	Real estate holdings&	Industry-adjusted value of real estate holdings. Real estate holdings are (fatb+fatl)/ppegt if available, and (ppenb+ppenl)/ppent otherwise.  Drop firms in 2 digit sics with $< 5$ firm or missing at or missing both ppent and ppegt. Subtract monthly industry-mean at the 2 digit SIC level.\\
\hline
RoE&	Haugen and Baker&	1996&	net income / book equity&	Net income (ni) over book value of equity (ceq). Exclude if price less than 5.\\
\hline
ShareRepurchase&	Ikenberry et al.&	1995	&Share repurchases&	Binary variable equal to 1 if stock repurchase indicated in cash flow statement (prstkc $> 0$), and 0 if prstkc = 0.\\
\hline
ShortInterest&	Dechow et al.&	2001&	Short Interest&	Short-interest from Compustat (shortint) scaled by shares outstanding (shrout). Short-interest data are available bi-weekly with a four-day lag. I use the mid-month observation to make sure data would be available in real time.\\
\hline
sinAlgo&	Hong and Kacperczyk&	2009&	Sin Stock (selection criteria)&	Using Compustat Segment data, sinAlgo is defined as a binary variable equal to 1 if at least one segment of a firm is listed as being in at least one of the following industries: sic $\geq$ 2100 \& sic $\leq$ 2199, sic $\geq$2080 \& sic $\leq$ 2085, NAICS in \{7132, 71312, 713210, 71329, 713290, 72112, 721120\}. As in the original paper, we assume that the sin stock indicator applies to the entire history and future of the identified firm. sinAlgo is equal to 0 if the firm is not identified in the CS Segment data as a sin stock and if the firm is in one of the following industries: (sic $\geq$ 2000 \& sic $\leq$ 2046) OR (sic $\geq$ 2050 \& sic $\leq$ 2063) OR (sic $\geq$ 2070 \& sic $\leq$ 2079) OR (sic $\geq$ 2090 \& sic $\leq$ 2092) OR (sic $\geq$ 2095 \& sic $\leq$ 2099) OR (sic $\geq$ 2064 \& sic $\leq$ 2068) OR (sic $\geq$ 2086 \& sic $\leq$ 2087) OR (sic $\geq$  920 \& sic $\leq$  999) OR (sic $\geq$ 3650 \& sic $\leq$ 3652) OR sic == 3732 OR  (sic $\geq$ 3931 \& sic $\leq$ 3932) OR (sic $\geq$ 3940 \& sic $\leq$ 3949) OR  (sic $\geq$ 7800 \& sic $\leq$ 7833) OR (sic $\geq$ 7840 \& sic $\leq$ 7841) OR (sic $\geq$ 7900 \& sic $\leq$ 7911) OR (sic $\geq$ 7920 \& sic $\leq$ 7933) OR (sic $\geq$ 7940 \& sic $\leq$ 7949) OR sic $==$ 7980 OR (sic $\geq$ 7990 \& sic $\leq$ 7999).\\
\hline
SP&	Barbee, Mukherji and Raines&	1996&	Sales-to-price&	Ratio of annual sales (sale) to market value of equity.\\
\hline
Spinoff&	Cusatis, Miles and Woolridge&	1993&	Spinoffs&	Spinoffs are identified as all observations in the CRSP acquisition file with valid acperm entry. Spinoff is a binary variable equal to 1 if a firm is identified in the CRSP Acquisition data and if it has at most two years of history in the CRSP stock return data. Spinoff is equal to 0 otherwise.\\
\hline
VarCF&	Haugen and Baker&	1996&	Cash-flow to price variance&	Rolling variance of (ib+dp)/mve\_c over the past 60 months (minimum 24 months data required).\\
\hline
VolMkt&	Haugen and Baker&	1996&	Volume to market equity&	Average monthly dollar trading volume (vol*abs(prc)) over the previous 12 months, scaled by market value of equity. Exclude if price less than 5.\\
\hline
VolumeTrend&	Haugen and Baker&	1996&	Volume Trend&	Rolling coefficient from regressing monthly trading volume on a linear time trend over a window of 60 months (require that at least 30 exist). Scale coefficient by 60-month average of trading volume.\\
\hline
\end{longtable}
\end{landscape}

\section{Hyperparameters used in this paper}\label{sec:CF_hyper_ch1}
\begin{table}[htbp]
\centering
\footnotesize
\resizebox{\textwidth}{!}{
\begin{tabular}{lcccccccc}
\hline
\textbf{Hyperparameter} & \textbf{fw5} & \textbf{fw4} & \textbf{fw3} & \textbf{fw2} & \textbf{fw1} & \textbf{GKX2020\_fw5} & \textbf{GKX2020\_fw4} & \textbf{GKX2020\_fw3} \\
\hline
Learning rate & [0.0007,0.001] & [0.0007,0.001] & [0.0007,0.001] & [0.0007,0.001] & [0.0007,0.001] & 0.02 & 0.017 & 0.01 \\
Batch size & \multicolumn{8}{c}{Full batch (entire training sample)} \\
Number of epochs & 8000 & 8000 & 8000 & 8000 & 8000 & 3000 & 5000 & 5000 \\
Early stopping patience & 500 & 500 & 500 & 500 & 500 & 500 & 500 & 500 \\
Regularization & L1 & L1 & L1 & L1 & L1 & L1 & L1 & L1 \\
Dropout rate & 0 & 0 & 0 & 0 & 0 & 0 & 0 & 0 \\
Seeds & 200 & 200 & 200 & 200 & 200 & 200 & 200 & 200 \\
\hline
\end{tabular}
}
\raggedright \textit{Note:} As each stock has its own optimal set of hyperparameters, more detailed parameter configurations can be provided upon request.
\end{table}

\begin{sidewaysfigure}[htbp!]
\section{Factor Importance plots of fw1, fw4 and fw5 in this chapter}\label{sec:CE_VI_plots_ch1}
\centering
\begin{subfigure}{0.32\textwidth}
  \includegraphics[width=\linewidth, height=0.22\textheight, keepaspectratio]{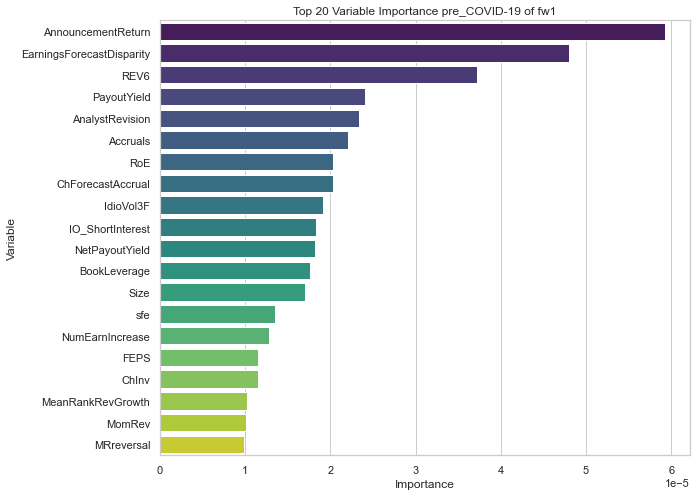}
  \caption{fw1(1911)}
\end{subfigure}
\hfill
\begin{subfigure}{0.32\textwidth}
  \includegraphics[width=\linewidth, height=0.22\textheight, keepaspectratio]{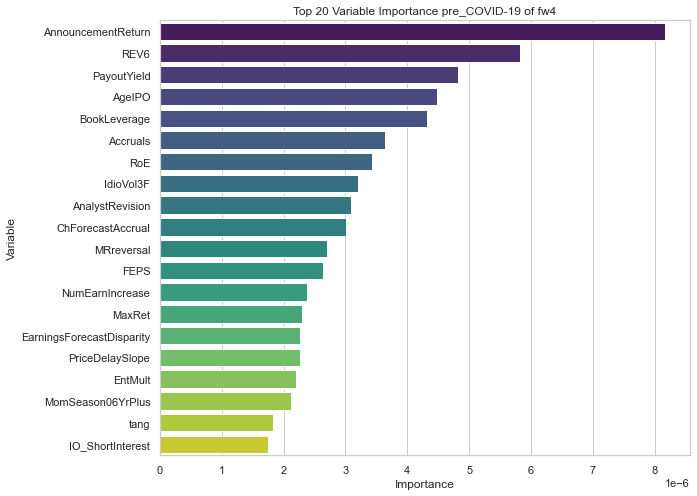}
  \caption{fw4(1911)}
\end{subfigure}
\hfill
\begin{subfigure}{0.32\textwidth}
  \includegraphics[width=\linewidth, height=0.22\textheight, keepaspectratio]{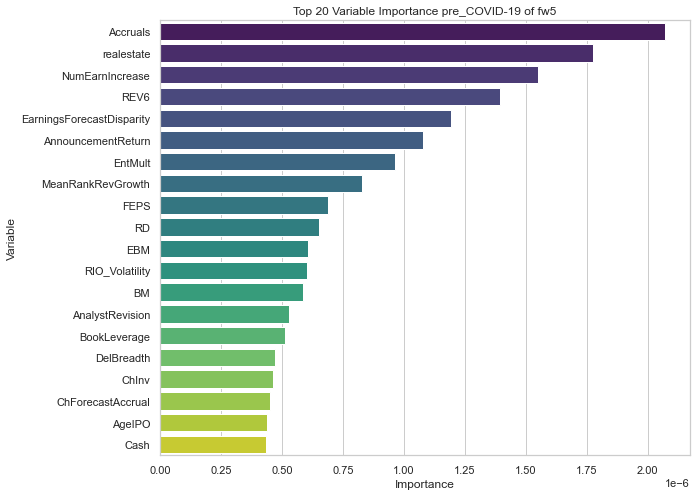}
  \caption{fw5(1911)}
\end{subfigure}

\vspace{0.1em} 

\begin{subfigure}{0.32\textwidth}
  \includegraphics[width=\linewidth, height=0.22\textheight, keepaspectratio]{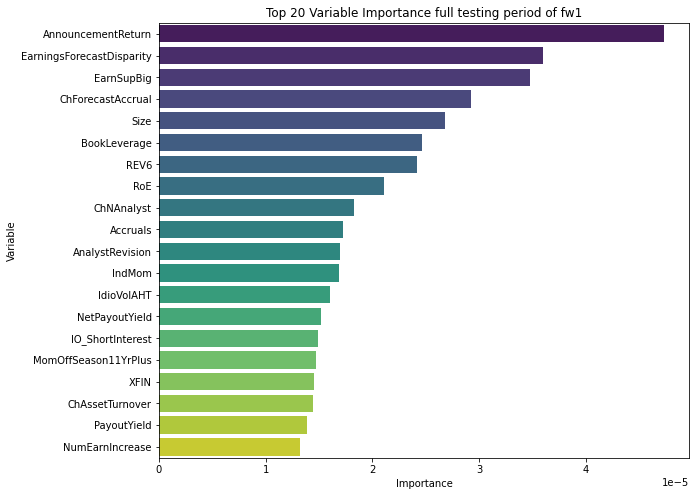}
  \caption{fw1(2112)}
\end{subfigure}
\hfill
\begin{subfigure}{0.32\textwidth}
  \includegraphics[width=\linewidth, height=0.22\textheight, keepaspectratio]{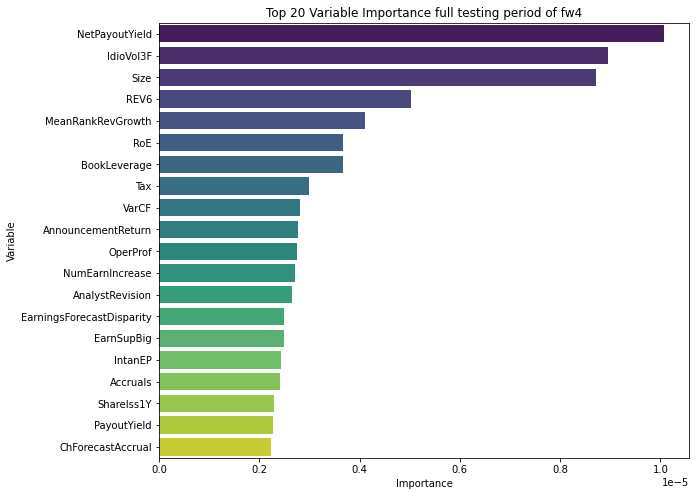}
  \caption{fw4(2112)}
\end{subfigure}
\hfill
\begin{subfigure}{0.32\textwidth}
  \includegraphics[width=\linewidth, height=0.22\textheight, keepaspectratio]{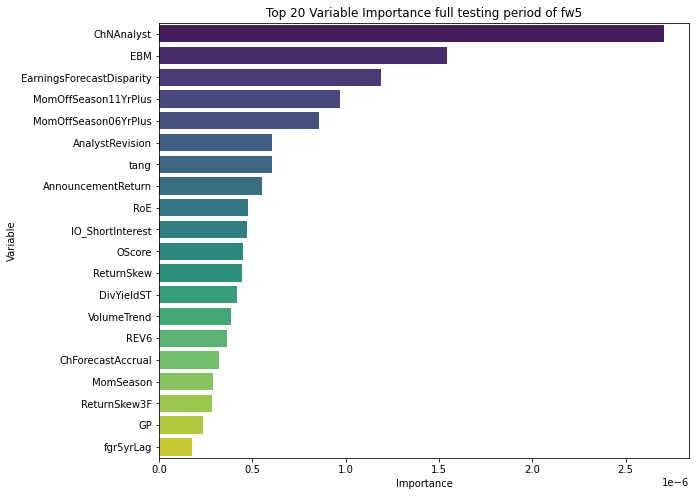}
  \caption{fw5(2112)}
\end{subfigure}

\vspace{0.1em} 

\begin{subfigure}{0.32\textwidth}
  \includegraphics[width=\linewidth, height=0.22\textheight, keepaspectratio]{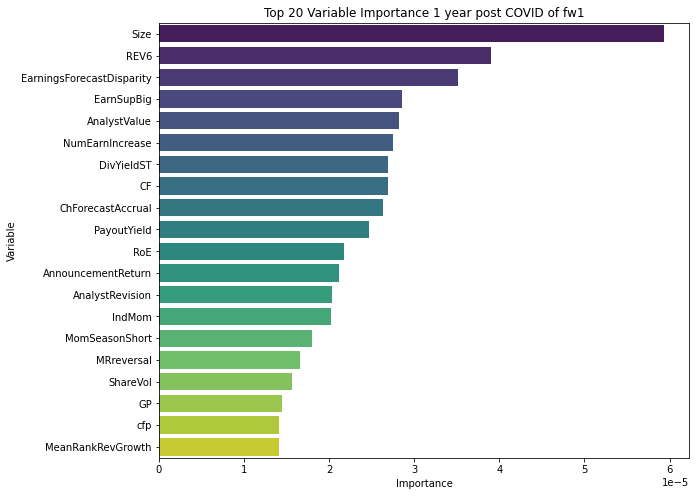}
  \caption{fw1(2212)}
\end{subfigure}
\hfill
\begin{subfigure}{0.32\textwidth}
  \includegraphics[width=\linewidth, height=0.22\textheight, keepaspectratio]{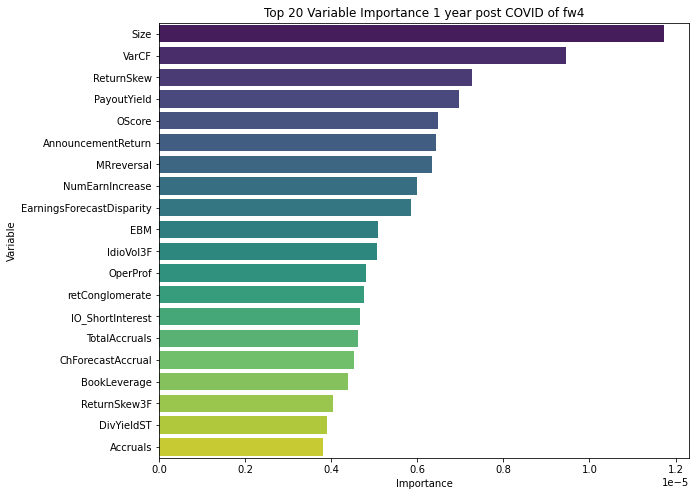}
  \caption{fw4(2212)}
\end{subfigure}
\hfill
\begin{subfigure}{0.32\textwidth}
  \includegraphics[width=\linewidth, height=0.22\textheight, keepaspectratio]{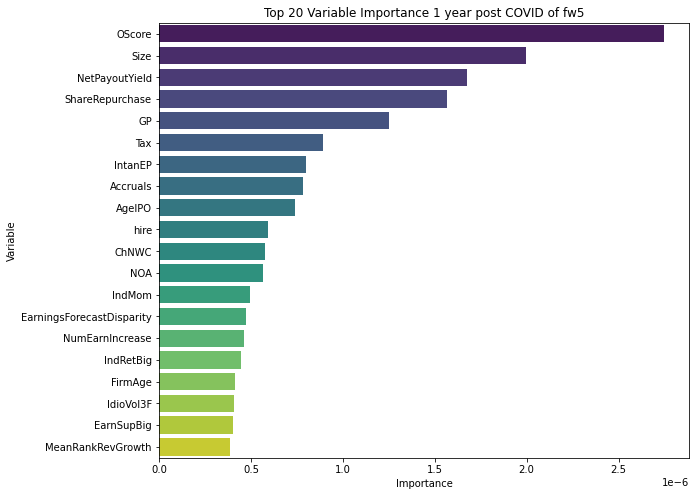}
  \caption{fw5(2212)}
\end{subfigure}
 \caption[Factor importance plots of fw1, fw4 and fw5.]{Factor importance plots of fw1, fw4 and fw5.}
  \label{fig:VI_plots_fw145_ch2}
\end{sidewaysfigure}

\clearpage
\section{Revised Python code of OOS \texorpdfstring{$R^2$}{R2} and MSE calculation}\label{sec:app_python_r2_mse_ch2}
Please note that the input actual return form and predicted return form, the row labels are the stock permno, the column labels are the time steps.\\

\begin{lstlisting}
import pandas as pd
import numpy as np
import os
import glob

path_history = r"x:\xxx\xxxx\xx.csv"
r_history_df = pd.read_csv(path_history, index_col=0)

stock_benchmarks = r_history_df.iloc[:, :-120].mean(axis=1)

path_actual_oos = r"x:\xxx\xxxx\xx.csv"
actual_all = pd.read_csv(path_actual_oos, index_col=0)

actual_sub_df = actual_all.iloc[:, -120:]

actual_values = actual_sub_df.values 

stock_benchmarks = stock_benchmarks.reindex(actual_sub_df.index)

benchmark_vector = stock_benchmarks.values.reshape(-1, 1)

input_folder = r"x:\xxx\xxxx\xx"

if not os.path.exists(details_folder):
    os.makedirs(details_folder)

summary_list = []

for file_name in os.listdir(input_folder):
    if file_name.endswith('.csv'):
        file_path = os.path.join(input_folder, file_name)
        
        pred_df = pd.read_csv(file_path, index_col=0)
        pred_df = pred_df.reindex(actual_sub_df.index)
        
        pred_df_sub = pred_df
        pred_values = pred_df_sub.values
        
        errors = actual_values - pred_values
        numerator = np.sum(errors**2, axis=1)
        
        benchmark_errors = actual_values - benchmark_vector
        
        denominator = np.sum(benchmark_errors**2, axis=1)

        stock_mse = np.mean(errors**2, axis=1)
        
        with np.errstate(divide='ignore', invalid='ignore'):
            stock_r2 = 1 - (numerator / denominator)
            stock_r2[~np.isfinite(stock_r2)] = np.nan
        
        stock_details = pd.DataFrame({
            'permno': actual_sub_df.index,
            'Benchmark_Used': stock_benchmarks.values,
            'OOS_R2': stock_r2,
            'OOS_MSE': stock_mse
        })
        
        detail_file_name = f"Detail_{file_name}"
        stock_details.to_csv(os.path.join(details_folder, detail_file_name), index=False)
        
        avg_r2 = np.nanmean(stock_r2)
        avg_mse = np.nanmean(stock_mse)
        
        summary_list.append({
            'Model': file_name,
            'Avg_OOS_R2': avg_r2,
            'Avg_OOS_MSE': avg_mse
        })
\end{lstlisting}

\end{appendices}

\clearpage
\bibliography{references}

\end{document}